\documentclass{aa}  

\usepackage{placeins}
\usepackage{graphicx}
\usepackage{lscape}
\usepackage{txfonts}
\usepackage{multicol}
\usepackage{multirow}
\usepackage{hyperref}
\hypersetup{
    colorlinks=true,
    linkcolor=blue,
    filecolor=magenta,      
    urlcolor=blue,
    citecolor=blue
    }
    
\begin{document}

   \title{The supermassive black hole population from seeding via collisions in Nuclear Star Clusters}

   \author{M. Liempi
          \inst{1,2}, D.R.G. Schleicher\inst{1,2}
          \and
          A. Benson\inst{3}
          \and
          A. Escala\inst{4}
          \and
          M.C. Vergara\inst{5}
          }

\institute{
         Dipartimento di Fisica, Sapienza Universit\`a di Roma, Piazzale Aldo Moro 5, 00185 Rome, Italy\\
        \email{mliempi2018@udec.cl}
        \and
        Departamento de Astronom\'ia, Facultad Ciencias F\'isicas y Matem\'aticas,    Universidad de Concepci\'on, Av. Esteban Iturra s/n Barrio Universitario, Casilla 160-C, Concepci\'on, Chile
         \and
             Carnegie Observatories, 813 Santa Barbara Street, Pasadena, California 91101, USA
         \and
             Departamento de Astronom\'ia, Universidad de Chile, Casilla 36-D, Santiago, Chile
         \and 
             Astronomisches Rechen-Institut, Zentrum für Astronomie, University of Heidelberg, Mönchhofstrasse 12-14, 69120, Heidelberg, Germany
             }

   \date{Received XXXX; accepted XXXX}

  \abstract
   {The coexistence of nuclear star clusters (NSCs) and supermassive black holes (SMBHs) in galaxies with stellar masses $\sim 10^{10}~$M$_\odot$, the scaling relations between their properties and properties of the host galaxy (e.g., $M_{\rm NSC}^{\rm stellar}-M_{\rm galaxy}^{\rm stellar}$, $M_{\rm BH}-M_{\rm galaxy}^{\rm stellar}$), and the fact that NSCs seem to take on the role of SMBHs in less massive galaxies and vice versa in the more massive ones, suggest that the origin of NSCs and SMBHs is related. In this study, we implement an `in-situ' NSC formation scenario, where NSCs are formed in the center of galaxies due to star formation in the accumulated gas. We explore the impact of the free parameter $A_{\rm res}$ which regulates the amount of gas transferred to the NSC reservoir, playing a crucial role in shaping the cluster's growth. Simultaneously, we include a BH seed formation recipe based on stellar collisions within NSCs in the Semi-Analytical Model (SAM) {\sc Galacticus} to explore the resulting population of SMBHs. We determine the parameter space  of the NSCs that form a BH seed  and find that in initially more compact NSCs the formation of these BH seeds is more favorable, leading to the formation of light, medium and heavy BH seeds which finally reach masses up to $\sim 10^9$~M$_\odot$ and is comparable with the observed SMBH mass function at masses above $10^8$~M$_\odot$. Additionally, we compare the resulting population of NSCs with a derived NSC mass function from the stellar mass function of galaxies from the GAMA survey at $z<0.06$ finding a well agreement in shape terms. We also find a considerable overlap in the observed scaling relations between the NSC mass, the stellar mass of the host galaxy, and the velocity dispersion, which is independent of the value of $A_{\rm res}$. However, the chi-square analysis suggests that the model requires further refinement to achieve better quantitative agreement.}

 \authorrunning{Liempi et. al}
   \maketitle


\section{Introduction}

Observations show that nuclear star clusters (NSCs) and/or supermassive black holes (SMBHs) are nearly ubiquitous in the centers of galaxies \citep{SETH2008,GRAHAM2009,NEUMAYER2011,NEUMAYER2012,NGUYEN2019}. Although the coexistence of both objects is observed in almost all galaxies with stellar masses $M_\mathrm{galaxy}^\mathrm{stellar}\simeq 10^{10}~$M$_\odot$ \citep{FILIPPENKO2003,GONZALEZ2008,SETH2008B,NEUMAYER2020}, more massive galaxies tend to host only SMBHs \citep{KORMENDY2013}, while less massive galaxies preferentially tend to host only NSCs. Sometimes therefore  the term central massive object (CMO) is used when both types of objects are present in the host galaxy \citep{FERRARESE2006}. 

It is interesting that NSCs and SMBHs follow similar scaling relations with the host-galaxy properties \citep[see e.g.,][]{WEHNER2006,GEORGIEV2016} suggesting that both undergo similar physical process in their formation and evolution. By way of illustration, previous works explored the $M_\mathrm{NSC}^\mathrm{stellar}-M_\mathrm{galaxy}^\mathrm{stellar}$ correlation \citep{BALCELLS2003,SCOTT2013,DENBROK2015,SANCHEZJANSSEN2019,GEORGIEV2016}, which is also observed for SMBHs as $M_{\rm BH}-M_\mathrm{galaxy}^\mathrm{stellar}$ \citep{DING2020, TANAKA2024}, and other correlations between  $M_\mathrm{NSC}^\mathrm{stellar}$, $M_{\rm BH}
$ and host galaxy luminosity or stellar velocity dispersion \citep{SETH2008,GULTEKIN2009,KORMENDY2013}.

Although the mechanisms for the formation of both objects are still unclear, there are different models to explain the environmental conditions from which they form.

Two principal models have been proposed to explain the formation of NSCs in the center of galaxies: (i) infall of globular clusters  into the nucleus due to dynamical friction and consecutive mergers in the center, building up a NSC \citep{TREMAINE1975,CAPUZZO1993,AGARWAL2011,ANTONINI2013,GRAY2024}, thereby explaining the deficit of massive globular clusters (GCs) in the inner part of the nucleated early-type galaxies \citep{JENNIFER2001,CAPUZZO2009}; (ii) `in-situ' star formation in the galactic nucleus, which happens when gas reaches the center of the galaxy, and NSCs are directly formed there \citep{LOOSE1982,ANTONINI2015}.

Even though the mechanism that drives the gas to the central parsecs of the galaxy is not well understood, there are several possible processes which can cause the accumulation of gas in the center of galaxies; (i) Bar-driven gas infall due to a non-axisymmetric potential  reaching the central pc \citep[e.g.,][]{SHLOSMAN1990}; (ii) the dissipative nucleation scenario, where the repeated merging of massive stellar and gaseous clumps developed from nuclear gaseous spiral arms as a consequence of the local gravitational instability leads to the accumulation of stars and gas in the nucleus \citep{BEKKI2006,BEKKI2007}; (iii)
 tidal forces in flat galaxies without NSCs act on the gas in the $0.1\%$ of the effective radius of the galaxy causing gas infall \citep{EMSELLEM2008}; (iv) the magneto-rotational instability of the disk composed of neutral gas under the effect of a weak magnetic field causes a radial gas transport towards the nucleus \citep{MILOSAVLJEVIC2004}.

Similar as for NSCs, different pathways have been proposed to explain the formation of SMBHs \citep[e.g.,][]{REES1984,VOLONTERI2010,WOODS2019,INAYOSHI2020}.  From observations of quasars at different redshifts we can conclude that the first BHs must have formed in the early universe. A direct example is the quasar survey of \cite{BANADOS2016}, containing more than 100 quasars at $5.6 \lesssim z \lesssim 6.7$ when the universe was about $1$~Gyr old. In a similar way, \cite{FAN2023} recently published a review with the quasar redshift frontier extended to $z\sim 7.6$ for more than 300 quasars.

One possible mechanism to explain the existence of those SMBHs, is the formation of light BH seeds from stellar remnants. If those seeds accrete at the Eddington rate with a standard efficiency $\epsilon \approx 0.1$, it takes at least $0.5$~Gyr to reach $\sim 10^{9}$~M$_\odot$ \citep{VOLONTERI2010}. Although these seeds can form SMBHs, the scenario is worst as most recently, the James Webb Space Telescope (JWST)\footnote{Webpage JWST:~ \url{https://webb.nasa.gov/}} detected a luminous galaxy at $z=10.6$ hosting a SMBH of $10^{6.2}$~M$_\odot$ which accretes at about five times the Eddington rate \citep{MAIOLINO2024}, providing strong evidence for the early formation of SMBHs and supporting the idea of SMBHs being the evolution of heavy BH seeds.

There are different ways to form more massive seeds: (i) the direct collapse of a massive metal-free gas cloud can lead to the formation of `heavy' BH seeds of the order of $\sim10^{5}~$M$_\odot$ \citep{REES1984,VOLONTERI2005,FERRARA2014,LATIF2015,REGAN2018}; (ii) the formation of light and medium seeds as remnants of Population III stars \citep{HAIMAN2004B,WHALEN2012,LATIF2013,LATIF2014,RIAZ2022,SINGH2023}, (iii) the formation and growth of BHs within dense NSCs due to runaway star collisions \citep{BEGELMAN1978,QUINLAN1987,EBISUZAKI2003,MILLER2002,PORTEGIES2002,PORTEGIES2004,GURKAN2004,FREITAG2006A,FREITAG2006B,ESCALA2021,VERGARA2023}.
Even though the detection of NSCs in nearby galaxies is usually through the analysis of the galaxy light profile, the detection of BHs within NSCs  requires indirect methods such as kinematic measurements of the orbits of the stars and/or gas near the BH. An example is the detection of the NSC \citep{BECKLIN1968} in the Milky Way, after which the presence of Sagitarius A$^*$ was inferred 6 years later \citep{BALICK1974}.

In this work, we study the expected population of SMBHs from the collision-based BH seeding scenario utilizing the semi-analytic model (SAM) for galaxy formation and evolution {\sc Galacticus} \citep{BENSON2012}. The advantage of SAMs is the fast exploration of the parameter space \citep{HENRIQUES2009,BENSON2010b,BOWER2010} with low computational cost compared to N-Body and/or hydrodynamical simulations. Of course, the degree of approximation depends on the implemented physics, but overall a good agreement can be achieved between semi-analytic and N-body/hydrodynamical models \citep{BENSON2001,HIRSCHMANN2012,COTE2018,MITCHELL2018}. In Sect.~\ref{Methodology}, we describe the implementation and the methodology of our study, where in subsection~\ref{NSCModel} we describe the NSC model implemented in {\sc Galacticus} and in subsection~\ref{BlackHoleFormation} the implementation of the BH seed formation scenario.  We compare our results with the observed data and summarize our findings in Sect.~\ref{results}. A final discussion and conclusions are provided in Sect.~\ref{discussion}.
\section{Methods} \label{Methodology}

We used the SAM {\sc Galacticus}\footnote{\url{https://github.com/galacticusorg/galacticus}} developed by \citet{BENSON2012}.
The functionality of {\sc Galacticus} is similar to other SAMs ---see for example {\sc Galform }\citep{COLE2000}, {\sc L-Galaxies} \citep{HENRIQUES2015}, {\sc Sage} \citep{CROTON2016} --- with the advantage of {\sc Galacticus} being highly modular. This key feature allows a rapid incorporation of new models related to galaxy formation and evolution processes as this field is constantly evolving. 

In this work, we built a distribution of dark matter (DM) merger trees in {\sc Galacticus} using the algorithm of \citet{COLE2000} which uses a modified version of the extended Press-Schechter (EPS) formalism and the branching probabilities for the merger rates of DM halos \citep{PARKINSON2008,BENSON2017}. The masses of the DM halos at $z=0$ are distributed uniformly from a power law mass distribution within the range $3.0\times 10^9-1.1\times 10^{15}~$M$_\odot$ calibrated to match the suite of MultiDark Planck cold dark matter N-body simulations\footnote{All the details about the calibration can be found here: \url{https://github.com/galacticusorg/galacticus/wiki/Constraints:-Dark-matter-halo-mass-function}}. In order to assure the resulting distribution of halo masses is representative of the expected halo mass function, {\sc Galacticus} assigns a weight for each merger tree generated, allowing us to construct corrected volume limited samples. 

The properties of DM halos are evolved forward in time until the mass resolution is reached, then the DM halos are populated with galaxies which are evolved over time in a $\Lambda$CDM universe with cosmological parameters equal to $H_0 = 67.36$ km s$^{-1}$\,Mpc$^{-1}$, $\Omega_b=0.0493$, $\Omega_\Lambda=0.6847$ and $\Omega_\mathrm{m} = 0.266 $ \citep{PLANCK2020}. We adopted a mass resolution equals to $4.86\times10^5$~M$_\odot$. We discuss the convergence of the solutions due to the mass resolution in Appendix \ref{MassResolutionTest}.

Furthermore, our simulations are based on the best match parameters constrained with observations.  We refer to best match parameters as the combination of parameters which better reproduce the observed datasets, for example, the stellar mass-halo mass relation of \citet{LEAUTHAUD2012}, the $z < 0.06$ stellar mass function of galaxies from the {\sc GAMA} survey \citep{BALDRY2012}, the $z = 2.5-3.0$ stellar mass function of galaxies from the {\sc ULTRAVISTA} survey \citep{MUZZIN2013} and the $z \cong 0.00$ black hole mass-bulge mass relation \citep{KORMENDY2013}. For a comprehensive description of the target datasets used to constrain the model, we encourage readers to read the work of \cite{KNEBE2018}.

We emphasize that the primary aim of this paper is to introduce a new model for NSC (and BH) formation and evolution within Galacticus, and explore the parameter space of NSCs which formed a BH seed under the proposed scenario. We also provide a brief comparison between the observed and predicted galaxy stellar mass functions in Appendix \ref{GalaxyMassFunctionComparison}. However, we emphasize also the preliminary nature of this comparison, considering the complex processes that contribute to the formation and evolution of NSCs and the formation of BHs within them. The comparison thus predominantly serves to determine whether the contribution is potentially relevant. For this purpose, we run a grid of parameters as detailed in Table~\ref{InitialParameters} that we will introduce jointly with the implementation in Sect.~ \ref{NSCModel}.

\subsection{Nuclear Star Cluster formation model}\label{NSCModel}

We assume an `in-situ' star formation scenario in the gas accumulated in the center of the galaxies, which is particularly well-justified for galaxies with stellar masses above $3\times10^{11}~$M$_\odot$ \citep{ANTONINI2015}. The gas mass accumulation rate  in the reservoir is correlated to the star formation rate in the bulge  \citep{GRANATO2004,HAIMAN2004,NEUMAYER2011,LAPI2014,ANTONINI2015} and is given by
\begin{equation}
\dot{M}^\mathrm{gas}_\mathrm{NSC} = A_\mathrm{res}\,\dot{M}_\mathrm{spheroid}^\mathrm{stellar}\;,\label{GasRate}
\end{equation}
where $\dot{M}_\mathrm{spheroid}^\mathrm{stellar}$ is the star formation rate in the spheroidal component of the galaxy and $A_\mathrm{res}$ is a free parameter previously reported of the order of $10^{-2}-10^{-3}$ \citep{ANTONINI2015}.

The size of the NSC scales with the square root of the dynamical mass of the system. This assumption is motivated by the size-luminosity (mass) scaling relation of the observed NSCs \citep{ANTONINI2012,ANTONINI2013}
\begin{equation}\label{massradius}
    r_\mathrm{NSC} = r_0 \cdot \sqrt{\frac{M_\mathrm{NSC}^\mathrm{dyn}}{10^{6}~{\rm M}_\odot}}\;,
\end{equation}
where $r_0$ is the mean radius of the observed NSCs set
 to $r_0=3.3$~pc \citep{NEUMAYER2020}, and $M_\mathrm{NSC}^\mathrm{dyn}=M_\mathrm{NSC}^\mathrm{gas}+M_\mathrm{NSC}^\mathrm{stellar}$ is the dynamical mass of the system.

To model the mass distribution of the NSCs we assume a S\'ersic profile \citep{SERSIC1963} with index $n=2.28$. This is motivated by the results of \citet{PECHETTI2020} who analyzed density profiles for 29 galaxies containing NSCs in a volume limited survey.

The star formation rate (SFR) of the NSC is assumed to follow the law of \citet{KRUMHOLZ2009}. 
The SFR happens in a `quiescent' mode and takes place on a timescale $t_\mathrm{SF}$ as described by \citet{KRUMHOLZ2009,SESANA2014,ANTONINI2015}, involving a fraction of cold gas ($f_c$) available for star formation:
\begin{equation}
\dot{M}_\mathrm{NSC}^\mathrm{stellar} =f_c\frac{M_\mathrm{NSC}^\mathrm{gas}}{t_\mathrm{SF}}.\label{eq:StarFormationRate}
\end{equation}

The fraction of cold gas available to form stars depends on the metallicity of the gas. At high metallicities ($Z>0.01~Z_\odot$) the fraction  is determined by the molecular gas. On the other hand, at lower metallicities ($Z<0.01~Z_\odot$), star formation takes place in the atomic phase \citep{KRUMHOLZ2012} and

\begin{equation}
    f_c = \mathrm{max}\left(2\%,1-\left[1+\left(\frac{3}{4}\frac{s}{1+\delta} \right)^{-5} \right]^{-1/5}\right),
\end{equation}
where 
\begin{eqnarray}
s  &=& \frac{\ln{(1+0.6\chi)}}{0.04\Sigma_1Z}, \\
\chi &=& 0.77(1+3.1Z^{0.365}),\\
\delta &=& 0.0712(0.1s^{-1}+0.675)^{-2.8},\\
\Sigma_1 &=& \frac{\Sigma_\mathrm{res}}{{\rm M}_\odot\,\mathrm{pc}^{-2}},\\
\Sigma_\mathrm{res} &=& \frac{M_\mathrm{NSC}^\mathrm{gas}}{4\pi r_\mathrm{NSC}^{2}}.
\end{eqnarray}

The timescale ($t_\mathrm{SF}$) is obtained assuming that star formation happens in clouds \citep{KRUMHOLZ2009,SESANA2014,ANTONINI2015} and is given by
\begin{equation}
t_\mathrm{SF}^{-1} = (2.6~\mathrm{Gyr})^{-1} \times \left\{ \begin{array}{cl}
\left( \frac{ \Sigma_\mathrm{res}}{\Sigma_\mathrm{th}}\right)^{-0.33}&,\; {\rm if }\;\; \Sigma_\mathrm{res} <\Sigma_\mathrm{th},\\
\left( \frac {\Sigma_\mathrm{res}}{\Sigma_\mathrm{th}}\right)^{+0.34}&,\; {\rm otherwise},
\end{array}
 \right.
\end{equation}
where $\Sigma_\mathrm{th}=85~$M$_\odot\,\mathrm{pc}^{-2}$.

Finally, we estimate the age of the system ($t_H$) using the mass weighted age due to the fact that young stars outshine older stars. This is a reasonable choice as the luminosity of NSCs is dominated by the light of the younger stars of the system \citep{BOEKER2003,WALCHER2006,DAMETTO2014}. The expression used to compute the age is given by the integral of the SFR of the NSC

\begin{equation}\label{eqAGE}
    t_H = t - \int_0^tdt^\prime t^\prime\dot{M}_\mathrm{NSC}^\mathrm{stellar}(t^\prime) \int_0^{t}dt^\prime\dot{M}_\mathrm{NSC}^\mathrm{stellar}(t^\prime),
\end{equation}
where $t$ is the present time and $\dot{M}_\mathrm{NSC}^\mathrm{stellar}(t^\prime)$ is the star formation rate  of the NSC at time $t^\prime$.

\subsection{Black hole formation and evolution model}\label{BlackHoleFormation}

As we mentioned before, the formation pathway of SMBHs is still unclear. Here in this paper we focus on the exploration of the formation pathway via collisions in NSCs, as motivated by the study of \citet{ESCALA2021}, who find that the observed NSCs are in a regime where collisions are not relevant over timescales corresponding to the lifetime of the system while well-resolved observed SMBHs are found in regimes where collisions are expected to be dynamically important, suggesting that SMBHs are formed from failed NSCs with average collision timescale shorter than the age of the system ($t_\mathrm{coll}\leq t_\mathrm{H}$). In the following subsections, we will describe the different ingredients of our model.

\subsubsection{Black hole seeding: Collision timescale and critical mass} \label{CollisionTimescale}

The formation of a massive BH within  dense stellar clusters has been discussed for a long time \citep[e.g.,][]{BEGELMAN1978,PORTEGIES2002,GLEBBEEK2009,DICARLO2021}. It is possible to estimate the collision timescale $t_\mathrm{coll}$ in any system with a large number of particles \citep{BINNEY2008}. The expression is given by 
\begin{equation}
    t_\mathrm{coll}=\frac{\lambda}{\sigma},
\end{equation}
where $\lambda$ is the mean free path and $\sigma$ is the characteristic velocity dispersion. The mean free path can be probabilistically defined as
\begin{equation}
    \lambda = \frac{1}{n\Sigma_0},
\end{equation}
where $n$ is the number density of stars and $\Sigma_0$ is the effective cross section \citep{LANDAU1980}. The velocity dispersion  of a virialized NSC is $\sigma=\sqrt{\frac{GM_\mathrm{NSC}^\mathrm{stellar}}{r_\mathrm{NSC}}}$, with $M_\mathrm{NSC}^\mathrm{stellar}$ the stellar mass of the NSC, $r_\mathrm{NSC}$ the radius of the NSC, and $G$ the  gravitational constant. If the NSC is composed of equal-mass stars with mass $M_\star$, and assuming a uniform distribution, the number density is $n=\frac{3}{4\pi}\frac{M_\mathrm{NSC}^\mathrm{stellar}}{ M_\star r_\mathrm{NSC}^3}$. This allows us to re-write the collision timescale as follows:
\begin{equation}
     t_\mathrm{coll}= \frac{4\pi }{3} \frac{M_\star r_\mathrm{NSC}^3}{\Sigma_0}\sqrt{\frac{r_\mathrm{NSC}}{GM_\mathrm{NSC}^\mathrm{stellar}}}. \label{collision}
\end{equation}
The Eq. \ref{collision} depends on the effective cross section ($\Sigma_0$) given by
\begin{equation}
    \Sigma_0 = 16\sqrt{\pi} R_\star^2(1+\Theta),
\end{equation}
where $R_\star$ is the radius of a single star, and $\Theta$ is the Safronov number \citep{BINNEY2008} given by
\begin{equation}
    \Theta = 9.54\left(\frac{M_\star ~ {\rm R}_\odot}{{\rm M}_\odot~ R_\star}\right)\left(\frac{100~\mathrm{km}\,\mathrm{s}^{-1}}{\sigma}\right)^2.
\end{equation}

It is possible to define a critical mass ($M_\mathrm{crit}$) as the mass for which the collision timescale becomes equal to the age of the system ($t_\mathrm{coll} = t_\mathrm{H}$), assuming stellar velocities corresponding to virial equilibrium \citep{VERGARA2023}:
\begin{equation}
    M_\mathrm{crit}(r_\mathrm{NSC}) = r_\mathrm{NSC}^{\frac{7}{3}}\left( \frac{4\pi M_\star}{3\Sigma_0 t_\mathrm{H}G^{\frac{1}{2}} } \right)^{\frac{2}{3}}.\label{CriticalMass}
\end{equation}
It was numerically determined by \citet{VERGARA2023} that even in systems where $t_\mathrm{relax} \leq t_\mathrm{coll}$ a significant fraction ($10-50\%$) of the final cluster mass is turned into a central massive object, when the ratio of cluster mass over critical mass is of order $1$ (i.e., for systems with collision timescale of the order of the age of the system).

We include an efficiency parameter $\epsilon_r$ in order to rescale the radius $r_\mathrm{NSC}$ as $r_\mathrm{NSC}\longrightarrow \epsilon_r r_\mathrm{NSC}$, with $0<\epsilon_r \leq 1$, and study the critical mass of initially more compact NSCs. This is motivated by observational and computational studies of young massive star clusters with initial sizes less than $0.3$~pc which can expand more than 10 times over their evolution \citep{BANERJEE2017}. Additionally, we assume our NSCs are composed by sun-like stars ($M_\star=1~{\rm M}_\odot$ and $R_\star = 1~{\rm R}_\odot$).

\subsubsection{Implementation of the black hole seeding and its evolution in {\sc Galacticus}} \label{BlackHoleFormationEvolution}

In the standard implementation of {\sc Galacticus}, it is assumed that each galaxy contains an initial  mass BH with mass $M_\bullet$ specified in the initial parameter file. As in this study we aim to explore the consequences of a physically motivated, collision based BH formation recipe, we fix the initial BH mass as $M_\bullet = 10~$M$_\odot$ but do not allow it to accrete gas until a new BH seed is created due to stellar collisions within NSCs, with the new seed then replacing the previous one.

A new BH seed in the center of the galaxy is formed if the stellar mass of the NSC is larger than the critical mass ($ M_\mathrm{NSC}^\mathrm{stellar} \geq M_\mathrm{crit}$), and the stellar mass of the NSC is larger than a mass threshold,  $M_\mathrm{NSC}^\mathrm{stellar} \geq M_\mathrm{threshold}$, introduced to avoid the collapse of nonphysical NSCs, as the concept of the critical mass implicitly assumes that the stellar cluster is well-sampled and we assume here that it requires a minimum of at least around 1000 stars. The mass of the newly formed seed is 
\begin{equation}
    M_\bullet = \epsilon_\bullet M_\mathrm{NSC}^\mathrm{stellar},
\end{equation}
where $\epsilon_\bullet=0.5$ is a free efficiency parameter. Note that each NSC is allowed to form only one BH seed  (unless it undergoes  a galaxy merger). The new BH seed merges with the initial seed and starts to accrete gas from a radiatively efficient thin disk as described by \cite{SHAKURA1973}. Thereupon the accretion is given by
\begin{equation}
    \dot{M}_\bullet = (1-\epsilon_\mathrm{rad}-\epsilon_\mathrm{jet})\dot{M}_0,
\end{equation}
where~$\epsilon_\mathrm{rad}$ is the radiative efficiency of the accretion flow,~ $\epsilon_\mathrm{jet}$ is the efficiency by which accretion power is converted to jet power, and $\dot{M}_0$ is the rest mass accretion rate computed assuming Bondi-Hoyle-Lyttleton accretion \citep{EDGAR2004}.

 For our accretion disk model, the~$\epsilon_\mathrm{rad}$ term is computed assuming that material falls into the BH without further energy loss from the innermost stable circular orbit (ISCO). The explicit equation is 
 \begin{equation}
     \epsilon_\mathrm{rad}= 1-E_\mathrm{ISCO},
 \end{equation}
 where $E_\mathrm{ISCO}$ is the specific energy (computed internally by {\sc Galacticus} in physical or gravitational units) of the BH. The jet power ($\epsilon_\mathrm{jet}$) is computed using the expressions 4 and 5 from \cite{MEIER2001}. We use equation 4 for (Schwarzchild) low spin BHs, while equation 5 is used for (Kerr) high spin BHs instead. 
 
 The rest mass accretion rate considers the contribution from the spheroid, the Circumgalactic Medium (CGM), and the NSC component, where each component is enhanced by a factor $\alpha$. The expression is given by
\begin{equation}
    \dot{M}_0 = \alpha_\mathrm{spheroid}\dot{M}_{\bullet, \mathrm{spheroid}} + \alpha_\mathrm{CGM}\dot{M}_{\bullet, \mathrm{CGM}} +\alpha_\mathrm{NSC}\dot{M}_{\bullet, \mathrm{NSC}},\label{eq:accretion}
\end{equation}
where $\alpha$ is equal to $5$ for the spheroid and the NSC, and $6$ for the CGM. Our Bondi-Hoyle-Lyttleton accretion model assumes an stationary accreting BH in a static uniform gaseous medium, which implies that the relative velocity is v$= 0$~km\,s$^{-1}$, and the sound speed of the gas must be estimated. In order to estimate the sound speed, the gas of all the components is assumed to follow the ideal gas law. For the spheroidal component, the assumed temperature is $T_\mathrm{spheroid}=100~$K due to the absence of significant heating mechanisms. The temperature of the gas in the NSC component has been estimated to be around $T_\mathrm{NSC}=100~$K \citep{TIELENS2010} from observations. Finally we compute the temperature of the CGM  assuming an isothermal profile for the halo with the temperature equal to the virial temperature. This allows us to estimate $c_s$ as
\begin{equation}
    c_s = \sqrt{\frac{5}{3}\frac{kT}{m}},
\end{equation}
with $k$ the Boltzmann constant, $T$ the temperature of the gas and $m$ the mean mass of a single molecule.

\begin{table}[!h]
\centering 
\caption{Initial parameters for our semi-analytic model.}
\begin{tabular}{c|c|c|c|c}  
\hline\hline  
\multicolumn{1}{c}{Model} & \multicolumn{1}{c}{$A_\mathrm{res}$}   & \multicolumn{1}{c}{$\epsilon_\bullet$} & \multicolumn{1}{c}{$\epsilon_r$ }& \multicolumn{1}{c}{$M_\mathrm{threshold}$} \\   
\multicolumn{1}{c}{-} & \multicolumn{1}{c}{-} & \multicolumn{1}{c}{-} &  \multicolumn{1}{c}{-}& \multicolumn{1}{c}{[M$_\odot$]}\\
\multicolumn{1}{c}{(1)} & \multicolumn{1}{c}{(2)}   & \multicolumn{1}{c}{(3)} & \multicolumn{1}{c}{(4) }& \multicolumn{1}{c}{(5)} \\  \hline                    
   A1 & \multirow{8}{*}{$1.0\cdot 10^{-1}$}& \multirow{8}{*}{$0.5$}  &1& \multirow{4}{*}{$10^3$} \\
   A2 & & & 0.5 &  \\
   A3 & & & 0.2 &  \\
   A4 & & & 0.1 &  \\ \cline{4 -5}
   A5 & & & 1   & \multirow{4}{*}{$10^4$}\\
   A6 & & & 0.5 &  \\
   A7 & & & 0.2 &  \\
   A8 & & & 0.1 &  \\
   \hline 
   B1 & \multirow{8}{*}{$9.0\cdot 10^{-2}$}& \multirow{8}{*}{$0.5$}& 1   & \multirow{4}{*}{$10^3$}   \\
   B2 & & & 0.5 & \\
   B3 & & & 0.2 & \\
   B4 & & & 0.1 & \\ \cline{4-5}
   B5 & & & 1   & \multirow{4}{*}{$10^4$} \\
   B6 & & & 0.5 &  \\
   B7 & & & 0.2 &  \\
   B8 & & & 0.1 &  \\
   \hline 
   C1 &  \multirow{8}{*}{$5.0\cdot 10^{-2}$}& \multirow{8}{*}{$0.5$} & 1   & \multirow{4}{*}{$10^3$} \\
   C2 & & & 0.5 & \\
   C3 & & & 0.2 & \\
   C4 & & & 0.1 & \\ \cline{4-5}
   C5 & & & 1 & \multirow{4}{*}{$10^4$} \\
   C6 & & & 0.5 & \\
   C7 & & & 0.2 & \\
   C8 & & & 0.1 & \\
   \hline 
   D1 & \multirow{8}{*}{$1.0\cdot 10^{-2}$}& \multirow{8}{*}{$0.5$} &1&  \multirow{4}{*}{$10^3$}\\
   D2 & & & 0.5 & \\
   D3 & & & 0.2 & \\
   D4 & & & 0.1 & \\ \cline{4-5}
   D5 & & & 1   &\multirow{4}{*}{$10^4$} \\
   D6 & & & 0.5 & \\
   D7 & & & 0.2 & \\
   D8 & & & 0.1 & \\
   \hline 
   E1 & \multirow{4}{*}{$9.0\cdot 10^{-3}$}& \multirow{4}{*}{$0.5$}  & 1   &  \multirow{4}{*}{$10^3$}\\
   E2 & & & 0.5 & \\
   E3 & & & 0.2 & \\
   E4 & & & 0.1 & \\
   \hline 
   F1 & \multirow{4}{*}{$5.0\cdot 10^{-3}$} &\multirow{4}{*}{$0.5$}& 1 & \multirow{4}{*}{$10^3$}\\
   F2 & & & 0.5 & \\
   F3 & & & 0.2 & \\
   F4 & & & 0.1 & \\
   \hline 
   G1 & \multirow{4}{*}{$1.0\cdot 10^{-3}$}& \multirow{4}{*}{$0.5$} & 1 &\multirow{4}{*}{$10^3$}\\
   G2 & & & 0.5 & \\
   G3 & & & 0.2 & \\
   G4 & & & 0.1 & \\
   \hline 
   H1 & \multirow{4}{*}{$1.0\cdot 10^{-4}$} & \multirow{4}{*}{$0.5$}& 1   & \multirow{4}{*}{$10^3$}\\
   H2 & & & 0.5 & \\
   H3 & & & 0.2 & \\
   H4 & & & 0.1 & \\
   \hline \hline           
\end{tabular}
\tablefoot{(1) Corresponds to the name of the model;(2) free parameter regulating the transfer of gas to the NSC gas reservoir; (3) efficiency parameter describing the fraction of the stellar mass converted into a BH seed once the cluster exceeds the critical mass; (4) efficiency parameter described in Sect. \ref{CollisionTimescale}; (5) denotes the minimum NSC mass adopted in the respective model.}\label{InitialParameters}
\end{table}

\section{Results} \label{results}

In this section we present the results of the simulations listed in Table~\ref{InitialParameters} and provide a comparison with observational results. First we explore the obtained scaling relations with the host galaxy in subsection \ref{nschost}. In subsection \ref{MF} we discuss the NSC mass function predicted with {\sc Galacticus} and the observed ones. In subsection \ref{ParameterSpace} we analyze the properties of the NSCs which satisfied the conditions to form a BH seed under this scenario within. We present the resulting black hole mass function in subsection \ref{BHMF}.

\subsection{Nuclear star clusters and host galaxies}\label{nschost}

\begin{figure}[!h]
    \centering
    \includegraphics[width=\hsize]{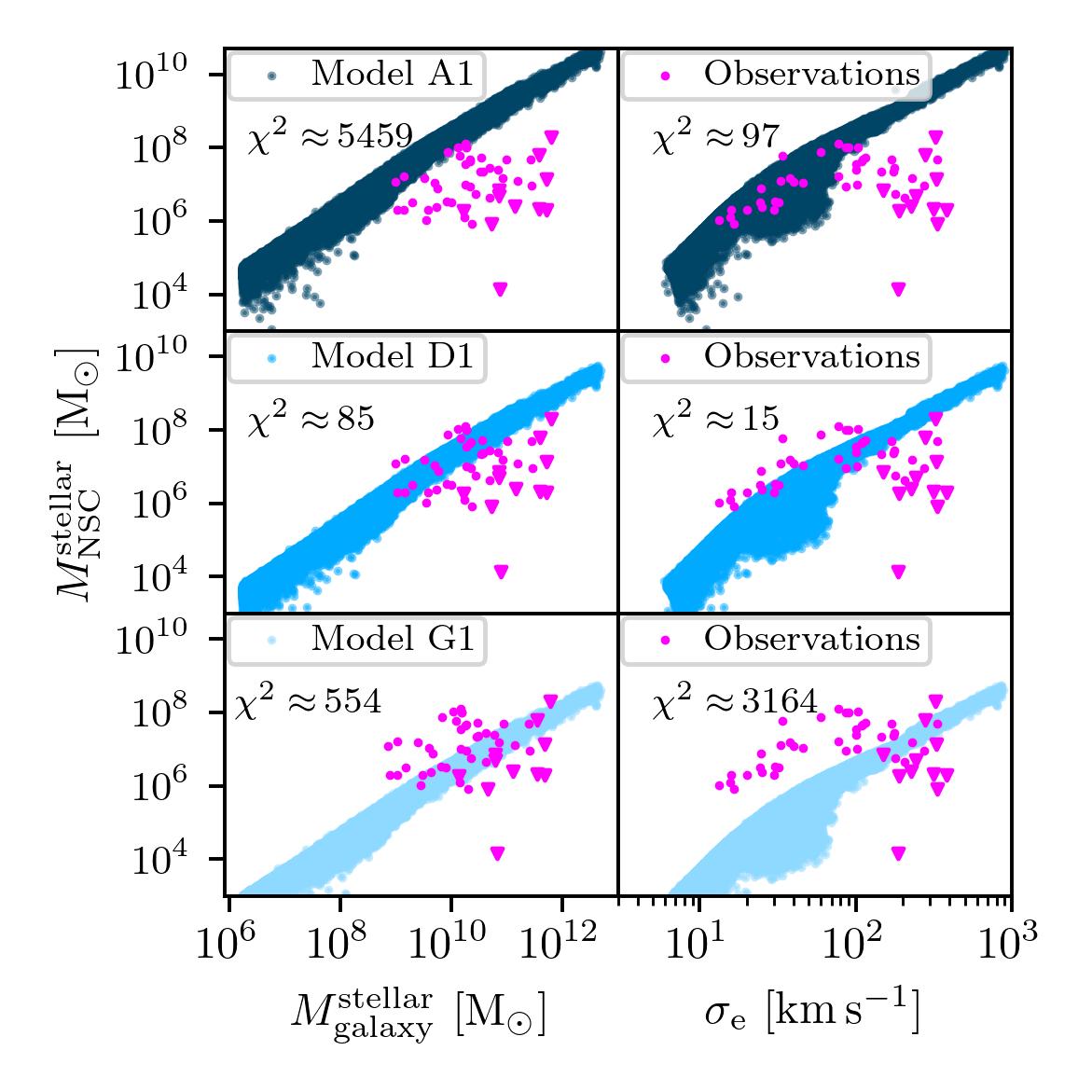}
    \caption{Left panel: Stellar mass of the NSC as a function of the stellar mass of the galaxy, for models A1, D1 and G1. Right panel: Stellar mass of the NSCs as a function of the stellar velocity dispersion of the galaxy for the same models. Magenta dots correspond to observed NSCs following Table 3 of  \citet{NEUMAYER2020}, where we removed galaxies where the stellar mass of the galaxy is not included and we include the stellar velocity dispersion available in the literature. Although, magenta dots corresponds to galaxies where the stellar mass of the NSC is well determined while magenta triangles represent an upper limit instead. The corresponding table is available in Appendix \ref{appendixD}.  Each panel includes the chi-squared test value of the model for reference.}
    \label{NSCGalaxyA}
\end{figure}

In this subsection we analyze the properties of NSCs hosted in galaxies with  $M_\mathrm{galaxy}^\mathrm{stellar} \geq 10^6~$M$_\odot$. In Fig. \ref{NSCGalaxyA}, we provide the stellar masses of NSCs as a function of the stellar mass of the host galaxy and as a function of the velocity dispersion of the bulge. For this purpose, we focus on the models A1, D1, and G1, corresponding to values of $A_{\rm res}=10^{-1}, 10^{-2}, 10^{-3}$, respectively. We checked that this is the main parameter affecting the NSC mass function and that other parameters like $\epsilon_\bullet$ or $\epsilon_r$ have a very minor effect in comparison.

In all models, we find a characteristic dependence of the stellar mass of the NSC on the stellar mass of the host galaxy, that approximately corresponds to a power-law  though with a spread of about one order of magnitude. In model A1 that we adopt as our reference case, the most massive NSCs with about $10^{10}$~M$_\odot$ are formed in galaxies with stellar masses of about $10^{12}$~M$_\odot$; NSCs in the range of $10^6-10^{7}$~M$_\odot$ form in galaxies with stellar masses of $\sim10^8-10^9$~M$_\odot$. Comparing models A1 with D1 and G1, we see that the NSC masses at a fixed stellar mass of the galaxy are roughly proportional to the parameter $A_{\rm res}$, so a decrease by an order of magnitude decreases the NSC masses by the same factor. 

On the right-hand side of the panel, the mass of the NSCs is shown as a function of the velocity dispersion of the host galaxy. For model A1,  we find a similar but somewhat steeper relation between the stellar mass of the NSC and the velocity dispersion of the host galaxy, with NSC masses of $10^{10}$~M$_\odot$ appearing for velocity dispersions of $\sim1000$~km~s$^{-1}$, while NSC masses in the range of $10^6-10^8$~M$_\odot$ are obtained for velocity dispersions around 20-40~km~s$^{-1}$. We also note an enhancement of the spread in NSC masses at low velocity dispersions. As on the left-hand side of the figure, we find that the NSC masses tend to decrease proportional to the parameter $A_{\rm res}$. 

We compare the model predictions with the observed relations of NSC clusters and their host galaxies as reported by \citet{NEUMAYER2020}. Within Table~\ref{galaxyprop} in our appendix, we have extended their Table 3 including the velocity dispersion of the host galaxy from the literature. By incorporating these data points in Fig.~\ref{NSCGalaxyA}, we find that our models overlap considerably with the observed population. The chi-squared value is reported in each panel. Model D1 best represents the observations. The values obtained suggest that the model is still somewhat oversimplified, as we neglect the growth via mergers with Globular Clusters, and even the `in-situ' channel itself may have an environmental dependence. An improvement of the model thus very likely requires an improved understanding of NSC formation itself. We further conclude that the observed population very likely does not correspond to one particular model set with a fixed value of $A_{\rm res}$, but rather the value of $A_{\rm res}$ may depend on the host galaxy and on its specific efficiency of transporting gas into the center, which potentially could be influenced by their galaxy type, the formation history as well as through effects from their environment. 

 Furthermore, we emphasize that fitting both relations well is effectively not feasible and shows some tension due to the current simplifications in our model. The contribution of mergers with globular clusters is currently not included, but may influence the velocity dispersions. The statistical nature of this process may further influence the dispersion overall.

\subsection{Nuclear Star Cluster mass function} \label{MF}

\begin{figure}[h!]
    \centering
\includegraphics[width=\hsize]{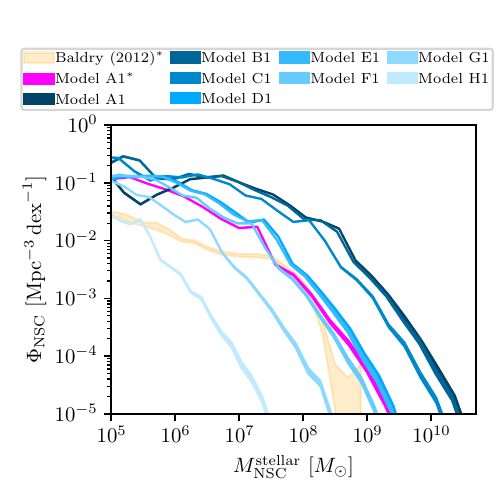}
    \caption{NSC stellar mass function for models listed in Table \ref{InitialParameters} compared to the observed NSC mass function derived from the $M_\mathrm{NSC}^\mathrm{stellar}-M_\mathrm{galaxy}^\mathrm{stellar}$ correlation. The orange region is the observed NSC mass function derived from the galaxy stellar mass function of \cite{BALDRY2012}, while the magenta line is the NSC mass function derived from the galaxy stellar mass function for model A1, both are denoted with $^*$ in the legend. The error in simulations is estimated as $\Phi_\mathrm{NSC}/\sqrt{N}$, where $N$ is the number of NSCs in the bin.}
    \label{Massfunction}
\end{figure}

The NSC mass function at $z=0$ obtained through {\sc Galacticus} via the models A1, B1, C1, D1, E1, F1, G1 and H1 is presented in Fig.~\ref{Massfunction}. As already mentioned in the previous subsection, the main parameter regulating the masses of the obtained NSCs is $A_{\rm res}$, which is varied here in a range from $10^{-4}$ up to $10^{-1}$. The shape of the NSC mass function is found to be approximately flat in the range of low and moderate NSC cluster masses up to a characteristic value $M_c$, 
which depends on the value of $A_{\rm res}$. More precisely, particularly in the NSC mass range of $10^5-10^6$~M$_\odot$, increasing the value of $A_\mathrm{res}$ decreases the population of NSCs in this mass range ($10^3-10^6$~M$_\odot$) as the NSCs tend to grow and shift to higher masses. For masses above $10^6$~M$_\odot$ up to $M_c$, the mass function scales roughly as $M_{\rm NSC}^{-1}$. Then for NSC masses above $M_c$, the mass function shows a typical scaling as a power law proportional to $M_{\rm NSC}^{-3/2}$. The value of $M_c$ corresponds to about $2\times10^8$~M$_\odot$ for $A_{\rm res}=10^{-1}$ and becomes as low as $2\times10^{5}$~M$_\odot$ for $A_{\rm res}=10^{-4}$.

Due to the lack of a uniform and complete sample of NSC in the Local Universe, we derive the NSC mass function from the stellar galaxy mass function using the $M_\mathrm{NSC}^\mathrm{stellar}-M_\mathrm{galaxy}^\mathrm{stellar}$ correlation. As mentioned in the introduction, NSCs are known to correlate with different properties of the host galaxy. Actually, the mass of the NSC scales with the stellar mass of the host galaxy as $M_\mathrm{NSC}^\mathrm{stellar}\approx 10^{-3} M_\mathrm{galaxy}^\mathrm{stellar}$ independent of the galaxy type \citep{GEORGIEV2016}. We used the stellar galaxy mass function of \cite{BALDRY2012} to derive the NSC mass function scaling the masses as $M_\mathrm{NSC}^\mathrm{stellar}= 10^{-3} M_\mathrm{galaxy}^\mathrm{stellar}$.

Their sample contains galaxies with stellar masses above $10^{8}$~M$_\odot$ in an area of 143 {\rm deg}$^2$ from the first three years of the {\rm GAMA} survey. The survey is limited to galaxies with  $r < 19.4~ \mathrm{mag}$ over two-thirds and $19.8 ~\mathrm{mag}$ over one-third of the area. The galaxy mass function is available in Table 1 of \cite{BALDRY2012}.  In addition to the magnitude limit,  the {\sc GAMA} survey contains an implicit  surface brightness limit which affects the faint end of the galaxy mass function   \citep{PHILLIPPS1986,CROSS2002}.  Specifically, \cite{GELLER2011} found a linear relation between the surface brightness and $M_r$ for the blue population. This relation falls in the {\sc GAMA } survey for $M^{\rm stellar}_{\rm galaxy} < 10^8$~M$_\odot$, where the authors assumed that their estimated mass function becomes a lower limit. Following this scheme, the derivation of the NSC mass function includes the $1-\sigma$ uncertainties in the orange area.

Model A1$^*$ is over the NSC mass function derived from the work of \cite{BALDRY2012} by a factor of $10$ at $10^5$~M$_\odot$. Although, this is not an issue, as the mass function should be considered as a lower limit for $M_{\rm NSC}^{\rm stellar} <10^5~$M$_\odot$. Furthermore, this method predicts maximum NSC masses about $\sim10^9$~M$_\odot$ which is consistent with observations and the high end of the mass function of our models D.

In general, we found that there is a shift of the NSC mas function derived from the galaxy stellar mass function (model A1$^*$ in Fig. \ref{Massfunction}), predicting less massive NSCs. This could suggest that the free parameter which regulates the gas transfer rate ($A_\mathrm{res}$) in the `in-situ' scenario is not constant over the time and may depend on environmental factors.

We overall find an underestimation between the observed NSC mass function including its uncertainties with the {\sc Galacticus} models. Although, there is an overlap between the observed and models A1 and B1 at $M_\mathrm{NSC}^\mathrm{stellar}\sim 5\times  10^7-5\times10^8$~M$_\odot$, but even the other models show relevant crosses with the observed NSC mass function in NSC masses about $5\times 10^{8}$~M$_\odot$. There is a clearly deviation at low masses, which however could very likely be explained by the absence of the GC migration formation mechanism which contributes mainly to the formation of less massive NSC \citep{NEUMAYER2020}. The lowest-mass NSC detected so far has $10^{4.15}$~M$_\odot$ \citep{NEUMAYER2012} and currently very few such objects are known. Also from a conceptual point of view, it may be somewhat arbitrary to define what should be considered as a NSC or not within the low-mass range. On the other hand, the most massive NSC is is NGC 4461  with a stellar mass of $\approx 2.5 \times 10^9$~M$_\odot$ \citep{SPENGLER2017}.  This does not necessarily indicate an upper limit on the mass of NSCs, though their abundance of course becomes increasingly rare towards higher masses.  For our present purposes, it is essentially important that the NSC mass function is reproduced well within the main mass range that is responsible for the formation of supermassive black holes and shows considerable overlap with the observed NSC mass function.

We estimated the discrepancy between models and observations using the same procedure applied to the galaxy mass function, as described in Appendix \ref{GalaxyMassFunctionComparison} and introduced earlier in Sect. \ref{Methodology}.

We find models A1, B1, and  C1 overestimate, on average, the mass function by factors of $1.14\pm 0.56$~dex, $1.17\pm0.51$~dex, and $1.08\pm0.40$~dex, respectively. Conversely, models D1, E1 and F1  overestimate the mass function by factors of less than $1$~dex, specifically  $0.75\pm 0.25$~dex, $0.72\pm 0.25$~dex, and $0.55\pm 0.34$~dex, respectively. Finally, models G1 and H1 underestimate the mass function,on average, by factors of $-0.05\pm 0.76$~dex, and $-0.98\pm1.24$~dex, respectively. 

Although no single model accurately reproduces the NSC mass function derived from the galaxy stellar mass function across all mass ranges, we highlight that models D1, E1, and F1 provide a good fit to the high-mass end of the NSC mass function. This aligns with the current understanding of NSC formation mechanisms in massive galaxies, where the `in-situ' formation scenario is typically invoked. 

\subsection{Nuclear Star Clusters at critical mass} \label{ParameterSpace}

\begin{table*}[!h]
\centering 
\caption{Range of BH masses formed in our semi-analytical models.}
\begin{tabular}{cccc}  
\hline\hline  
Model & $M_\mathrm{seed}$ & Model &$M_\mathrm{seed}$  \\
- & [M$_\odot$] & - & [M$_\odot$]\\
(1) & (2) & (1) &(2) \\
\hline                       
   A1 & $-$                             & D1 & $-$                             \\
   A2 & $5.00\cdot 10^2-1.41\cdot 10^3$ & D2 & $-$ \\
   A3 & $5.00\cdot 10^2-1.67\cdot 10^4$ & D3 & $5.00\cdot 10^2-1.93\cdot 10^4$ \\
   A4 & $5.00\cdot 10^2-1.70\cdot 10^5$ & D4 & $5.00\cdot 10^2-1.71\cdot 10^5$ \\ 
   A5 & $-$                             & D5 & $-$ \\
   A6 & $-$                             & D6 & $-$ \\
   A7 & $5.00\cdot 10^3-1.99\cdot 10^4$ & D7 & $5.00\cdot 10^3-1.97\cdot 10^4$ \\
   A8 & $5.00\cdot 10^3-1.70\cdot 10^5$ & D8 & $5.00\cdot 10^3-1.71\cdot 10^5$ \\ \hline
   B1 & $-$                             & E1 & $-$                            \\
   B2 & $5.00\cdot 10^2-1.23\cdot 10^3$ & E2 & $5.00\cdot 10^2-1.13\cdot 10^3$\\
   B3 & $5.00\cdot 10^2-2.02\cdot 10^4$& E3 & $5.00\cdot 10^2-1.94\cdot 10^4$\\
   B4 & $5.00\cdot 10^2-1.69\cdot 10^5$& E4 & $5.00\cdot 10^2-1.60\cdot
   10^5$ \\ 
   B5 & $-$                             & F1 & $-$                            \\
   B6 & $-$                             & F2 & $5.00\cdot 10^2-4.79\cdot 10^3$\\
   B7 & $5.00\cdot 10^3-2.17\cdot 10^4$ & F3 & $5.00\cdot 10^2-1.73\cdot 10^4$\\ 
   B8 & $5.00\cdot 10^3-1.75\cdot 10^5$ & F4 & $5.00\cdot 10^2-1.56\cdot 10^5$\\ \hline 
   C1 & $-$                             & G1 & $-$                            \\
   C2 & $5.00\cdot 10^2-1.04\cdot 10^3$ & G2 & $5.00\cdot 10^2-1.38\cdot 10^3$\\
   C3 & $5.00\cdot 10^2-1.94\cdot 10^4$ & G3 & $5.00\cdot 10^2-1.56\cdot 10^4$\\
   C4 & $5.00\cdot 10^2-1.73\cdot 10^5$ & G4 & \\
   C5 & $-$                             & H1 & $-$                            \\
   C6 & $-$                             & H2 & $5.00\cdot 10^2-1.09\cdot 10^3$\\
   C7 & $5.00\cdot 10^3-2.12\cdot 10^4$ & H3 & $5.00\cdot 10^2-1.49\cdot 10^4$\\
   C8 & $5.00\cdot 10^3-1.73\cdot 10^5$ & H4 & $5.00\cdot 10^2-1.31\cdot 10^5$\\
   \hline \hline   
\end{tabular}
\tablefoot{(1) Name of the model; (2) minimum and maximum mass of the BH seed formed in solar masses.}   
\label{SEEDMASSES}
\end{table*}

To understand the implications of NSCs for the formation of massive black holes, we analyze the properties of NSCs at the moment when their stellar mass is equal to the critical mass. We recall that the  critical mass $M_{\rm crit}$ (see Eq.~\ref{CriticalMass}) depends on key characteristics of the NSC such as the radius $r_\mathrm{NSC}$, velocity dispersion $\sigma$ and the age $t_\mathrm{H}$ of the star cluster, where we use Eq.~\ref{eqAGE} to obtain the mass-weighted age of the cluster considering its star formation history. By analyzing these properties at the moment of BH seed formation, we obtain insights into the conditions driving the formation of BHs.

\begin{figure}[!h]
    \centering
\includegraphics[width=\hsize]{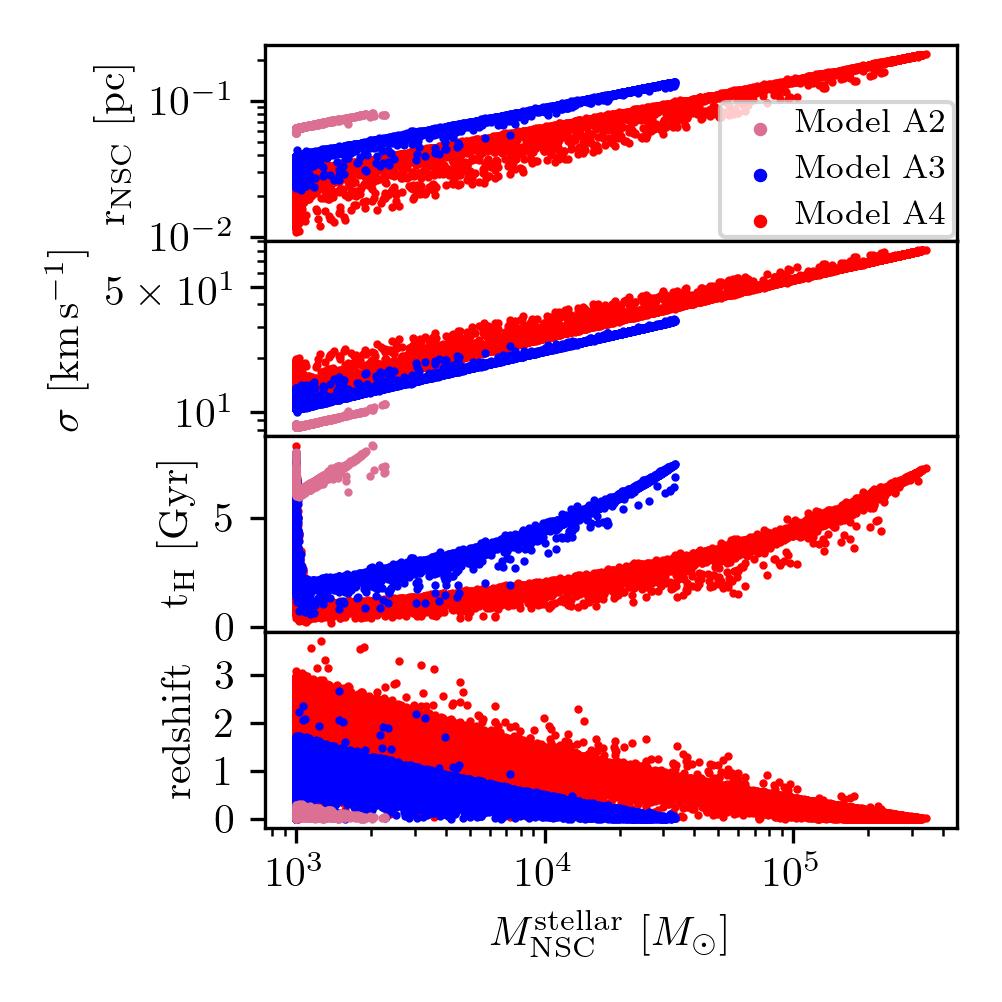}
    \caption{ Radius of the NSC ($r_\mathrm{NSC}$) multiplied by the respective efficiency $\epsilon_r$, velocity dispersion ($\sigma$) at $r_\mathrm{NSC}$, age of the system ($t_\mathrm{H}$), and the redshift in function of the stellar mass of the NSC for models A2 (rose dots), A3 (blue dots), and A4 (red dots), where  $M_\mathrm{threshold}=10^3$~M$_\odot$ and $A_\mathrm{res}=10^{-1}$ for all the models. The values of  $\epsilon_r$ are $0.5$, $0.2$, and $0.1$ for models A2, A3 and A4 respectively, as listed in Table~\ref{InitialParameters}. These properties are extracted at the moment in which the conditions to form a black hole seed are fulfilled ($M_\mathrm{crit} \leq M_\mathrm{NSC}^\mathrm{stellar}$ and $M_\mathrm{threshold} \leq M_\mathrm{NSC}^\mathrm{stellar}$). Model A1 does not form any black hole seed.}
    \label{fig:PSModelA1}
\end{figure}

\begin{figure}[!h] 
    \centering
\includegraphics[width=\hsize]{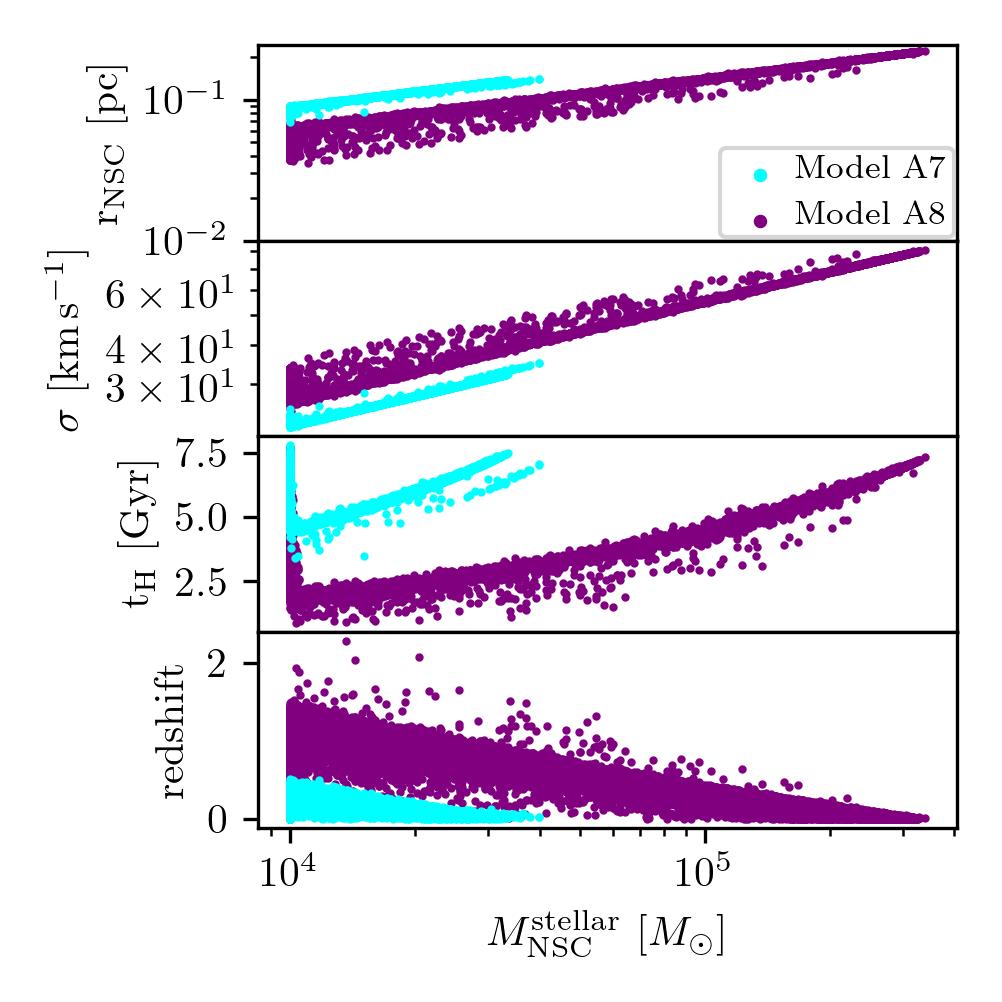}
    \caption{ Same as Fig. \ref{fig:PSModelA1} but for models A7 (cyan dots) and A8 (purple dots), with $M_\mathrm{threshold}=10^4$~M$_\odot$  and $\epsilon_r=0.2$, and $0.1$, respectively, as listed in Table~\ref{InitialParameters}. Models A5 and A6 do not form any black hole seed as the conditions for seeding ($M_\mathrm{crit} \leq M_\mathrm{NSC}^\mathrm{stellar}$ and $M_\mathrm{threshold} \leq M_\mathrm{NSC}^\mathrm{stellar}$) are not fulfilled for these models.}
  \label{fig:PSModelA2} 
\end{figure}

\begin{figure}[!h]
    \centering
\includegraphics[width=\hsize]{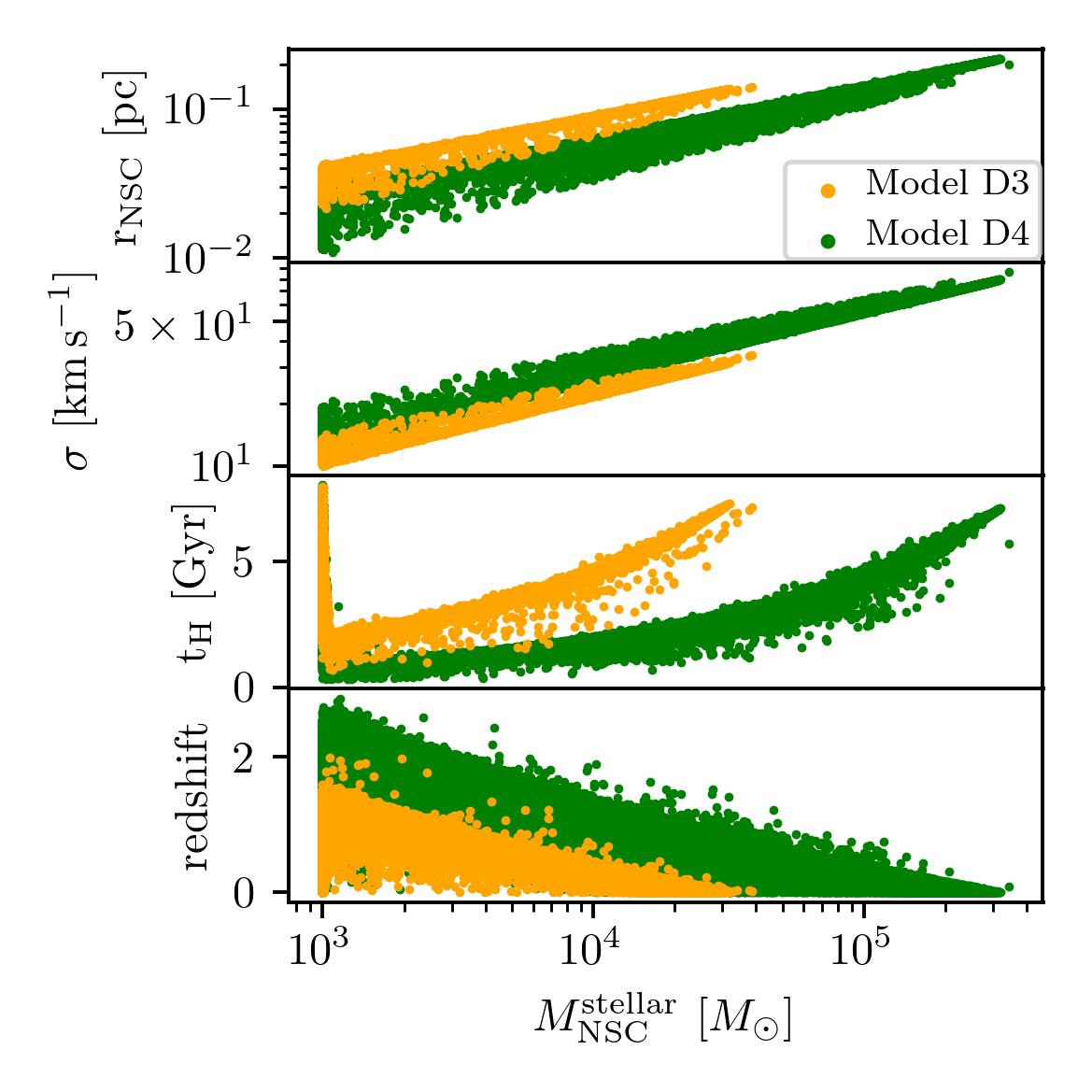}
    \caption{ 
      Radius of the NSC ($r_\mathrm{NSC}$) multiplied by the respective efficiency $\epsilon_r$, velocity dispersion ($\sigma$) at $r_\mathrm{NSC}$, age of the system ($t_\mathrm{H}$), and the redshift in function of the stellar mass of the NSC for models D3 (orange dots), and D4 (green dots), for $M_\mathrm{threshold}=10^3$~M$_\odot$ $10^3$~M$_\odot$ and $A_\mathrm{res}= 10^{-2}$. The values of  $\epsilon_r$ for each model are $0.5$, $0.2$, and $0.1$, respectively, as listed in Table~\ref{InitialParameters}. Models D1 and D2  do not form any black hole seed as the conditions for seeding ($M_\mathrm{crit} \leq M_\mathrm{NSC}^\mathrm{stellar}$ and $M_\mathrm{threshold} \leq M_\mathrm{NSC}^\mathrm{stellar}$) are not fulfilled for these models. }
    \label{fig:PSModelD1}
\end{figure}

\begin{figure}[!h]
    \centering
\includegraphics[width=\hsize]{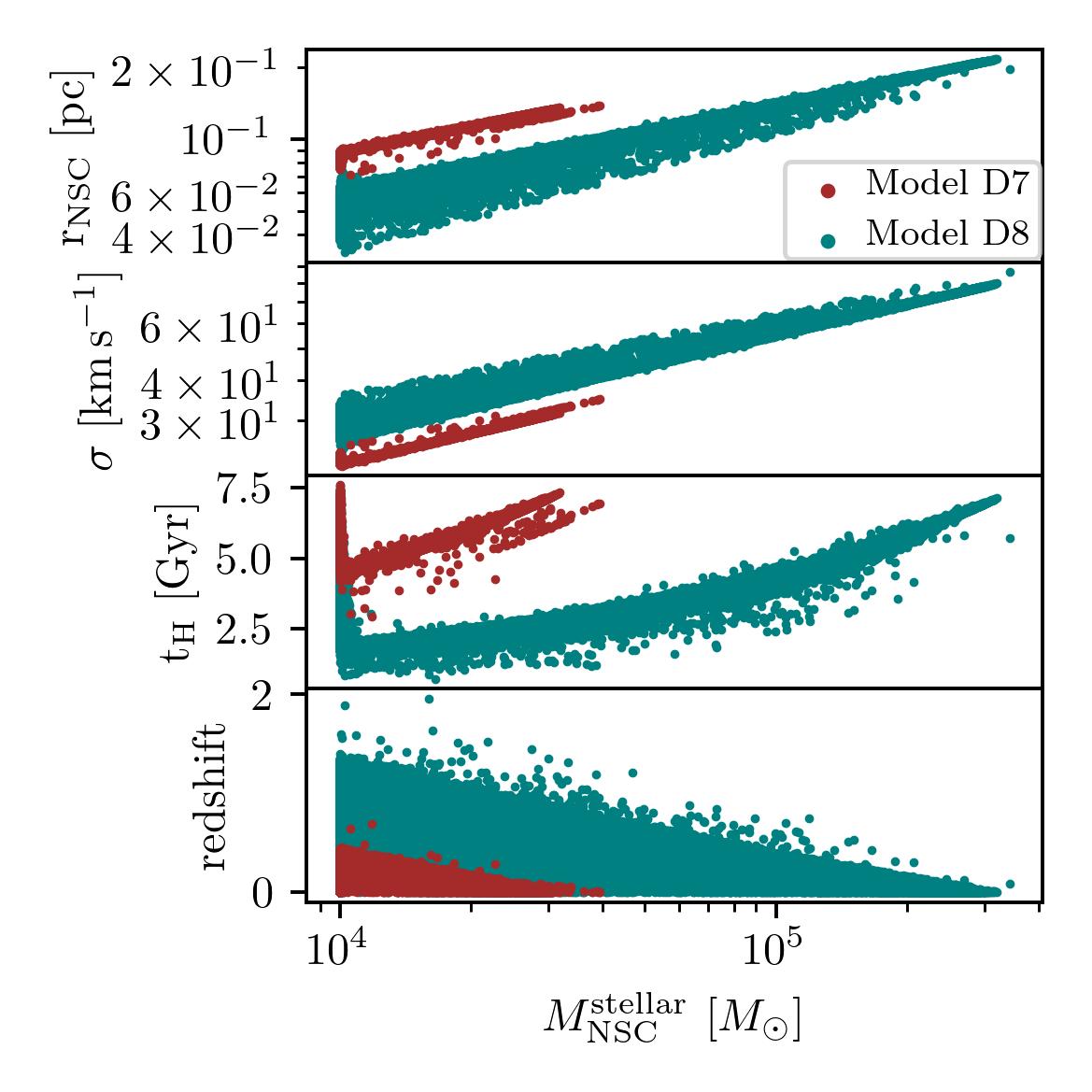}
    \caption{ Same as Fig. \ref{fig:PSModelD1} but for models D7 (brown dots) and D4 (teal dots), with  $M_\mathrm{threshold}=10^4$~M$_\odot$ and $\epsilon_r=0.5$ and $0.1$, respectively, as listed in Table~ \ref{InitialParameters}. Models D5 and D6 do not form any black hole seed as the conditions for seeding ($M_\mathrm{crit} \leq M_\mathrm{NSC}^\mathrm{stellar}$ and $M_\mathrm{threshold} \geq M_\mathrm{NSC}^\mathrm{stellar}$) are not fulfilled for these models.}
    \label{fig:PSModelD2}
\end{figure}

Figures \ref{fig:PSModelA1}, \ref{fig:PSModelA2}, \ref{fig:PSModelD1} and \ref{fig:PSModelD2} show the the typical properties of NSCs which satisfies the conditions to form a BH seed. This includes particularly their radius, velocity dispersion, the mass-weighted age and also the redshift when their mass is equal to the critical mass, which are all being shown as a function of the stellar mass of the NSC. In Fig.\ref{fig:PSModelA1} we focus on models A2, A3 and A4, which share the same parameter $A_{\rm res}=10^{-1}$ and consider a minimum NSC mass of $10^3$~M$_\odot$. The models however employ different mass - radius relations for the NSCs, considering values of $\epsilon_r$ between $0.1$ and $1$. Model A1 is not being shown  as in this case the radii are too large for the NSCs to reach the critical mass. For model A2 with $\epsilon_r=0.5$, some NSCs reach the critical mass, with their masses ranging between $10^3$~M$_\odot$ and $\sim 2\times10^3$~M$_\odot$. The radii of the NSCs are on the somewhat lower end of the general mass-radius relation (Eq.~\ref{massradius}), as the stellar masses of the NSCs are typically below $10^6$~M$_\odot$. The radii of the clusters considered here are slightly enhanced, as the mass-radius relation from Eq.~\ref{massradius} considers the dynamical mass including the gas, leading to the presence of some scatter in the relation between NSC radius and stellar mass. 

The number of NSCs for which the NSC mass becomes equal to the critical mass increases significantly in model A3 where $\epsilon_r=0.2$, as the critical mass decreases strongly with decreasing radius. This becomes even more pronounced for model A4 with $\epsilon_r=0.1$. For NSCs with masses above $10^5$~M$_\odot$, the typical radii are in the range of $0.1-1$~pc, and shift towards $\sim0.1$~pc for even lower-mass clusters. The velocity dispersions are in the range of $10$~km/s for low-mass clusters and more of order $100$~km/s for intermediate-mass ones, again with spread depending on the presence of gas in the system. The spread is enhanced in models A3 and A4 as the decrease in the general mass-radius relation implies that even NSCs with larger amounts of gas (and thereby radii enhanced by the presence of gas) can still become unstable to collisions.  The stellar mass of the NSCs forms a power law with the velocity dispersion of the NSCs as M$_\mathrm{NSC}^\mathrm{stellar }\sim \sigma^\frac{3}{2}$ though with a decrease in the overall spread as the stellar mass increases. Also in terms of the ages of the NSCs, there is a significant variation; for model A2 most of the NSCs that form black holes are older than $6$~Gyrs, while in models A3 and A4 there is a significant spread over the ages ranging from systems with ages of $7.5$~Gyrs up to even relatively young but more compact NSCs. As a result the black hole seeds form over a range of redshift, between redshift $\sim4$ and redshift $0$. We note that there is a wide spread in formation redshifts for the lower-mass NSCs, while the more massive ones form around redshift $0$.

In Fig.~\ref{fig:PSModelA2}, we show the models A7 and A8 with parameters $\epsilon_r=0.2$ and $0.1$, but a mass threshold of $10^4$~M$_\odot$. We do not show models A5 and A6 because the critical mass is not reached due to their radii. The behavior is overall similar to Fig.~\ref{fig:PSModelA1}; the main difference is a somewhat larger spread in the correlations at the threshold mass $\sim 10^4~$M$_\odot$ but otherwise the same type of behavior and relations. The increase of the threshold effectively also represents a delay for the formation of the first black hole seeds, as larger structures first have to form for them to potentially come into place.

The models shown in Figs.~\ref{fig:PSModelD1}-\ref{fig:PSModelD2}, D3, D4, D7 and D8, correspond to a diminished rate of gas transference to the reservoir ($A_\mathrm{res} = 10^{-2}$) in comparison with models A. This results in less gaseous material available to form stars in the nuclear region leading to lower star formation rates, leading to a delay in the formation of massive NSCs. This is clearly visible in model D4 where the   mass threshold is equals to $10^3$~M$_\odot$, and the efficiency $\epsilon_r$ is $0.1$. A direct comparison with model A4, which shares the same parameters except for the enhanced value of $A_\mathrm{res}$  shows that in A4 the formation of the seeds starts at  earlier times ($z\sim 4$) when the value of $A_\mathrm{res}$ is higher.

An important difference is model D2 do not form any BH seed as model A2, this is caused due to the relative low gas transfer, which impacts directly in the radius of the NSCs and avoid the formation of BHs. Instead, models D3 and D4, with $\epsilon_r$ equals to $0.2$ and $0.1$, form seeds for a variety of ages ranging from very young systems and up to ages of $\sim7.5$~Gyrs. The more massive NSCs are typically older and will form more massive black holes, though at a correspondingly lower redshift. For example, heavy seeds with masses of the order of $10^5$~M$_\odot$ start to form at redshift $z\sim 0.2$, while the formation of light-medium seeds starts at $z \sim 2.4$ in model D4. 

Increasing the value of $M_\mathrm{threshold}$ to $10^4$~M$_\odot$ just delay the formation of seeds in models D7 and D8, similar to what was seen for models A7 and A8. Again model D8 with $\epsilon_r=0.1$ starts to form seeds at earlier times than model D7 ($\epsilon_r=0.2$) due to the more compact radius. The behavior is very similar to the behavior already noted above. The most massive seeds are formed in older systems and at low redshifts as the most massive NSCs form at those redshift. The first seeds are formed in model D8 at $z\sim 1.8$, with masses of the order of $5\times10^3$~M$_\odot$, while the formation of seeds in NSCs with stellar masses up to $10^5$~M$_\odot$ results in seeds larger than $5\times 10^4$~M$_\odot$ from $z\sim 0.75$.  Model D7 starts forming the lightest seeds at $z\sim 0.7$ and the most massive seeds ($\sim10^4$~M$_\odot$) at $z=0$. 

\subsection{Black hole mass function}\label{BHMF}

In this subsection we present the black hole mass function (BHMF) obtained via the different models through {\sc Galacticus} for redshift $z=0$ hosted in galaxies with stellar masses above $10^{6}$~M$_\odot$ and coexisting with NSC with stellar masses larger than $10^{3}$~M$_\odot$. Thereby, we compare the model predictions with the BHMF from observations of the Local Universe estimated by \citet{VIKA2009}. This particularly allows us to  study the impact of $A_{\rm res}$, $\epsilon_r$ and $M_\mathrm{threshold}$ on the population at $z=0$  through the proposed scenario previously introduced in subsection~\ref{BlackHoleFormation} and implemented in {\sc Galacticus} as described in subsection \ref{BlackHoleFormationEvolution}.

\begin{figure*}
    \centering
\includegraphics[width=\hsize]{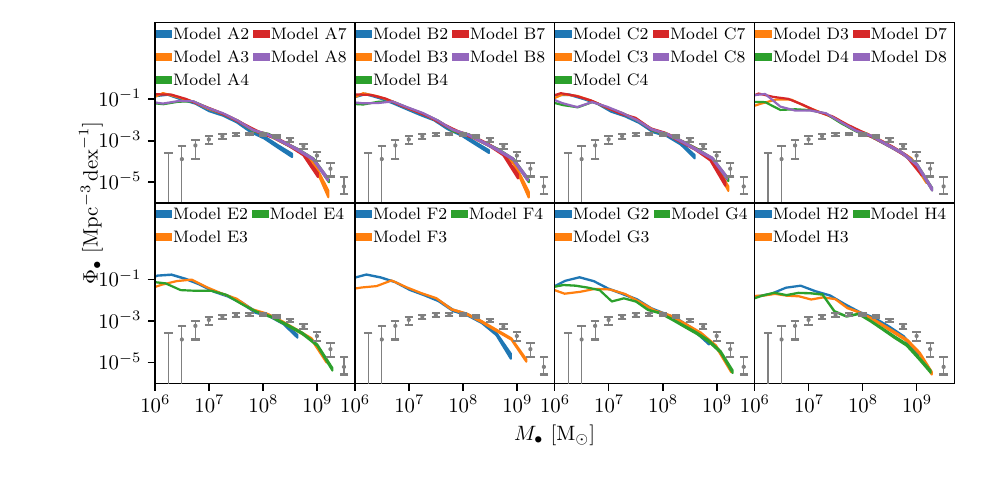}
    \caption{Comparison between the BHMF determined by \citet{VIKA2009} in gray dots with our semi analytic models at $z=0$ for galaxies with stellar masses larger than $10^6$~M$_\odot$ and hosting NSCs with stellar masses above $10^3$~M$_\odot$. 
     The gray dots include the $\pm 1 \sigma$ error bars, while the thickness of our lines denotes the $\pm 1\sigma$ error for our semi-analytic models estimated as $\Phi_\bullet/\sqrt{N}$, where $N$ is the number of SMBHs contained in the bin.}
    \label{fig:BHMF}
\end{figure*}

 In Fig. \ref{fig:BHMF} we provide the BHMF $(\Phi_\bullet)$ for all the models listed in Table~\ref{SEEDMASSES}, except for the models which do not form any seeds (due to the NSCs not reaching the critical mass). In general, models A2, B2, C2, E2, F2, G2, and H2, with parameters $M_\mathrm{threshold}=10^3$~M$_\odot$ and $\epsilon_r=0.5$ have a BHMF that starts with an initial increase from $\Phi_\bullet \sim 10^{-2}$~[Mpc$^{-3}$\,dex$^{-1}$] at $M_{\rm BH}=10^5$~M$_\odot$ to $\Phi_\bullet \sim 2\times 10^{-1}$~[Mpc$^{-3}$\,dex$^{-1}$] at $M_{\rm BH}=3\times10^6$~M$_\odot$, presumably as the black holes rapidly grow in mass and do not stay in that mass range for a significant amount of time. At higher masses, all the models (including models with $\epsilon_r = 0.2,\, 0.1$, and $M_\mathrm{threshold}=10^4~$M$_\odot$) show an increase in the population of SMBHs until they reach a peak on the population at masses of the order $10^6$~M$_\odot$ where $\Phi_\bullet \sim 5~[\mathrm{Mpc}^{-1}\, \mathrm{dex}^{-1}]$. From this peak towards higher masses, $\Phi_\bullet$ decreases monotonically until  reaching the maximum BH mass that is still realized in one of the dark matter halos modeled via {\sc Galacticus}, which in general is up to a few times $10^{9}$~M$_\odot$, with $\Phi_\bullet$ oscillating between $\sim 10^{-5}-10^{-4}~[\mathrm{Mpc}^{-3}\, \mathrm{dex}^{-1}]$.  
 
In all of our models it is possible to distinguish differences in the BHMF when comparing different values of $\epsilon_r$ at a fixed $A_\mathrm{res}$.  In models E,  $\Phi_\bullet$ is enhanced by a factor of $\sim 10$ in model E2 compared to E3 for SMBH masses in the range $10^5-10^7$~M$_\odot$, while at higher masses the models give comparable predictions for the mass function. Such differences at lower masses are not seen in models with fast gas transfer (high $A_{\rm res}$), where models with $\epsilon_r=0.1$ and $0.2$ are essentially indistinguishable.
 
 The decrease  of $\Phi_\bullet$ for SMBHs with masses larger than $> 10^7$~M$_\odot$ indicates that such massive objects are less frequent compared to the total population predicted in our models. Models with $\epsilon_r = 0.1$ (A4, A8, B4, B8, C4, C8, E4, F4, G4, H4) indeed include the formation of SMBHs with masses up to a few times $ 10^{9}~$M$_\odot$ in some of the halos. Combining this information with Figs. \ref{fig:PSModelA2}-\ref{fig:PSModelD2}, and Table~\ref{SEEDMASSES}, we note that these models form heavier seeds than models A2, A3, and D3, which form at most medium seeds at earlier times, allowing those seeds to accrete more material from their environment.

The BHMF from  \citet{VIKA2009} is comparable with the high-end mass in almost all our models for BH masses above $10^8$~M$_\odot$. Particularly, at below masses of $10^8$~M$_\odot$, the models tend to overpredict the observed population by at least a factor of $10$. The overestimation is about $10^2$ times in masses in the order of  $10^6- 10^7$~M$_\odot$, where the uncertainties of the BHMF becomes larger due to the observational limits.

The underestimation at higher masses does not represent a problem, as here we consider only one seeding mechanism, and other mechanisms could still change the overall picture, including even the possibility of mergers of some of the SMBHs formed here with other seeds. The comparison also shows that quite a significant number of BHs seeds can be formed in this manner, providing a relevant contribution to the general population.

 For a detailed quantitative analysis of the over- or underestimation of the BHMF in our models, we kindly refer readers to our Appendix \ref{GalaxyMassFunctionComparison}.

\subsection{Nuclear star clusters, supermassive black holes and host galaxies}\label{scaling}

We explore the scaling relations between the global properties of the NSCs and their host galaxies. We include in our analysis the masses of NSCs and SMBHs coexisting in the observed galactic  nuclei from  \citet{SETH2008,GRAHAM2009,NEUMAYER2012,GEORGIEV2016, NGUYEN2018}. The resulting  sample contains early (late)-type galaxies and stripped galaxy nuclei. We remove the UCDs for which the information of the stellar mass of the galaxy is not available.

We make a power-law fit for the correlation in the observed sample consideration a relation between $M_\mathrm{BH}/M_\mathrm{NSC}^\mathrm{stellar}$ and $M_\mathrm{galaxy}^\mathrm{stellar}$ of the form 
\begin{equation}
    \log{\left(\frac{M_\mathrm{BH}}{M_\mathrm{NSC}^\mathrm{stellar}}\right)} = c_1+c_2\log{\left(\frac{M_\mathrm{galaxy}^\mathrm{stellar}}{1~{\rm M}_\odot}\right)}, \label{log}
\end{equation}

where $M_\mathrm{BH}$ is the mass of the central SMBH, $M_\mathrm{NSC}^\mathrm{stellar}$ is the stellar mass of the NSC,   $M_\mathrm{galaxy}^\mathrm{stellar}$ is the stellar mass of the host galaxy and $c_1$ and $c_2$ are the free coefficients to fit. 

\begin{figure*}[h]
    \centering
    \includegraphics[width=\hsize]{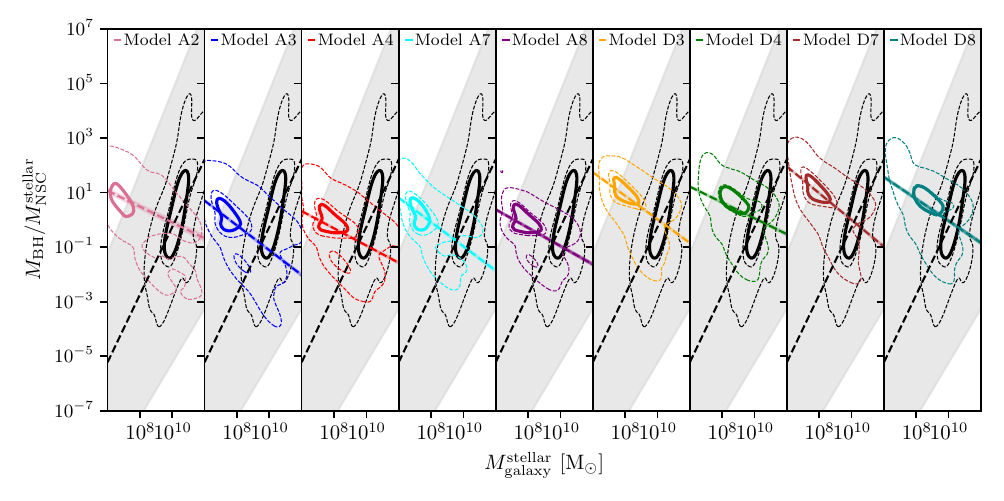}
    \caption{$M_\mathrm{galaxy}^\mathrm{stellar}$ versus $M_\mathrm{BH}/M_\mathrm{NSC}^\mathrm{stellar}$. The 2D PDF is shown as contour  lines where the thick solid line mark the $1\sigma$ of the 2D PDFs. The best fit of the data is represented by dashed lines.  From left to right, the contours indicate model A2 (light rose), A3 (blue), A4 (red), A7 (cyan), A8 (purple), D3 (orange), D4 (green), D7 (brown) and  model D8 (teal), while black lines correspond to the observational sample.  The filled light areas correspond to the $1\sigma$ error of the fit slope and intercept.}
    \label{Fig:2PDs}
\end{figure*}

In the 2D PDF corresponding to the observed sample, the ratio $M_\mathrm{BH}/M_\mathrm{NSC}$ for galaxies of low  stellar masses is dominated by the mass of the NSC, and dominated by the SMBH  at the high end of galaxy stellar masses.  Combining this figures with the results of the best fits given in Table~\ref{bestfit}, we can note an interesting trend: as $\epsilon_r$ decreases, the slope of the fit approaches the slope of the observed fit, suggesting that as $\epsilon_r$ decreases, NSCs form more massive seeds at early times allowing them to reach higher final masses. In Fig. \ref{Fig:2PDs} we show the $M_\mathrm{BH}/M_\mathrm{NSC}^\mathrm{stellar}$-$M_\mathrm{galaxy}^\mathrm{stellar}$ correlation predicted by {\sc Galacticus} for galaxies with stellar masses larger than $10^6$~M$_\odot$ hosting NSCs and SMBHs with masses larger than $10^{3}$~M$_\odot$ and $10^{5}$~M$_\odot$ respectively. 

We found that NSCs and SMBHs coexist in models A2, A3, A4, A7 and A8. We include the contour lines of the 2D PDF as dashed lines and the thick solid line corresponds to the $1\sigma$ of the 2D-PDF. The results of the different fits are included as dashed lines, where the shadow areas indicate the $1\sigma$ error of the fit slope and intercept coefficient. We encourage readers to refer to Appendix~\ref{PDF} for further details on the PDF. 

Model A2 shows an overlap with the observed correlation for galaxies with stellar masses from $\sim10^{8}$~M$_\odot$ to $\sim10^{11}$~M$_\odot$ with a $M_\mathrm{BH}/M_\mathrm{NSC}^\mathrm{stellar}$ ranging from $10^{-1}$ to $10^{2}$. Although there is an overlap between our predictions and the observations, {\sc Galacticus} shows  a big spread in the $M_\mathrm{BH}/M_\mathrm{NSC}^\mathrm{stellar}$ ratio for galaxies with stellar masses below $10^{10}$~M$_\odot$. In particular, almost all the galaxies are contained outside the overlap with the observations  at $\sim 10^7$~M$_\odot$ hosting a SMBH with $M_\mathrm{BH}/M_\mathrm{NSC}^\mathrm{stellar}=10^{0}-10^{1}$. Furthermore, {\sc Galacticus} predicts the existence of a population of galaxies with masses of the order of $10^{11}$~M$_\odot$ which overlap well with observations. This population corresponds to galaxies which recently formed a BH seed under the proposed scenario in NSCs with stellar masses in the order of $\sim10^{9}$~M$_\odot$ to $\sim 10^{10}$~M$_\odot$, forming seeds with similar masses.

Models A3 and A7 ($\epsilon_r=0.2$ and $M_\mathrm{threshold}=10^3,10^4$~M$_\odot$, respectively) predominantly include galaxies with stellar masses of the order of $10^{8}-10^{11}$~M$_\odot$, with $M_\mathrm{BH}/M_\mathrm{NSC}^\mathrm{stellar}=10^{-3}-10^{2}$, thus providing a significant overlap between the models and observations at the contour lines. Models A4 and A8 ($\epsilon_r=0.1$ and $M_\mathrm{threshold}=10^3,10^4$~M$_\odot$, respectively), on the other hand, show the best  overlap around $M_\mathrm{galaxy}^\mathrm{stellar}=10^{10}$~M$_\odot$, and the values of  $M_\mathrm{BH}/M_\mathrm{NSC}^\mathrm{stellar}$ are less spread in comparison to the models  A3 and A7. Furthermore, the distribution of the galaxies is moving toward the observed sample contours, but with a different slope as listed in Table~\ref{bestfit}.

We previously found that the models labeled with D provide better predictions for the maximum stellar masses of NSCs, as demonstrated by the comparison of predicted and observed NSC mass function in Fig.~\ref{Massfunction}. In Fig. \ref{Fig:2PDs} we observe a similar trend as models A.  It is clear that models D4 and D8, with $M_\mathrm{threshold}=10^3$~M$_\odot$ and $M_\mathrm{threshold}=10^4$~M$_\odot$, respectively, and $\epsilon_r=0.1$, show an overlap from $M_\mathrm{galaxy}^\mathrm{stellar}=10^{8}$ to $M_\mathrm{galaxy}^\mathrm{stellar}=10^{11}$  with observations, quite similar to models A4 and A8. The values of  $M_\mathrm{BH}/M_\mathrm{NSC}^\mathrm{stellar}$ range from $10^{-2}$ to $\sim10^{2}$, meaning that NSCs and SMBHs have similar masses or at most there is a mass difference by two orders of magnitude.  The impact of $\epsilon_r$ is reflected in the value of $c_1$ and $c_2$ given in the best fit in Table~\ref{bestfit}. As the value of $\epsilon_r$ decreases,  the values of the free parameters fitted to our models approach the values of the observed correlation.

\begin{figure}[h]
    \centering
\includegraphics[width=\hsize]{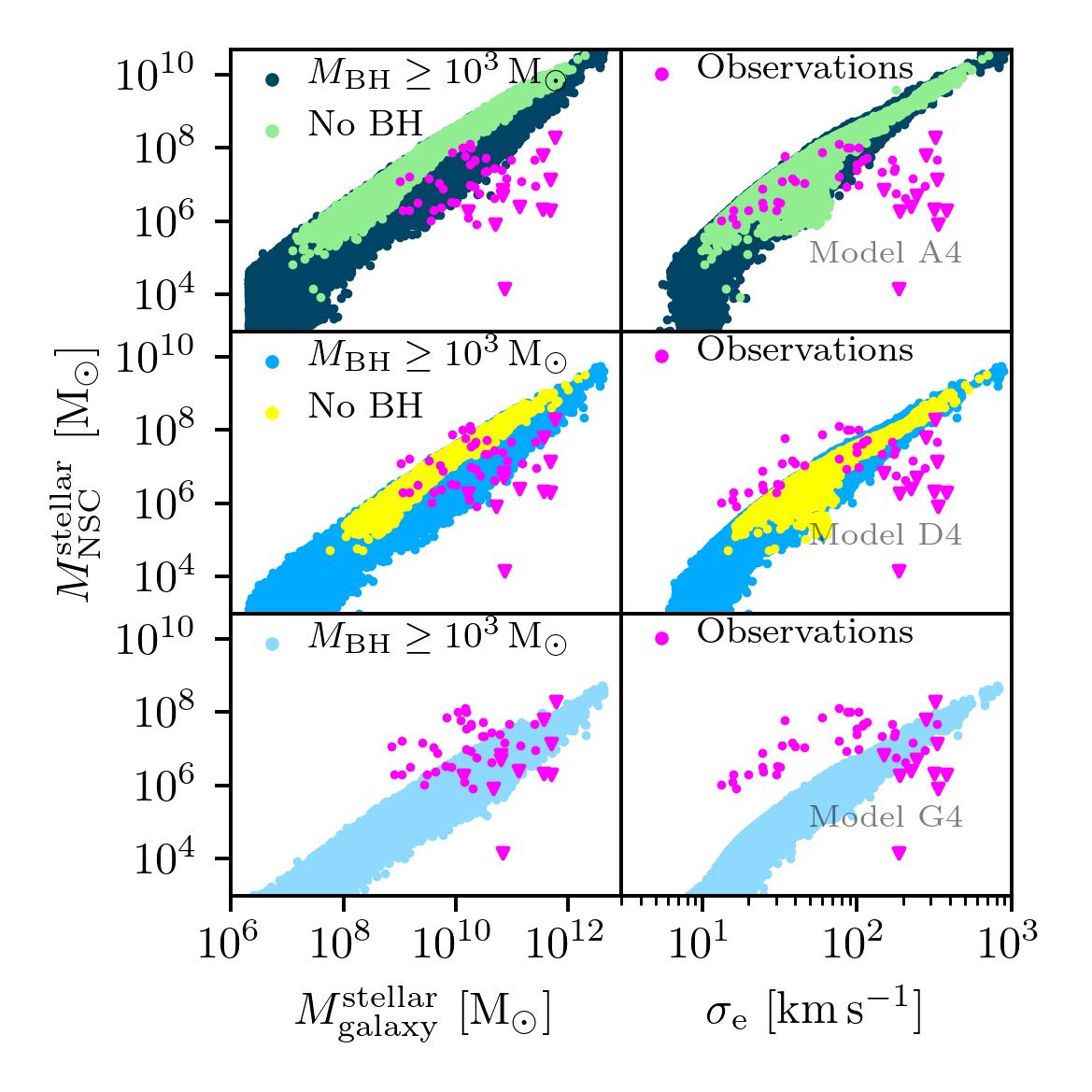}
    \caption{Same as Fig. \ref{NSCGalaxyA} but for models with BH formation (A4, D4 and, G4). We made a distinction for each model in galaxies hosting (or not) a SMBH in their center. Triangles indicate upper limits for the stellar mass estimation of the NSCs. }
    \label{fig:blackhole-NSC-GAL}
\end{figure}

Figure \ref{fig:blackhole-NSC-GAL} shows the same as Fig.~\ref{NSCGalaxyA}, but we made a distinction between galaxies with NSCs hosting (or not hosting) SMBHs. In model G4, all the galaxies which host a NSC with stellar masses larger than $10^4$~M$_\odot$ host a SMBH in their center. For models A4 and A8 where $\epsilon_r=0.1$,   $58\%$ and  $54\%$ of the galaxies host both types of objects (NSC and SMBH, respectively). For models with larger values of $\epsilon_r$, the fraction drops as the formation of seeds becomes less efficient. In consequence, the number of galaxies hosting a SMBH decreases for models A2, A3, and A7, where the SMBH  occupation fraction is $1\%$, $20\%$, and $12\%$, respectively. For models D, on the other hand, we find SMBH occupation fractions of $90\%$ and $88\%$, respectively, for models D4 and D8, and $2\%$, $52\%$, and $35\%$ for models  D3 and D7.

We find that there is  a scaling relation between the stellar mass of the NSC and the stellar mass of the galaxy in all our models as well as in the observed population. In our models, galaxies with stellar masses between $10^9$~M$_\odot$ and $10^{11}$~M$_\odot$ show a scatter in the stellar mass of the NSC in galaxies hosting a SMBH. A possible explanation for this is the fact of the SMBH accreting material from the NSC gas reservoir, removing a fraction of the available gas for star formation in the NSC. This loss of gas directly affects the star formation rate of the NSC as  $\dot{M}_\mathrm{NSC}^\mathrm{stellar}\propto M_\mathrm{NSC}^\mathrm{gas}$, suggesting that BH suppress the `in-situ' star formation in NSCs.

\begin{table}[!h]
    \centering
    \caption{Values of the best fit parameters $c_1$ and $c_2$.}
    \begin{tabular}{ccc}
    \hline \hline
    \multicolumn{1}{c}{Model} &\multicolumn{1}{c}{$c_1$} &\multicolumn{1}{c}{$c_2$} \\
    \multicolumn{1}{c}{(1)} &\multicolumn{1}{c}{(2)} &\multicolumn{1}{c}{(3)} \\ \hline
    Observed &$0.539\pm 0.100$ & $-12.661\pm 2.518$ \\ 
    A2 &$-0.126\pm 0.002$ &$2.812\pm0.041$\\
    A3 &$-0.197\pm 0.001$ &$3.426\pm0.018$\\
    A4 &$-0.137\pm 0.001$ &$2.216\pm0.014$\\
    A7 &$-0.194\pm 0.001$ &$3.485\pm0.020$\\
    A8 &$-0.146\pm 0.001$ &$2.393\pm0.014$\\
    D3 &$-0.187\pm 0.001$ &$4.341\pm0.015$\\ 
    D4 &$-0.126\pm 0.001$ &$2.964\pm0.019$\\
    D7 &$-0.213\pm 0.001$ &$4.893\pm0.015$ \\ 
    D8 &$-0.175\pm 0.001$ &$3.983\pm0.018$  \\ \hline
    \hline
    \end{tabular}
    \tablefoot{  (1) Name of the model; (2, 3) intercept coefficient and slope coefficient appearing in Eq. \ref{log} with the respective $\pm 1\sigma$ value. }
    \label{bestfit}
\end{table}

\section{Discussion and conclusions}\label{discussion}

In this work we employed the semi-analytic model {\sc Galacticus} developed by \citet{BENSON2012} to explore the seeding of SMBHs via collisions in NSCs, considering the framework proposed by \citet{ESCALA2021},  where SMBHs are formed from (partially) failed NSCs with average collision timescale shorter than the age of the system, thus implying the efficient formation of a central massive object via runaway collisions. Particularly, we employ here the concept of the critical mass scale, for which the average timescale for collisions within the NSC (of a given radius) is equal to the age of the cluster, that was introduced by \citet{VERGARA2023} in order to numerically test this framework (with successful results) and later extended its validity to a range of stellar systems (globular clusters and ultra-compact dwarf galaxies, in addition to NSCs; \citealp{Vergara2024}). The formation of NSCs themselves is modeled here via the `in-situ' mechanism, which is adequate for the more massive galaxy population and thus for the formation of SMBH in more massive galaxies \citep{ANTONINI2012, ANTONINI2015}. 

Within this framework, we find that the formation of NSCs and their mass function is regulated through the efficiency parameter that describes the gas mass transfer to the center of the galaxy, which here is taken to be proportional to the global star formation rate. The mass function we obtain in our models is comparable in terms of shape with the NSC mass function constructed using the $M_\mathrm{NSC}^\mathrm{stellar}-M_\mathrm{galaxy}^\mathrm{stellar}$ and the fitted  galaxy stellar mass function derived by \citet{BALDRY2012}. There is also considerable overlap in the observed scaling relations between NSC mass and stellar mass of the galaxy. The comparison however suggests that in real galaxies there is not one fixed value of the efficiency to transport gas into the center, but the efficiency may depend on the galaxy and its environment. The  maximum observed stellar masses observed are of the order of $10^{9}$~M$_\odot$. The efficiency parameters we employ to find agreement with observations are consistent with those reported by \cite{ANTONINI2015} and used in their semi-analytic model. 

Although there is an agreement in NSC masses about $\sim 10^{8}$~M$_\odot$, our models in general overestimate the population of NSCs by a factor of $10$. As mention in subsection, this is not an issue as in low NSC masses the observed NSC mass function is a lower limit. We found that the population of NSCs in lower regime  depends strongly on the value of $A_\mathrm{res}$. 

Although we do not have access to a direct comparison with the observed NSC, and our estimation is reliable in masses above $10^5$~M$_\odot$ due to the limit of observations, our models in general overestimate the population of NSC. We also remark that derive a NSC mass function from observations often present difficulties at low end NSC masses, due to the low resolution that is insufficient to study remote and distant NSCs in detail \citep{COTE2006}. Additionally, there is an ongoing issue with the criteria for defining what constitutes an NSC, where the definition of NSCs could be biased due ambiguity in classification, that is some clusters might be classified as NSCs or not depending on the definition employed. Furthermore, NSCs can be difficult to distinguish from the central bulges of galaxies, where the background light of the most luminous bulges plays an important role in galaxies with a Hubble type $t\geq 3.5$ \citep{GEORGIEV2014}. Other issues are present in elliptical and lenticular galaxies where the stellar density gradient can be smooth and to delineate where the bulge ends and the NSC begins is challenging as the galactic nuclei does not show a clear discontinuity  especially in galaxies where the NSC and bulge have a similar brightness profile \citep{CAROLLO1997,COTE2006,BOKER2010, NEUMAYER2020}

The overestimation of the population may indicate that the `in-situ' star formation scenario is enough to explain the population of the most massive NSCs but it is not the main mechanism in the low mass regime, suggesting that the invocation of other formation mechanisms such as the migration of globular clusters \citep[e.g.,][]{TREMAINE1975,CAPUZZO1993} is also required and the value of $A_\mathrm{res}$ is not constant for all galaxies, depending on environmental factors. Some previous studies suggest a correlation between the morphological type of the host galaxy and the formation scenario of the NSC, to be specific, globular cluster infall scenario is related to early type galaxies \citep{TREMAINE1975,CAPUZZO1993}, while in massive late type galaxies the `in-situ' formation dominates the growth of the NSC \citep{PINNA2021,HANNAH2021}. In our models, we assume that the gas transfer to the nuclear reservoir is related to the star formation in the spheroidal component of the galaxies. This is consistent with galaxies which typically have abundant gas and provides a rich gas reservoir for `in-situ' star formation within the nucleus of the galaxies. As our model does not account for the population of NSCs formed in bulgeless galaxies or those without a significant spheroidal component, where dynamical processes are the main formation mechanism, we are missing part of the population of NSCs formed via  the cluster infall scenario. This scenario is particularly predominant in early-type galaxies, where NSCs are more likely to form through the accumulation and merging of pre-existing globular clusters that migrate to the center over time as previously explored by \citet{CAPUZZO1993} in a triaxial galaxy model, N-body simulations \citep{MILOSAVLJEVIC2004} and more recently in self-consistent numerical simulations 
\citep{BEKKI2010}.

For the formation of massive black holes via collisions, a crucial aspect is the mass - radius correlation of NSCs. Here we do not explore this correlation from first principles, but rather adopt the observed correlation from \citet{NEUMAYER2020} and vary it via a parameter $\epsilon_r$ to explore its variation, considering both that there is a range in the observed radii as well as the possibility of evolution over cosmic time. As expected, more compact NSCs result in earlier formation of more massive BH seeds due to them being more favorable for stellar collisions, and the probability to form a seed increases affecting directly the fraction of galaxies hosting both NSCs and SMBHs. Models with $\epsilon_r=0.1$ are able to form more massive seeds at earlier times, specifically, redshifts of 3 or 4. As we mentioned in subsection \ref{CollisionTimescale}, this is consistent with the study of \citet{BANERJEE2017}, showing the evolution of the effective radius of massive star clusters by more than a factor of $10$. Our models show a significant correlation between the NSC mass and the stellar mass (and velocity dispersion) of the host galaxy, following a power-law distribution. This correlation holds also when varying the efficiency parameter for the mass transport to the center of the galaxy, and the correlation is independent of the potential presence of a SMBH. The spread in the stellar mass of the NSCs at a fixed galaxy stellar mass however increases when hosting an SMBH. As mentioned, probably less massive NSCs could be due to a decrease of the star formation rate as the BH seed accretes gaseous material from the nuclear reservoir. 

In general, all our models are able to produce BH seeds, except those with $\epsilon_r=1$, and only few of these form in models with $\epsilon_r=0.5$. This may suggest that also in the observed population, the mechanism operates preferably in the more compact part of the population, or possibly also at early stages where the NSCs were still more compact.  Overall, the models are able to produce a range of BHs from light seeds ($\sim10^2$~M$_\odot$) to heavy seeds ($\sim10^5$~M$_\odot$). The formation of the most massive seeds ($\sim10^5$~M$_\odot$) is independent of the mass threshold we employ for the NSCs, but only happens in models where $\epsilon_r=0.1$. Furthermore, those seeds are formed in  systems with ages of $\gtrsim 4$~Gyrs and velocity dispersions higher than $70$~km\,s$^{-1}$. The formation of these massive seeds however happens relatively late, that is at redshifts $z\sim0.35$ within the halos modeled here. We cannot exclude that in some cases, this may also happen early, and in this respect it might be important to study the halo merger history of the first quasars forming around $z\sim6$, to obtain more insight into their specific histories and to understand whether this channel might be contributing there or not. In this context, also the presence of gas within the NSCs might be even more relevant, and it could be important to include the extension of the critical mass concept in the presence of gas as recently proposed by \citet{Vergara2024}. In addition, it is clear that other seeding scenarios may also contribute to the general population, and the interaction between different seeding mechanisms (i.e., due to mergers) needs to be explored. This includes, for example, seeding via the direct collapse \citep{Bromm2003,Koushiappas2004, Wise2008, LATIF2013, Latif2022} as well as mixed scenarios considering the interaction of collisions and accretion \citep{Boekholt2018, Chon2020, Schleicher2022, Reinoso2023, Schleicher2023}.

 The models successfully reproduce the  shape of the observed mass functions of NSCs and SMBHs, showing a strong match within the mass range relevant to SMBH formation. However, they overpredict the NSC and BH mass functions at lower masses, likely due to an excess (by a factor of $\sim 10$) in the number of galaxies predicted by Galacticus compared to the \cite{BALDRY2012} sample.

 Despite this discrepancy, our results demonstrate that the proposed scenario effectively forms black hole seeds capable of growing to masses up to $10^9$~M$_\odot$. A logical next step is to enhance the statistical robustness of our predictions. Unfortunately, rescaling the stellar mass function is not straightforward. One potential approach is to employ the Markov Chain Monte Carlo (MCMC) functionality built into {\sc Galacticus}. While MCMC can improve statistical predictions, it requires a significant computational investment to achieve reliable results.

The predicted scaling relations also show considerable overlap with observed data. Around BH masses of $10^7$~M$_\odot$, our model overpredicts the observed population (constructed from the sample of \citealt{VIKA2009}) by roughly a factor of $10$, while at higher masses we tend to underpredict the observed population. However, this of course may change when additional black hole formation mechanisms are being considered, and some of the BH seeds predicted by our model may also merge with seeds produced via different channels \citep{Sassano2021, Trinca2022}.

Nonetheless, the comparison shows that the collision-based channel may contribute in a relevant way to the overall population. This could include a mechanism that is related but still different, for example, the formation of massive black holes in dark cores that are formed in the center of NSCs \citep[e.g.,][]{Davies2011, Lupi2014, Kroupa2020, Gaete2024}. We expect that more light can be shed on the seeds of SMBHs with upcoming observations both with the JWST and the Extremely Large Telescope (ELT)\footnote{Webpage ELT:~\url{https://elt.eso.org/}}, and together with the emission of gravitational waves detected by current interferometers, such as the Laser Interferometer Gravitational-Wave Observatory (LIGO)\footnote{Webpage LIGO:~\url{https://www.ligo.caltech.edu} }/Virgo\footnote{Webpage Virgo:~\url{https://www.virgo-gw.eu}}/ Kamioka Gravitational Wave Detector (KAGRA)\footnote{Webpage KRAGA:~ \url{https://gwcenter.icrr.u-tokyo.ac.jp/}} collaboration \citep{ABBOTT2024}, and in the future with LISA \citep{AMARO2017} and Einstein Telescope\footnote{Webpage ET:~\url{https://www.et-gw.eu}} \citep{PUNTURO2010} both methods in the local and the high-redshift Universe.

\begin{acknowledgements}
We gratefully acknowledge support by the ANID BASAL project FB21003, Fondecyt Regular (project code 1201280) and ANID-Quimal 220002. DRGS thanks for funding via the  Alexander von Humboldt - Foundation, Bonn, Germany.
MCV acknowledges funding through ANID (Doctorado acuerdo bilateral DAAD/62210038) and DAAD (funding program number 57600326).
ML would also like to express his gratitude to Dr. Nadine Neumayer for her valuable discussions and comments during his visit to MPIA, Heidelberg, and to Dr. Enrico Barausse for his insightful feedback during my time at SISSA, Trieste.
\end{acknowledgements}

\bibliographystyle{aa}
\bibliography{References}

\begin{thebibliography}{187}
\expandafter\ifx\csname natexlab\endcsname\relax\def\natexlab#1{#1}\fi

\bibitem[{{Abbott} {et~al.}(2024){Abbott}, {Abbott}, {Acernese}, {Ackley},
  {Adams}, {Adhikari}, {Adhikari}, {Adya}, {Affeldt}, {Agarwal}, {Agathos},
  {Agatsuma}, {Aggarwal}, {Aguiar}, {Aiello}, {Ain}, {Ajith}, {Albanesi},
  {Allocca}, {Altin}, {Amato}, {Anand}, {Anand}, {Ananyeva}, {Anderson},
  {Anderson}, {Andrade}, {Andres}, {Andri{\'c}}, {Angelova}, {Ansoldi},
  {Antelis}, {Antier}, {Appert}, {Arai}, {Araya}, {Areeda}, {Ar{\`e}ne},
  {Arnaud}, {Aronson}, {Arun}, {Asali}, {Ashton}, {Assiduo}, {Aston}, {Astone},
  {Aubin}, {Austin}, {Babak}, {Badaracco}, {Bader}, {Badger}, {Bae}, {Baer},
  {Bagnasco}, {Bai}, {Baird}, {Ball}, {Ballardin}, {Ballmer}, {Balsamo},
  {Baltus}, {Banagiri}, {Bankar}, {Barayoga}, {Barbieri}, {Barish}, {Barker},
  {Barneo}, {Barone}, {Barr}, {Barsotti}, {Barsuglia}, {Barta}, {Bartlett},
  {Barton}, {Bartos}, {Bassiri}, {Basti}, {Bawaj}, {Bayley}, {Baylor},
  {Bazzan}, {B{\'e}csy}, {Bedakihale}, {Bejger}, {Belahcene}, {Benedetto},
  {Beniwal}, {Bennett}, {Bentley}, {BenYaala}, {Bergamin}, {Berger},
  {Bernuzzi}, {Berry}, {Bersanetti}, {Bertolini}, {Betzwieser}, {Beveridge},
  {Bhandare}, {Bhardwaj}, {Bhattacharjee}, {Bhaumik}, {Bilenko}, {Billingsley},
  {Bini}, {Birney}, {Birnholtz}, {Biscans}, {Bischi}, {Biscoveanu}, {Bisht},
  {Biswas}, {Bitossi}, {Bizouard}, {Blackburn}, {Blair}, {Blair}, {Blair},
  {Bobba}, {Bode}, {Boer}, {Bogaert}, {Boldrini}, {Bonavena}, {Bondu},
  {Bonilla}, {Bonnand}, {Booker}, {Boom}, {Bork}, {Boschi}, {Bose}, {Bose},
  {Bossilkov}, {Boudart}, {Bouffanais}, {Bozzi}, {Bradaschia}, {Brady},
  {Bramley}, {Branch}, {Branchesi}, {Brau}, {Breschi}, {Briant}, {Briggs},
  {Brillet}, {Brinkmann}, {Brockill}, {Brooks}, {Brooks}, {Brown}, {Brunett},
  {Bruno}, {Bruntz}, {Bryant}, {Bulik}, {Bulten}, {Buonanno}, {Buscicchio},
  {Buskulic}, {Buy}, {Byer}, {Cadonati}, {Cagnoli}, {Cahillane}, {Bustillo},
  {Callaghan}, {Callister}, {Calloni}, {Cameron}, {Camp}, {Canepa},
  {Canevarolo}, {Cannavacciuolo}, {Cannon}, {Cao}, {Capote}, {Carapella},
  {Carbognani}, {Carlin}, {Carney}, {Carpinelli}, {Carrillo}, {Carullo},
  {Carver}, {Diaz}, {Casentini}, {Castaldi}, {Caudill}, {Cavagli{\`a}},
  {Cavalier}, {Cavalieri}, {Ceasar}, {Cella}, {Cerd{\'a}-Dur{\'a}n},
  {Cesarini}, {Chaibi}, {Chakravarti}, {Subrahmanya}, {Champion}, {Chan},
  {Chan}, {Chan}, {Chan}, {Chandra}, {Chanial}, {Chao}, {Charlton}, {Chase},
  {Chassande-Mottin}, {Chatterjee}, {Chatterjee}, {Chatterjee},
  {Chattopadhyay}, {Chaturvedi}, {Chaty}, {Chatziioannou}, {Chen}, {Chen},
  {Chen}, {Chen}, {Chen}, {Cheng}, {Cheong}, {Cheung}, {Chia}, {Chiadini},
  {Chiarini}, {Chierici}, {Chincarini}, {Chiofalo}, {Chiummo}, {Cho}, {Cho},
  {Choudhary}, {Choudhary}, {Christensen}, {Chu}, {Chua}, {Chung}, {Ciani},
  {Ciecielag}, {Cie{\'s}lar}, {Cifaldi}, {Ciobanu}, {Ciolfi}, {Cipriano},
  {Cirone}, {Clara}, {Clark}, {Clark}, {Clarke}, {Clearwater}, {Clesse},
  {Cleva}, {Coccia}, {Codazzo}, {Cohadon}, {Cohen}, {Cohen}, {Colleoni},
  {Collette}, {Colombo}, {Colpi}, {Compton}, {Constancio}, {Conti}, {Cooper},
  {Corban}, {Corbitt}, {Cordero-Carri{\'o}n}, {Corezzi}, {Corley}, {Cornish},
  {Corre}, {Corsi}, {Cortese}, {Costa}, {Cotesta}, {Coughlin}, {Coulon},
  {Countryman}, {Cousins}, {Couvares}, {Coward}, {Cowart}, {Coyne}, {Coyne},
  {Creighton}, {Creighton}, {Criswell}, {Croquette}, {Crowder}, {Cudell},
  {Cullen}, {Cumming}, {Cummings}, {Cunningham}, {Cuoco}, {Cury{\l}o},
  {Dabadie}, {Canton}, {Dall'Osso}, {D{\'a}lya}, {Dana},
  {DaneshgaranBajastani}, {D'Angelo}, {Danila}, {Danilishin}, {D'Antonio},
  {Danzmann}, {Darsow-Fromm}, {Dasgupta}, {Datrier}, {Datta}, {Dattilo},
  {Dave}, {Davier}, {Davies}, {Davis}, {Davis}, {Daw}, {Dean}, {DeBra},
  {Deenadayalan}, {Degallaix}, {De Laurentis}, {Del{\'e}glise}, {Del Favero},
  {De Lillo}, {De Lillo}, {Del Pozzo}, {DeMarchi}, {De Matteis}, {D'Emilio},
  {Demos}, {Dent}, {Depasse}, {De Pietri}, {De Rosa}, {De Rossi}, {DeSalvo},
  {De Simone}, {Dhurandhar}, {D{\'\i}az}, {Diaz-Ortiz}, {Didio}, {Dietrich},
  {Di Fiore}, {Di Fronzo}, {Di Giorgio}, {Di Giovanni}, {Di Giovanni}, {Di
  Girolamo}, {Di Lieto}, {Ding}, {Di Pace}, {Di Palma}, {Di Renzo},
  {Divakarla}, {Divyajyoti}, {Doctor}, {D'Onofrio}, {Donovan}, {Dooley},
  {Doravari}, {Dorrington}, {Drago}, {Driggers}, {Drori}, {Ducoin}, {Dupej},
  {Durante}, {D'Urso}, {Duverne}, {Dwyer}, {Eassa}, {Easter}, {Ebersold},
  {Eckhardt}, {Eddolls}, {Edelman}, {Edo}, {Edy}, {Effler}, {Eichholz},
  {Eikenberry}, {Eisenmann}, {Eisenstein}, {Ejlli}, {Engelby}, {Errico},
  {Essick}, {Estell{\'e}s}, {Estevez}, {Etienne}, {Etzel}, {Evans}, {Evans},
  {Ewing}, {Fafone}, {Fair}, {Fairhurst}, {Fanning}, {Farah}, {Farinon},
  {Farr}, {Farr}, {Farrow}, {Fauchon-Jones}, {Favaro}, {Favata}, {Fays},
  {Fazio}, {Feicht}, {Fejer}, {Fenyvesi}, {Ferguson}, {Fernandez-Galiana},
  {Ferrante}, {Ferreira}, {Fidecaro}, {Figura}, {Fiori}, {Fishbach}, {Fisher},
  {Fittipaldi}, {Fiumara}, {Flaminio}, {Floden}, {Fong}, {Font}, {Fornal},
  {Forsyth}, {Franke}, {Frasca}, {Frasconi}, {Frederick}, {Freed}, {Frei},
  {Freise}, {Frey}, {Fritschel}, {Frolov}, {Fronz{\'e}}, {Fulda}, {Fyffe},
  {Gabbard}, {Gabella}, {Gadre}, {Gair}, {Gais}, {Galaudage}, {Gamba},
  {Ganapathy}, {Ganguly}, {Gaonkar}, {Garaventa}, {Garc{\'\i}a},
  {Garc{\'\i}a-N{\'u}{\~n}ez}, {Garc{\'\i}a-Quir{\'o}s}, {Garufi}, {Gateley},
  {Gaudio}, {Gayathri}, {Gemme}, {Gennai}, {George}, {George}, {Gerberding},
  {Gergely}, {Gewecke}, {Ghonge}, {Ghosh}, {Ghosh}, {Ghosh}, {Ghosh},
  {Giacomazzo}, {Giacoppo}, {Giaime}, {Giardina}, {Gibson}, {Gier}, {Giesler},
  {Giri}, {Gissi}, {Glanzer}, {Gleckl}, {Godwin}, {Goetz}, {Goetz}, {Gohlke},
  {Goncharov}, {Gonz{\'a}lez}, {Gopakumar}, {Gosselin}, {Gouaty}, {Gould},
  {Grace}, {Grado}, {Granata}, {Granata}, {Grant}, {Gras}, {Grassia}, {Gray},
  {Gray}, {Greco}, {Green}, {Green}, {Gretarsson}, {Gretarsson}, {Griffith},
  {Griffiths}, {Griggs}, {Grignani}, {Grimaldi}, {Grimm}, {Grote}, {Grunewald},
  {Gruning}, {Guerra}, {Guidi}, {Guimaraes}, {Guix{\'e}}, {Gulati}, {Guo},
  {Guo}, {Gupta}, {Gupta}, {Gupta}, {Gustafson}, {Gustafson}, {Guzman},
  {Haegel}, {Halim}, {Hall}, {Hamilton}, {Hammond}, {Haney}, {Hanks}, {Hanna},
  {Hannam}, {Hannuksela}, {Hansen}, {Hansen}, {Hanson}, {Harder}, {Hardwick},
  {Haris}, {Harms}, {Harry}, {Harry}, {Hartwig}, {Haskell}, {Hasskew},
  {Haster}, {Haughian}, {Hayes}, {Healy}, {Heidmann}, {Heidt}, {Heintze},
  {Heinze}, {Heinzel}, {Heitmann}, {Hellman}, {Hello}, {Helmling-Cornell},
  {Hemming}, {Hendry}, {Heng}, {Hennes}, {Hennig}, {Hennig}, {Hernandez},
  {Vivanco}, {Heurs}, {Hild}, {Hill}, {Hines}, {Hochheim}, {Hofman}, {Hohmann},
  {Holcomb}, {Holland}, {Holley-Bockelmann}, {Hollows}, {Holmes}, {Holt},
  {Holz}, {Hopkins}, {Hough}, {Hourihane}, {Howell}, {Hoy}, {Hoyland},
  {Hreibi}, {Hsu}, {Huang}, {H{\"u}bner}, {Huddart}, {Hughey}, {Hui}, {Husa},
  {Huttner}, {Huxford}, {Huynh-Dinh}, {Idzkowski}, {Iess}, {Ingram}, {Isi},
  {Isleif}, {Iyer}, {JaberianHamedan}, {Jacqmin}, {Jadhav}, {Jadhav}, {James},
  {Jan}, {Jani}, {Janquart}, {Janssens}, {Janthalur}, {Jaranowski}, {Jariwala},
  {Jaume}, {Jenkins}, {Jenner}, {Jeunon}, {Jia}, {Johns}, {Johnson-McDaniel},
  {Jones}, {Jones}, {Jones}, {Jones}, {Jones}, {Jonker}, {Ju}, {Junker},
  {Juste}, {Kalaghatgi}, {Kalogera}, {Kamai}, {Kandhasamy}, {Kang}, {Kanner},
  {Kao}, {Kapadia}, {Kapasi}, {Karat}, {Karathanasis}, {Karki}, {Kashyap},
  {Kasprzack}, {Kastaun}, {Katsanevas}, {Katsavounidis}, {Katzman}, {Kaur},
  {Kawabe}, {K{\'e}f{\'e}lian}, {Keitel}, {Key}, {Khadka}, {Khalili}, {Khan},
  {Khazanov}, {Khetan}, {Khursheed}, {Kijbunchoo}, {Kim}, {Kim}, {Kim}, {Kim},
  {Kim}, {Kimball}, {Kinley-Hanlon}, {Kirchhoff}, {Kissel}, {Kleybolte},
  {Klimenko}, {Knee}, {Knowles}, {Knyazev}, {Koch}, {Koekoek}, {Koley},
  {Kolitsidou}, {Kolstein}, {Komori}, {Kondrashov}, {Kontos}, {Koper},
  {Korobko}, {Kovalam}, {Kozak}, {Kringel}, {Krishnendu}, {Kr{\'o}lak},
  {Kuehn}, {Kuei}, {Kuijer}, {Kulkarni}, {Kumar}, {Kumar}, {Kumar}, {Kumar},
  {Kuns}, {Kuwahara}, {Lagabbe}, {Laghi}, {Lalande}, {Lam}, {Lamberts},
  {Landry}, {Lane}, {Lang}, {Lange}, {Lantz}, {La Rosa}, {Lartaux-Vollard},
  {Lasky}, {Laxen}, {Lazzarini}, {Lazzaro}, {Leaci}, {Leavey}, {Lecoeuche},
  {Lee}, {Lee}, {Lee}, {Lee}, {Lehmann}, {Lema{\^\i}tre}, {Leroy}, {Letendre},
  {Levesque}, {Levin}, {Leviton}, {Leyde}, {Li}, {Li}, {Li}, {Li}, {Li},
  {Linde}, {Linker}, {Linley}, {Littenberg}, {Liu}, {Liu}, {Liu}, {Llamas},
  {Llorens-Monteagudo}, {Lo}, {Lockwood}, {London}, {Longo}, {Lopez},
  {Portilla}, {Lorenzini}, {Loriette}, {Lormand}, {Losurdo}, {Lott}, {Lough},
  {Lousto}, {Lovelace}, {Lucaccioni}, {L{\"u}ck}, {Lumaca}, {Lundgren},
  {Lynam}, {Macas}, {MacInnis}, {Macleod}, {MacMillan}, {Macquet}, {Hernandez},
  {Magazz{\`u}}, {Magee}, {Maggiore}, {Magnozzi}, {Mahesh}, {Majorana},
  {Makarem}, {Maksimovic}, {Maliakal}, {Malik}, {Man}, {Mandic}, {Mangano},
  {Mango}, {Mansell}, {Manske}, {Mantovani}, {Mapelli}, {Marchesoni}, {Marion},
  {Mark}, {M{\'a}rka}, {M{\'a}rka}, {Markakis}, {Markosyan}, {Markowitz},
  {Maros}, {Marquina}, {Marsat}, {Martelli}, {Martin}, {Martin}, {Martinez},
  {Martinez}, {Martinez}, {Martinovic}, {Martynov}, {Marx}, {Masalehdan},
  {Mason}, {Massera}, {Masserot}, {Massinger}, {Masso-Reid}, {Mastrogiovanni},
  {Matas}, {Mateu-Lucena}, {Matichard}, {Matiushechkina}, {Mavalvala},
  {McCann}, {McCarthy}, {McClelland}, {McClincy}, {McCormick}, {McCuller},
  {McGhee}, {McGuire}, {McIsaac}, {McIver}, {McRae}, {McWilliams}, {Meacher},
  {Mehmet}, {Mehta}, {Meijer}, {Melatos}, {Melchor}, {Mendell},
  {Menendez-Vazquez}, {Menoni}, {Mercer}, {Mereni}, {Merfeld}, {Merilh},
  {Merritt}, {Merzougui}, {Meshkov}, {Messenger}, {Messick}, {Meyers},
  {Meylahn}, {Mhaske}, {Miani}, {Miao}, {Michaloliakos}, {Michel}, {Middleton},
  {Milano}, {Miller}, {Miller}, {Miller}, {Millhouse}, {Mills}, {Milotti},
  {Minazzoli}, {Minenkov}, {Mir}, {Miravet-Ten{\'e}s}, {Mishra}, {Mishra},
  {Mistry}, {Mitra}, {Mitrofanov}, {Mitselmakher}, {Mittleman}, {Mo}, {Moguel},
  {Mogushi}, {Mohapatra}, {Mohite}, {Molina}, {Molina-Ruiz}, {Mondin},
  {Montani}, {Moore}, {Moraru}, {Morawski}, {More}, {Moreno}, {Moreno},
  {Morisaki}, {Mours}, {Mow-Lowry}, {Mozzon}, {Muciaccia}, {Mukherjee},
  {Mukherjee}, {Mukherjee}, {Mukherjee}, {Mukherjee}, {Mukund}, {Mullavey},
  {Munch}, {Mu{\~n}iz}, {Murray}, {Musenich}, {Muusse}, {Nadji}, {Nagar},
  {Napolano}, {Nardecchia}, {Naticchioni}, {Nayak}, {Nayak}, {Neil}, {Neilson},
  {Nelemans}, {Nelson}, {Nery}, {Neubauer}, {Neunzert}, {Ng}, {Ng}, {Nguyen},
  {Nguyen}, {Nguyen}, {Nichols}, {Nissanke}, {Nitoglia}, {Nocera}, {Norman},
  {North}, {Nuttall}, {Oberling}, {O'Brien}, {O'Dell}, {Oelker}, {Oganesyan},
  {Oh}, {Oh}, {Ohme}, {Ohta}, {Okada}, {Olivetto}, {Oram}, {O'Reilly},
  {Ormiston}, {Ormsby}, {Ortega}, {O'Shaughnessy}, {O'Shea}, {Ossokine},
  {Osthelder}, {Ottaway}, {Overmier}, {Pace}, {Pagano}, {Page}, {Pagliaroli},
  {Pai}, {Pai}, {Palamos}, {Palashov}, {Palomba}, {Pan}, {Panda}, {Pang},
  {Pankow}, {Pannarale}, {Pant}, {Panther}, {Paoletti}, {Paoli}, {Paolone},
  {Park}, {Parker}, {Pascucci}, {Pasqualetti}, {Passaquieti}, {Passuello},
  {Patel}, {Pathak}, {Patricelli}, {Patron}, {Patrone}, {Paul}, {Payne},
  {Pedraza}, {Pegoraro}, {Pele}, {Penn}, {Perego}, {Pereira}, {Pereira},
  {Perez}, {P{\'e}rigois}, {Perkins}, {Perreca}, {Perri{\`e}s}, {Petermann},
  {Petterson}, {Pfeiffer}, {Pham}, {Phukon}, {Piccinni}, {Pichot},
  {Piendibene}, {Piergiovanni}, {Pierini}, {Pierro}, {Pillant}, {Pillas},
  {Pilo}, {Pinard}, {Pinto}, {Pinto}, {Piotrzkowski}, {Pirello}, {Pitkin},
  {Placidi}, {Planas}, {Plastino}, {Pluchar}, {Poggiani}, {Polini}, {Pong},
  {Ponrathnam}, {Popolizio}, {Porter}, {Poulton}, {Powell}, {Pracchia},
  {Pradier}, {Prajapati}, {Prasai}, {Prasanna}, {Pratten}, {Principe}, {Prodi},
  {Prokhorov}, {Prosposito}, {Prudenzi}, {Puecher}, {Punturo}, {Puosi},
  {Puppo}, {P{\"u}rrer}, {Qi}, {Quetschke}, {Quitzow-James}, {Raab},
  {Raaijmakers}, {Radkins}, {Radulesco}, {Raffai}, {Rail}, {Raja}, {Rajan},
  {Ramirez}, {Ramirez}, {Ramos-Buades}, {Rana}, {Rapagnani}, {Rapol}, {Ray},
  {Raymond}, {Raza}, {Razzano}, {Read}, {Rees}, {Regimbau}, {Rei}, {Reid},
  {Reid}, {Reitze}, {Relton}, {Renzini}, {Rettegno}, {Reza}, {Rezac}, {Ricci},
  {Richards}, {Richardson}, {Richardson}, {Riemenschneider}, {Riles},
  {Rinaldi}, {Rink}, {Rizzo}, {Robertson}, {Robie}, {Robinet}, {Rocchi},
  {Rodriguez}, {Rolland}, {Rollins}, {Romanelli}, {Romano}, {Romel},
  {Romero-Rodr{\'\i}guez}, {Romero-Shaw}, {Romie}, {Ronchini}, {Rosa}, {Rose},
  {Rosell}, {Rosi{\'n}ska}, {Ross}, {Rowan}, {Rowlinson}, {Roy}, {Roy}, {Roy},
  {Rozza}, {Ruggi}, {Ruiz-Rocha}, {Ryan}, {Sachdev}, {Sadecki}, {Sadiq},
  {Sakellariadou}, {Salafia}, {Salconi}, {Saleem}, {Salemi}, {Samajdar},
  {Sanchez}, {Sanchez}, {Sanchez}, {Sanchis-Gual}, {Sanders}, {Sanuy},
  {Saravanan}, {Sarin}, {Sassolas}, {Satari}, {Sauter}, {Savage}, {Sawant},
  {Sawant}, {Sayah}, {Schaetzl}, {Scheel}, {Scheuer}, {Schiworski}, {Schmidt},
  {Schmidt}, {Schnabel}, {Schneewind}, {Schofield}, {Sch{\"o}nbeck}, {Schulte},
  {Schutz}, {Schwartz}, {Scott}, {Scott}, {Seglar-Arroyo}, {Sellers},
  {Sengupta}, {Sentenac}, {Seo}, {Sequino}, {Sergeev}, {Setyawati}, {Shaffer},
  {Shahriar}, {Shams}, {Sharma}, {Sharma}, {Shawhan}, {Shcheblanov},
  {Shikauchi}, {Shoemaker}, {Shoemaker}, {ShyamSundar}, {Sieniawska}, {Sigg},
  {Singer}, {Singh}, {Singh}, {Singha}, {Sintes}, {Sipala}, {Skliris},
  {Slagmolen}, {Slaven-Blair}, {Smetana}, {Smith}, {Smith}, {Soldateschi},
  {Somala}, {Son}, {Soni}, {Soni}, {Sordini}, {Sorrentino}, {Sorrentino},
  {Soulard}, {Souradeep}, {Sowell}, {Spagnuolo}, {Spencer}, {Spera},
  {Srinivasan}, {Srivastava}, {Srivastava}, {Staats}, {Stachie}, {Steer},
  {Steinhoff}, {Steinlechner}, {Steinlechner}, {Stevenson}, {Stops}, {Stover},
  {Strain}, {Strang}, {Stratta}, {Strunk}, {Sturani}, {Stuver}, {Sudhagar},
  {Sudhir}, {Suh}, {Summerscales}, {Sun}, {Sun}, {Sunil}, {Sur}, {Suresh},
  {Sutton}, {Swinkels}, {Szczepa{\'n}czyk}, {Szewczyk}, {Tacca}, {Tait},
  {Talbot}, {Talbot}, {Tanasijczuk}, {Tanner}, {Tao}, {Tao}, {Mart{\'\i}n},
  {Taranto}, {Tasson}, {Tenorio}, {Terhune}, {Terkowski},
  {Thirugnanasambandam}, {Thomas}, {Thomas}, {Thomas}, {Thompson}, {Thondapu},
  {Thorne}, {Thrane}, {Tiwari}, {Tiwari}, {Tiwari}, {Toivonen}, {Toland},
  {Tolley}, {Tonelli}, {Torres-Forn{\'e}}, {Torrie}, {e Melo}, {T{\"o}yr{\"a}},
  {Trapananti}, {Travasso}, {Traylor}, {Trevor}, {Tringali}, {Tripathee},
  {Troiano}, {Trovato}, {Trozzo}, {Trudeau}, {Tsai}, {Tsai}, {Tsang}, {Tse},
  {Tso}, {Tsukada}, {Tsuna}, {Tsutsui}, {Turbang}, {Turconi}, {Ubhi}, {Udall},
  {Ueno}, {Unnikrishnan}, {Urban}, {Utina}, {Vahlbruch}, {Vajente}, {Vajpeyi},
  {Valdes}, {Valentini}, {Valsan}, {van Bakel}, {van Beuzekom}, {van den
  Brand}, {Van Den Broeck}, {Vander-Hyde}, {van der Schaaf}, {van Heijningen},
  {Vanosky}, {van Remortel}, {Vardaro}, {Vargas}, {Varma}, {Vas{\'u}th},
  {Vecchio}, {Vedovato}, {Veitch}, {Veitch}, {Venneberg}, {Venugopalan},
  {Verkindt}, {Verma}, {Verma}, {Veske}, {Vetrano}, {Vicer{\'e}}, {Vidyant},
  {Viets}, {Vijaykumar}, {Villa-Ortega}, {Vinet}, {Virtuoso}, {Vitale}, {Vo},
  {Vocca}, {von Reis}, {von Wrangel}, {Vorvick}, {Vyatchanin}, {Wade}, {Wade},
  {Wagner}, {Walet}, {Walker}, {Wallace}, {Wallace}, {Walsh}, {Wang}, {Wang},
  {Ward}, {Warner}, {Was}, {Washington}, {Watchi}, {Weaver}, {Webster},
  {Weinert}, {Weinstein}, {Weiss}, {Weller}, {Weller}, {Wellmann}, {Wen},
  {We{\ss}els}, {Wette}, {Whelan}, {White}, {Whiting}, {Whittle}, {Wilken},
  {Williams}, {Williams}, {Williamson}, {Willis}, {Willke}, {Wilson},
  {Winkler}, {Wipf}, {Wlodarczyk}, {Woan}, {Woehler}, {Wofford}, {Wong}, {Wu},
  {Wysocki}, {Xiao}, {Yamamoto}, {Yang}, {Yang}, {Yang}, {Yang}, {Yap},
  {Yeeles}, {Yelikar}, {Ying}, {Yoo}, {Yu}, {Yu}, {Zadro{\.Z}ny}, {Zanolin},
  {Zelenova}, {Zendri}, {Zevin}, {Zhang}, {Zhang}, {Zhang}, {Zhang}, {Zhao},
  {Zhao}, {Zhao}, {Zhou}, {Zhou}, {Zhu}, {Zimmerman}, {Zlochower}, {Zucker},
  {Zweizig}, {LIGO Scientific Collaboration}, \& {the Virgo
  Collaboration}}]{ABBOTT2024}
{Abbott}, R., {Abbott}, T.~D., {Acernese}, F., {et~al.} 2024, \prd, 109, 022001

\bibitem[{{Agarwal} \& {Milosavljevi{\'c}}(2011)}]{AGARWAL2011}
{Agarwal}, M. \& {Milosavljevi{\'c}}, M. 2011, \apj, 729, 35

\bibitem[{{Amaro-Seoane} {et~al.}(2017){Amaro-Seoane}, {Audley}, {Babak},
  {Baker}, {Barausse}, {Bender}, {Berti}, {Binetruy}, {Born}, {Bortoluzzi},
  {Camp}, {Caprini}, {Cardoso}, {Colpi}, {Conklin}, {Cornish}, {Cutler},
  {Danzmann}, {Dolesi}, {Ferraioli}, {Ferroni}, {Fitzsimons}, {Gair}, {Gesa
  Bote}, {Giardini}, {Gibert}, {Grimani}, {Halloin}, {Heinzel}, {Hertog},
  {Hewitson}, {Holley-Bockelmann}, {Hollington}, {Hueller}, {Inchauspe},
  {Jetzer}, {Karnesis}, {Killow}, {Klein}, {Klipstein}, {Korsakova}, {Larson},
  {Livas}, {Lloro}, {Man}, {Mance}, {Martino}, {Mateos}, {McKenzie},
  {McWilliams}, {Miller}, {Mueller}, {Nardini}, {Nelemans}, {Nofrarias},
  {Petiteau}, {Pivato}, {Plagnol}, {Porter}, {Reiche}, {Robertson},
  {Robertson}, {Rossi}, {Russano}, {Schutz}, {Sesana}, {Shoemaker}, {Slutsky},
  {Sopuerta}, {Sumner}, {Tamanini}, {Thorpe}, {Troebs}, {Vallisneri},
  {Vecchio}, {Vetrugno}, {Vitale}, {Volonteri}, {Wanner}, {Ward}, {Wass},
  {Weber}, {Ziemer}, \& {Zweifel}}]{AMARO2017}
{Amaro-Seoane}, P., {Audley}, H., {Babak}, S., {et~al.} 2017, arXiv e-prints,
  arXiv:1702.00786, submitted to ESA

\bibitem[{{Antonini}(2013)}]{ANTONINI2013}
{Antonini}, F. 2013, \apj, 763, 62

\bibitem[{{Antonini} {et~al.}(2015){Antonini}, {Barausse}, \&
  {Silk}}]{ANTONINI2015}
{Antonini}, F., {Barausse}, E., \& {Silk}, J. 2015, \apj, 812, 72

\bibitem[{{Antonini} \& {Perets}(2012)}]{ANTONINI2012}
{Antonini}, F. \& {Perets}, H.~B. 2012, \apj, 757, 27

\bibitem[{{Ba{\~n}ados} {et~al.}(2016){Ba{\~n}ados}, {Venemans}, {Decarli},
  {Farina}, {Mazzucchelli}, {Walter}, {Fan}, {Stern}, {Schlafly}, {Chambers},
  {Rix}, {Jiang}, {McGreer}, {Simcoe}, {Wang}, {Yang}, {Morganson}, {De Rosa},
  {Greiner}, {Balokovi{\'c}}, {Burgett}, {Cooper}, {Draper}, {Flewelling},
  {Hodapp}, {Jun}, {Kaiser}, {Kudritzki}, {Magnier}, {Metcalfe}, {Miller},
  {Schindler}, {Tonry}, {Wainscoat}, {Waters}, \& {Yang}}]{BANADOS2016}
{Ba{\~n}ados}, E., {Venemans}, B.~P., {Decarli}, R., {et~al.} 2016, \apjs, 227,
  11

\bibitem[{{Balcells} {et~al.}(2003){Balcells}, {Graham},
  {Dom{\'\i}nguez-Palmero}, \& {Peletier}}]{BALCELLS2003}
{Balcells}, M., {Graham}, A.~W., {Dom{\'\i}nguez-Palmero}, L., \& {Peletier},
  R.~F. 2003, \apjl, 582, L79

\bibitem[{{Baldry} {et~al.}(2012){Baldry}, {Driver}, {Loveday}, {Taylor},
  {Kelvin}, {Liske}, {Norberg}, {Robotham}, {Brough}, {Hopkins}, {Bamford},
  {Peacock}, {Bland-Hawthorn}, {Conselice}, {Croom}, {Jones}, {Parkinson},
  {Popescu}, {Prescott}, {Sharp}, \& {Tuffs}}]{BALDRY2012}
{Baldry}, I.~K., {Driver}, S.~P., {Loveday}, J., {et~al.} 2012, \mnras, 421,
  621

\bibitem[{{Balick} \& {Brown}(1974)}]{BALICK1974}
{Balick}, B. \& {Brown}, R.~L. 1974, \apj, 194, 265

\bibitem[{{Banerjee} \& {Kroupa}(2017)}]{BANERJEE2017}
{Banerjee}, S. \& {Kroupa}, P. 2017, \aap, 597, A28

\bibitem[{{Barth} {et~al.}(2002){Barth}, {Ho}, \& {Sargent}}]{BARTH2002}
{Barth}, A.~J., {Ho}, L.~C., \& {Sargent}, W. L.~W. 2002, \aj, 124, 2607

\bibitem[{{Barth} {et~al.}(2009){Barth}, {Strigari}, {Bentz}, {Greene}, \&
  {Ho}}]{BARTH2009}
{Barth}, A.~J., {Strigari}, L.~E., {Bentz}, M.~C., {Greene}, J.~E., \& {Ho},
  L.~C. 2009, \apj, 690, 1031

\bibitem[{{Becklin} \& {Neugebauer}(1968)}]{BECKLIN1968}
{Becklin}, E.~E. \& {Neugebauer}, G. 1968, \apj, 151, 145

\bibitem[{{Begelman} \& {Rees}(1978)}]{BEGELMAN1978}
{Begelman}, M.~C. \& {Rees}, M.~J. 1978, \mnras, 185, 847

\bibitem[{{Bekki}(2007)}]{BEKKI2007}
{Bekki}, K. 2007, \pasa, 24, 77

\bibitem[{{Bekki}(2010)}]{BEKKI2010}
{Bekki}, K. 2010, \mnras, 401, 2753

\bibitem[{{Bekki} {et~al.}(2006){Bekki}, {Couch}, \& {Shioya}}]{BEKKI2006}
{Bekki}, K., {Couch}, W.~J., \& {Shioya}, Y. 2006, \apjl, 642, L133

\bibitem[{{Bender} {et~al.}(2005){Bender}, {Kormendy}, {Bower}, {Green},
  {Thomas}, {Danks}, {Gull}, {Hutchings}, {Joseph}, {Kaiser}, {Lauer},
  {Nelson}, {Richstone}, {Weistrop}, \& {Woodgate}}]{BENDER2005}
{Bender}, R., {Kormendy}, J., {Bower}, G., {et~al.} 2005, \apj, 631, 280

\bibitem[{{Benson}(2012)}]{BENSON2012}
{Benson}, A.~J. 2012, \na, 17, 175

\bibitem[{{Benson}(2017)}]{BENSON2017}
{Benson}, A.~J. 2017, \mnras, 467, 3454

\bibitem[{{Benson} \& {Bower}(2010)}]{BENSON2010b}
{Benson}, A.~J. \& {Bower}, R. 2010, \mnras, 405, 1573

\bibitem[{{Benson} {et~al.}(2001){Benson}, {Pearce}, {Frenk}, {Baugh}, \&
  {Jenkins}}]{BENSON2001}
{Benson}, A.~J., {Pearce}, F.~R., {Frenk}, C.~S., {Baugh}, C.~M., \& {Jenkins},
  A. 2001, \mnras, 320, 261

\bibitem[{{Binney} \& {Tremaine}(2008)}]{BINNEY2008}
{Binney}, J. \& {Tremaine}, S. 2008, {Galactic Dynamics: Second Edition}
  (Princeton University Press)

\bibitem[{{Bland-Hawthorn} \& {Gerhard}(2016)}]{BLAND2016}
{Bland-Hawthorn}, J. \& {Gerhard}, O. 2016, \araa, 54, 529

\bibitem[{{Boeker} {et~al.}(2003){Boeker}, {van der Marel}, {Gerssen},
  {Walcher}, {Rix}, {Shields}, \& {Ho}}]{BOEKER2003}
{Boeker}, T., {van der Marel}, R.~P., {Gerssen}, J., {et~al.} 2003, in Society
  of Photo-Optical Instrumentation Engineers (SPIE) Conference Series, Vol.
  4834, Discoveries and Research Prospects from 6- to 10-Meter-Class Telescopes
  II, ed. P.~{Guhathakurta}, 57--65

\bibitem[{{Boekholt} {et~al.}(2018){Boekholt}, {Schleicher}, {Fellhauer},
  {Klessen}, {Reinoso}, {Stutz}, \& {Haemmerl{\'e}}}]{Boekholt2018}
{Boekholt}, T.~C.~N., {Schleicher}, D.~R.~G., {Fellhauer}, M., {et~al.} 2018,
  \mnras, 476, 366

\bibitem[{{B{\"o}ker}(2010)}]{BOKER2010}
{B{\"o}ker}, T. 2010, in Star Clusters: Basic Galactic Building Blocks
  Throughout Time and Space, ed. R.~{de Grijs} \& J.~R.~D. {L{\'e}pine}, Vol.
  266, 58--63

\bibitem[{{B{\"o}ker} {et~al.}(1999){B{\"o}ker}, {van der Marel}, \&
  {Vacca}}]{BOKER1999}
{B{\"o}ker}, T., {van der Marel}, R.~P., \& {Vacca}, W.~D. 1999, \aj, 118, 831

\bibitem[{{Bower} {et~al.}(2001){Bower}, {Green}, {Bender}, {Gebhardt},
  {Lauer}, {Magorrian}, {Richstone}, {Danks}, {Gull}, {Hutchings}, {Joseph},
  {Kaiser}, {Weistrop}, {Woodgate}, {Nelson}, \& {Malumuth}}]{BOWER2001}
{Bower}, G.~A., {Green}, R.~F., {Bender}, R., {et~al.} 2001, \apj, 550, 75

\bibitem[{{Bower} {et~al.}(1998){Bower}, {Green}, {Danks}, {Gull}, {Heap},
  {Hutchings}, {Joseph}, {Kaiser}, {Kimble}, {Kraemer}, {Weistrop}, {Woodgate},
  {Lindler}, {Hill}, {Malumuth}, {Baum}, {Sarajedini}, {Heckman}, {Wilson}, \&
  {Richstone}}]{BOWER1998}
{Bower}, G.~A., {Green}, R.~F., {Danks}, A., {et~al.} 1998, \apjl, 492, L111

\bibitem[{{Bower} {et~al.}(2010){Bower}, {Vernon}, {Goldstein}, {Benson},
  {Lacey}, {Baugh}, {Cole}, \& {Frenk}}]{BOWER2010}
{Bower}, R.~G., {Vernon}, I., {Goldstein}, M., {et~al.} 2010, \mnras, 407, 2017

\bibitem[{{Bromm} \& {Loeb}(2003)}]{Bromm2003}
{Bromm}, V. \& {Loeb}, A. 2003, \apj, 596, 34

\bibitem[{{Burbidge} {et~al.}(1961){Burbidge}, {Burbidge}, \&
  {Fish}}]{BURBIDGE1961}
{Burbidge}, E.~M., {Burbidge}, G.~R., \& {Fish}, R.~A. 1961, \apj, 134, 251

\bibitem[{{Cappellari} {et~al.}(2013){Cappellari}, {McDermid}, {Alatalo},
  {Blitz}, {Bois}, {Bournaud}, {Bureau}, {Crocker}, {Davies}, {Davis}, {de
  Zeeuw}, {Duc}, {Emsellem}, {Khochfar}, {Krajnovi{\'c}}, {Kuntschner},
  {Morganti}, {Naab}, {Oosterloo}, {Sarzi}, {Scott}, {Serra}, {Weijmans}, \&
  {Young}}]{CAPPELLARI2013}
{Cappellari}, M., {McDermid}, R.~M., {Alatalo}, K., {et~al.} 2013, \mnras, 432,
  1862

\bibitem[{{Cappellari} {et~al.}(2002){Cappellari}, {Verolme}, {van der Marel},
  {Verdoes Kleijn}, {Illingworth}, {Franx}, {Carollo}, \& {de
  Zeeuw}}]{CAPPELLARI2002}
{Cappellari}, M., {Verolme}, E.~K., {van der Marel}, R.~P., {et~al.} 2002,
  \apj, 578, 787

\bibitem[{{Capuzzo-Dolcetta}(1993)}]{CAPUZZO1993}
{Capuzzo-Dolcetta}, R. 1993, \apj, 415, 616

\bibitem[{{Capuzzo-Dolcetta} \& {Mastrobuono-Battisti}(2009)}]{CAPUZZO2009}
{Capuzzo-Dolcetta}, R. \& {Mastrobuono-Battisti}, A. 2009, \aap, 507, 183

\bibitem[{{Carollo} {et~al.}(1997){Carollo}, {Stiavelli}, {de Zeeuw}, \&
  {Mack}}]{CAROLLO1997}
{Carollo}, C.~M., {Stiavelli}, M., {de Zeeuw}, P.~T., \& {Mack}, J. 1997, \aj,
  114, 2366

\bibitem[{{Chon} \& {Omukai}(2020)}]{Chon2020}
{Chon}, S. \& {Omukai}, K. 2020, \mnras, 494, 2851

\bibitem[{{Cole} {et~al.}(2000){Cole}, {Lacey}, {Baugh}, \& {Frenk}}]{COLE2000}
{Cole}, S., {Lacey}, C.~G., {Baugh}, C.~M., \& {Frenk}, C.~S. 2000, \mnras,
  319, 168

\bibitem[{{C{\^o}t{\'e}} {et~al.}(2018){C{\^o}t{\'e}}, {Silvia}, {O'Shea},
  {Smith}, \& {Wise}}]{COTE2018}
{C{\^o}t{\'e}}, B., {Silvia}, D.~W., {O'Shea}, B.~W., {Smith}, B., \& {Wise},
  J.~H. 2018, \apj, 859, 67

\bibitem[{{C{\^o}t{\'e}} {et~al.}(2006){C{\^o}t{\'e}}, {Piatek}, {Ferrarese},
  {Jord{\'a}n}, {Merritt}, {Peng}, {Ha{\c{s}}egan}, {Blakeslee}, {Mei}, {West},
  {Milosavljevi{\'c}}, \& {Tonry}}]{COTE2006}
{C{\^o}t{\'e}}, P., {Piatek}, S., {Ferrarese}, L., {et~al.} 2006, \apjs, 165,
  57

\bibitem[{{Cretton} \& {van den Bosch}(1999)}]{CRETTON1999}
{Cretton}, N. \& {van den Bosch}, F.~C. 1999, \apj, 514, 704

\bibitem[{Cross \& Driver(2002)}]{CROSS2002}
Cross, N. \& Driver, S.~P. 2002, Monthly Notices of the Royal Astronomical
  Society, 329, 579

\bibitem[{{Croton} {et~al.}(2016){Croton}, {Stevens}, {Tonini}, {Garel},
  {Bernyk}, {Bibiano}, {Hodkinson}, {Mutch}, {Poole}, \&
  {Shattow}}]{CROTON2016}
{Croton}, D.~J., {Stevens}, A. R.~H., {Tonini}, C., {et~al.} 2016, \apjs, 222,
  22

\bibitem[{Dametto {et~al.}(2014)Dametto, Riffel, Pastoriza, Rodr'iguez-Ardila,
  Hernandez-Jimenez, Astronomia, do~Rio Grande~do Sul, Brasil., Astrof'isica,
  de~Itajub'a, Rei, \& do~Sudeste~de Minas}]{DAMETTO2014}
Dametto, N., Riffel, R., Pastoriza, M., {et~al.} 2014, Monthly Notices of the
  Royal Astronomical Society, 443, 1754

\bibitem[{{Davies} {et~al.}(2011){Davies}, {Miller}, \&
  {Bellovary}}]{Davies2011}
{Davies}, M.~B., {Miller}, M.~C., \& {Bellovary}, J.~M. 2011, \apjl, 740, L42

\bibitem[{{de Francesco} {et~al.}(2006){de Francesco}, {Capetti}, \&
  {Marconi}}]{DEFRANCESCO2006}
{de Francesco}, G., {Capetti}, A., \& {Marconi}, A. 2006, \aap, 460, 439

\bibitem[{{den Brok} {et~al.}(2015){den Brok}, {Seth}, {Barth}, {Carson},
  {Neumayer}, {Cappellari}, {Debattista}, {Ho}, {Hood}, \&
  {McDermid}}]{DENBROK2015}
{den Brok}, M., {Seth}, A.~C., {Barth}, A.~J., {et~al.} 2015, \apj, 809, 101

\bibitem[{{Devereux} {et~al.}(2003){Devereux}, {Ford}, {Tsvetanov}, \&
  {Jacoby}}]{DEVEREUX2003}
{Devereux}, N., {Ford}, H., {Tsvetanov}, Z., \& {Jacoby}, G. 2003, \aj, 125,
  1226

\bibitem[{{Di Carlo} {et~al.}(2021){Di Carlo}, {Mapelli}, {Pasquato},
  {Rastello}, {Ballone}, {Dall'Amico}, {Giacobbo}, {Iorio}, {Spera},
  {Torniamenti}, \& {Haardt}}]{DICARLO2021}
{Di Carlo}, U.~N., {Mapelli}, M., {Pasquato}, M., {et~al.} 2021, \mnras, 507,
  5132

\bibitem[{{Ding} {et~al.}(2020){Ding}, {Silverman}, {Treu}, {Schulze},
  {Schramm}, {Birrer}, {Park}, {Jahnke}, {Bennert}, {Kartaltepe}, {Koekemoer},
  {Malkan}, \& {Sanders}}]{DING2020}
{Ding}, X., {Silverman}, J., {Treu}, T., {et~al.} 2020, \apj, 888, 37

\bibitem[{{Ebisuzaki}(2003)}]{EBISUZAKI2003}
{Ebisuzaki}, T. 2003, in Astrophysical Supercomputing using Particle
  Simulations, ed. J.~{Makino} \& P.~{Hut}, Vol. 208, 157

\bibitem[{Edgar(2004)}]{EDGAR2004}
Edgar, R. 2004, New Astronomy Reviews, 48, 843

\bibitem[{{Emsellem} {et~al.}(1999){Emsellem}, {Dejonghe}, \&
  {Bacon}}]{EMSELLEM1999}
{Emsellem}, E., {Dejonghe}, H., \& {Bacon}, R. 1999, \mnras, 303, 495

\bibitem[{{Emsellem} \& {van de Ven}(2008)}]{EMSELLEM2008}
{Emsellem}, E. \& {van de Ven}, G. 2008, \apj, 674, 653

\bibitem[{{Escala}(2021)}]{ESCALA2021}
{Escala}, A. 2021, \apj, 908, 57

\bibitem[{{Faber} \& {Jackson}(1976)}]{FABER1976}
{Faber}, S.~M. \& {Jackson}, R.~E. 1976, \apj, 204, 668

\bibitem[{{Fan} {et~al.}(2023){Fan}, {Ba{\~n}ados}, \& {Simcoe}}]{FAN2023}
{Fan}, X., {Ba{\~n}ados}, E., \& {Simcoe}, R.~A. 2023, \araa, 61, 373

\bibitem[{{Ferrara} {et~al.}(2014){Ferrara}, {Salvadori}, {Yue}, \&
  {Schleicher}}]{FERRARA2014}
{Ferrara}, A., {Salvadori}, S., {Yue}, B., \& {Schleicher}, D. 2014, \mnras,
  443, 2410

\bibitem[{{Ferrarese} {et~al.}(2006){Ferrarese}, {C{\^o}t{\'e}}, {Dalla
  Bont{\`a}}, {Peng}, {Merritt}, {Jord{\'a}n}, {Blakeslee}, {Ha{\c{s}}egan},
  {Mei}, {Piatek}, {Tonry}, \& {West}}]{FERRARESE2006}
{Ferrarese}, L., {C{\^o}t{\'e}}, P., {Dalla Bont{\`a}}, E., {et~al.} 2006,
  \apjl, 644, L21

\bibitem[{{Ferrarese} {et~al.}(1996){Ferrarese}, {Ford}, \&
  {Jaffe}}]{FERRARESE1996}
{Ferrarese}, L., {Ford}, H.~C., \& {Jaffe}, W. 1996, \apj, 470, 444

\bibitem[{{Ferrarese} \& {Merritt}(2000)}]{FERRARESE2000}
{Ferrarese}, L. \& {Merritt}, D. 2000, \apjl, 539, L9

\bibitem[{{Filippenko} \& {Ho}(2003)}]{FILIPPENKO2003}
{Filippenko}, A.~V. \& {Ho}, L.~C. 2003, \apjl, 588, L13

\bibitem[{{Freitag} {et~al.}(2006{\natexlab{a}}){Freitag}, {G{\"u}rkan}, \&
  {Rasio}}]{FREITAG2006A}
{Freitag}, M., {G{\"u}rkan}, M.~A., \& {Rasio}, F.~A. 2006{\natexlab{a}},
  \mnras, 368, 141

\bibitem[{{Freitag} {et~al.}(2006{\natexlab{b}}){Freitag}, {Rasio}, \&
  {Baumgardt}}]{FREITAG2006B}
{Freitag}, M., {Rasio}, F.~A., \& {Baumgardt}, H. 2006{\natexlab{b}}, \mnras,
  368, 121

\bibitem[{{Gaete} {et~al.}(2024){Gaete}, {Schleicher}, {Lupi}, {Reinoso},
  {Fellhauer}, \& {Vergara}}]{Gaete2024}
{Gaete}, B., {Schleicher}, D. R.~G., {Lupi}, A., {et~al.} 2024, arXiv e-prints,
  arXiv:2406.13072, accepted for publication with A\&A

\bibitem[{{Gebhardt} {et~al.}(2011){Gebhardt}, {Adams}, {Richstone}, {Lauer},
  {Faber}, {G{\"u}ltekin}, {Murphy}, \& {Tremaine}}]{GEBHARDT2011}
{Gebhardt}, K., {Adams}, J., {Richstone}, D., {et~al.} 2011, \apj, 729, 119

\bibitem[{{Gebhardt} {et~al.}(2001){Gebhardt}, {Lauer}, {Kormendy}, {Pinkney},
  {Bower}, {Green}, {Gull}, {Hutchings}, {Kaiser}, {Nelson}, {Richstone}, \&
  {Weistrop}}]{GEBHARDT2001}
{Gebhardt}, K., {Lauer}, T.~R., {Kormendy}, J., {et~al.} 2001, \aj, 122, 2469

\bibitem[{{Gebhardt} {et~al.}(2003){Gebhardt}, {Richstone}, {Tremaine},
  {Lauer}, {Bender}, {Bower}, {Dressler}, {Faber}, {Filippenko}, {Green},
  {Grillmair}, {Ho}, {Kormendy}, {Magorrian}, \& {Pinkney}}]{GEBHARDT2003}
{Gebhardt}, K., {Richstone}, D., {Tremaine}, S., {et~al.} 2003, \apj, 583, 92

\bibitem[{{Geha} {et~al.}(2002){Geha}, {Guhathakurta}, \& {van der
  Marel}}]{GEHA2002}
{Geha}, M., {Guhathakurta}, P., \& {van der Marel}, R.~P. 2002, \aj, 124, 3073

\bibitem[{{Geller} {et~al.}(2012){Geller}, {Diaferio}, {Kurtz}, {Dell'Antonio},
  \& {Fabricant}}]{GELLER2011}
{Geller}, M.~J., {Diaferio}, A., {Kurtz}, M.~J., {Dell'Antonio}, I.~P., \&
  {Fabricant}, D.~G. 2012, \aj, 143, 102

\bibitem[{{Genzel} {et~al.}(2010){Genzel}, {Eisenhauer}, \&
  {Gillessen}}]{GENZEL2010}
{Genzel}, R., {Eisenhauer}, F., \& {Gillessen}, S. 2010, Reviews of Modern
  Physics, 82, 3121

\bibitem[{{Georgiev} \& {B{\"o}ker}(2014)}]{GEORGIEV2014}
{Georgiev}, I.~Y. \& {B{\"o}ker}, T. 2014, \mnras, 441, 3570

\bibitem[{{Georgiev} {et~al.}(2016){Georgiev}, {B{\"o}ker}, {Leigh},
  {L{\"u}tzgendorf}, \& {Neumayer}}]{GEORGIEV2016}
{Georgiev}, I.~Y., {B{\"o}ker}, T., {Leigh}, N., {L{\"u}tzgendorf}, N., \&
  {Neumayer}, N. 2016, \mnras, 457, 2122

\bibitem[{{Glebbeek} {et~al.}(2009){Glebbeek}, {Gaburov}, {de Mink}, {Pols}, \&
  {Portegies Zwart}}]{GLEBBEEK2009}
{Glebbeek}, E., {Gaburov}, E., {de Mink}, S.~E., {Pols}, O.~R., \& {Portegies
  Zwart}, S.~F. 2009, \aap, 497, 255

\bibitem[{{Gonz{\'a}lez Delgado} {et~al.}(2008){Gonz{\'a}lez Delgado},
  {P{\'e}rez}, {Cid Fernandes}, \& {Schmitt}}]{GONZALEZ2008}
{Gonz{\'a}lez Delgado}, R.~M., {P{\'e}rez}, E., {Cid Fernandes}, R., \&
  {Schmitt}, H. 2008, \aj, 135, 747

\bibitem[{{Graham} \& {Soria}(2019)}]{GRAHAM2019}
{Graham}, A.~W. \& {Soria}, R. 2019, \mnras, 484, 794

\bibitem[{{Graham} \& {Spitler}(2009)}]{GRAHAM2009}
{Graham}, A.~W. \& {Spitler}, L.~R. 2009, \mnras, 397, 2148

\bibitem[{{Granato} {et~al.}(2004){Granato}, {De Zotti}, {Silva}, {Bressan}, \&
  {Danese}}]{GRANATO2004}
{Granato}, G.~L., {De Zotti}, G., {Silva}, L., {Bressan}, A., \& {Danese}, L.
  2004, \apj, 600, 580

\bibitem[{{Gray} {et~al.}(2024){Gray}, {Read}, {Taylor}, {Orkney}, {Rey},
  {Yates}, {Kim}, {No{\"e}l}, {Agertz}, {Andersson}, \& {Pontzen}}]{GRAY2024}
{Gray}, E.~I., {Read}, J.~I., {Taylor}, E., {et~al.} 2024, arXiv e-prints,
  arXiv:2405.19286

\bibitem[{{G{\"u}ltekin} {et~al.}(2009{\natexlab{a}}){G{\"u}ltekin},
  {Richstone}, {Gebhardt}, {Lauer}, {Pinkney}, {Aller}, {Bender}, {Dressler},
  {Faber}, {Filippenko}, {Green}, {Ho}, {Kormendy}, \& {Siopis}}]{GULTEKIN2009}
{G{\"u}ltekin}, K., {Richstone}, D.~O., {Gebhardt}, K., {et~al.}
  2009{\natexlab{a}}, \apj, 695, 1577

\bibitem[{{G{\"u}ltekin} {et~al.}(2009{\natexlab{b}}){G{\"u}ltekin},
  {Richstone}, {Gebhardt}, {Lauer}, {Tremaine}, {Aller}, {Bender}, {Dressler},
  {Faber}, {Filippenko}, {Green}, {Ho}, {Kormendy}, {Magorrian}, {Pinkney}, \&
  {Siopis}}]{GULTEKIN2009b}
{G{\"u}ltekin}, K., {Richstone}, D.~O., {Gebhardt}, K., {et~al.}
  2009{\natexlab{b}}, \apj, 698, 198

\bibitem[{{G{\"u}rkan} {et~al.}(2004){G{\"u}rkan}, {Freitag}, \&
  {Rasio}}]{GURKAN2004}
{G{\"u}rkan}, M.~A., {Freitag}, M., \& {Rasio}, F.~A. 2004, \apj, 604, 632

\bibitem[{{Haiman}(2004)}]{HAIMAN2004B}
{Haiman}, Z. 2004, \apj, 613, 36

\bibitem[{{Haiman} {et~al.}(2004){Haiman}, {Ciotti}, \&
  {Ostriker}}]{HAIMAN2004}
{Haiman}, Z., {Ciotti}, L., \& {Ostriker}, J.~P. 2004, \apj, 606, 763

\bibitem[{{Hannah} {et~al.}(2021){Hannah}, {Seth}, {Nguyen}, {Dumont},
  {Kacharov}, {Neumayer}, \& {den Brok}}]{HANNAH2021}
{Hannah}, C.~H., {Seth}, A.~C., {Nguyen}, D.~D., {et~al.} 2021, \aj, 162, 281

\bibitem[{{Henriques} {et~al.}(2009){Henriques}, {Thomas}, {Oliver}, \&
  {Roseboom}}]{HENRIQUES2009}
{Henriques}, B. M.~B., {Thomas}, P.~A., {Oliver}, S., \& {Roseboom}, I. 2009,
  \mnras, 396, 535

\bibitem[{{Henriques} {et~al.}(2015){Henriques}, {White}, {Thomas}, {Angulo},
  {Guo}, {Lemson}, {Springel}, \& {Overzier}}]{HENRIQUES2015}
{Henriques}, B. M.~B., {White}, S. D.~M., {Thomas}, P.~A., {et~al.} 2015,
  \mnras, 451, 2663

\bibitem[{Hirschmann {et~al.}(2012)Hirschmann, Naab, Somerville, Burkert, \&
  Oser}]{HIRSCHMANN2012}
Hirschmann, M., Naab, T., Somerville, R.~S., Burkert, A., \& Oser, L. 2012,
  Monthly Notices of the Royal Astronomical Society, 419, 3200

\bibitem[{{Inayoshi} {et~al.}(2020){Inayoshi}, {Visbal}, \&
  {Haiman}}]{INAYOSHI2020}
{Inayoshi}, K., {Visbal}, E., \& {Haiman}, Z. 2020, \araa, 58, 27

\bibitem[{{Jarrett} {et~al.}(2003){Jarrett}, {Chester}, {Cutri}, {Schneider},
  \& {Huchra}}]{JARRET2003}
{Jarrett}, T.~H., {Chester}, T., {Cutri}, R., {Schneider}, S.~E., \& {Huchra},
  J.~P. 2003, \aj, 125, 525

\bibitem[{{Karachentsev} {et~al.}(2013){Karachentsev}, {Makarov}, \&
  {Kaisina}}]{KARACHENTSEV2013}
{Karachentsev}, I.~D., {Makarov}, D.~I., \& {Kaisina}, E.~I. 2013, \aj, 145,
  101

\bibitem[{{Knebe} {et~al.}(2018){Knebe}, {Stoppacher}, {Prada}, {Behrens},
  {Benson}, {Cora}, {Croton}, {Padilla}, {Ruiz}, {Sinha}, {Stevens},
  {Vega-Mart{\'\i}nez}, {Behroozi}, {Gonzalez-Perez}, {Gottl{\"o}ber},
  {Klypin}, {Yepes}, {Enke}, {Libeskind}, {Riebe}, \& {Steinmetz}}]{KNEBE2018}
{Knebe}, A., {Stoppacher}, D., {Prada}, F., {et~al.} 2018, \mnras, 474, 5206

\bibitem[{{Kormendy} \& {Ho}(2013)}]{KORMENDY2013}
{Kormendy}, J. \& {Ho}, L.~C. 2013, \araa, 51, 511

\bibitem[{{Koushiappas} {et~al.}(2004){Koushiappas}, {Bullock}, \&
  {Dekel}}]{Koushiappas2004}
{Koushiappas}, S.~M., {Bullock}, J.~S., \& {Dekel}, A. 2004, \mnras, 354, 292

\bibitem[{{Krajnovi{\'c}} {et~al.}(2018){Krajnovi{\'c}}, {Cappellari},
  {McDermid}, {Thater}, {Nyland}, {de Zeeuw}, {Falc{\'o}n-Barroso}, {Khochfar},
  {Kuntschner}, {Sarzi}, \& {Young}}]{KRAJNOVIC2018}
{Krajnovi{\'c}}, D., {Cappellari}, M., {McDermid}, R.~M., {et~al.} 2018,
  \mnras, 477, 3030

\bibitem[{{Kroupa} {et~al.}(2020){Kroupa}, {Subr}, {Jerabkova}, \&
  {Wang}}]{Kroupa2020}
{Kroupa}, P., {Subr}, L., {Jerabkova}, T., \& {Wang}, L. 2020, \mnras, 498,
  5652

\bibitem[{{Krumholz}(2012)}]{KRUMHOLZ2012}
{Krumholz}, M.~R. 2012, \apj, 759, 9

\bibitem[{{Krumholz} {et~al.}(2009){Krumholz}, {McKee}, \&
  {Tumlinson}}]{KRUMHOLZ2009}
{Krumholz}, M.~R., {McKee}, C.~F., \& {Tumlinson}, J. 2009, \apj, 699, 850

\bibitem[{Landau {et~al.}(1980)Landau, Lifshitz, \& Pitaevskii}]{LANDAU1980}
Landau, L., Lifshitz, E., \& Pitaevskii, L. 1980, Course of Theoretical Physics
  (Pergamon Press)

\bibitem[{{Lapi} {et~al.}(2014){Lapi}, {Raimundo}, {Aversa}, {Cai}, {Negrello},
  {Celotti}, {De Zotti}, \& {Danese}}]{LAPI2014}
{Lapi}, A., {Raimundo}, S., {Aversa}, R., {et~al.} 2014, \apj, 782, 69

\bibitem[{{Latif} \& {Schleicher}(2015)}]{LATIF2015}
{Latif}, M.~A. \& {Schleicher}, D.~R.~G. 2015, \aap, 578, A118

\bibitem[{{Latif} {et~al.}(2014){Latif}, {Schleicher}, {Bovino}, {Grassi}, \&
  {Spaans}}]{LATIF2014}
{Latif}, M.~A., {Schleicher}, D.~R.~G., {Bovino}, S., {Grassi}, T., \&
  {Spaans}, M. 2014, \apj, 792, 78

\bibitem[{{Latif} {et~al.}(2013){Latif}, {Schleicher}, {Schmidt}, \&
  {Niemeyer}}]{LATIF2013}
{Latif}, M.~A., {Schleicher}, D.~R.~G., {Schmidt}, W., \& {Niemeyer}, J. 2013,
  \apjl, 772, L3

\bibitem[{{Latif} {et~al.}(2022){Latif}, {Whalen}, {Khochfar}, {Herrington}, \&
  {Woods}}]{Latif2022}
{Latif}, M.~A., {Whalen}, D.~J., {Khochfar}, S., {Herrington}, N.~P., \&
  {Woods}, T.~E. 2022, \nat, 607, 48

\bibitem[{{Lauer} {et~al.}(2005){Lauer}, {Faber}, {Gebhardt}, {Richstone},
  {Tremaine}, {Ajhar}, {Aller}, {Bender}, {Dressler}, {Filippenko}, {Green},
  {Grillmair}, {Ho}, {Kormendy}, {Magorrian}, {Pinkney}, \&
  {Siopis}}]{LAUER2005}
{Lauer}, T.~R., {Faber}, S.~M., {Gebhardt}, K., {et~al.} 2005, \aj, 129, 2138

\bibitem[{{Leauthaud} {et~al.}(2012){Leauthaud}, {Tinker}, {Bundy}, {Behroozi},
  {Massey}, {Rhodes}, {George}, {Kneib}, {Benson}, {Wechsler}, {Busha},
  {Capak}, {Cort{\^e}s}, {Ilbert}, {Koekemoer}, {Le F{\`e}vre}, {Lilly},
  {McCracken}, {Salvato}, {Schrabback}, {Scoville}, {Smith}, \&
  {Taylor}}]{LEAUTHAUD2012}
{Leauthaud}, A., {Tinker}, J., {Bundy}, K., {et~al.} 2012, \apj, 744, 159

\bibitem[{{Loose} {et~al.}(1982){Loose}, {Kruegel}, \& {Tutukov}}]{LOOSE1982}
{Loose}, H.~H., {Kruegel}, E., \& {Tutukov}, A. 1982, \aap, 105, 342

\bibitem[{{Lotz} {et~al.}(2001){Lotz}, {Telford}, {Ferguson}, {Miller},
  {Stiavelli}, \& {Mack}}]{JENNIFER2001}
{Lotz}, J.~M., {Telford}, R., {Ferguson}, H.~C., {et~al.} 2001, \apj, 552, 572

\bibitem[{{Lupi} {et~al.}(2014){Lupi}, {Colpi}, {Devecchi}, {Galanti}, \&
  {Volonteri}}]{Lupi2014}
{Lupi}, A., {Colpi}, M., {Devecchi}, B., {Galanti}, G., \& {Volonteri}, M.
  2014, \mnras, 442, 3616

\bibitem[{{Maiolino} {et~al.}(2024){Maiolino}, {Scholtz}, {Witstok},
  {Carniani}, {D'Eugenio}, {de Graaff}, {{\"U}bler}, {Tacchella},
  {Curtis-Lake}, {Arribas}, {Bunker}, {Charlot}, {Chevallard}, {Curti},
  {Looser}, {Maseda}, {Rawle}, {Rodr{\'\i}guez del Pino}, {Willott}, {Egami},
  {Eisenstein}, {Hainline}, {Robertson}, {Williams}, {Willmer}, {Baker},
  {Boyett}, {DeCoursey}, {Fabian}, {Helton}, {Ji}, {Jones}, {Kumari},
  {Laporte}, {Nelson}, {Perna}, {Sandles}, {Shivaei}, \& {Sun}}]{MAIOLINO2024}
{Maiolino}, R., {Scholtz}, J., {Witstok}, J., {et~al.} 2024, \nat, 627, 59

\bibitem[{{McConnachie}(2012)}]{MCCONNACHIE2012}
{McConnachie}, A.~W. 2012, \aj, 144, 4

\bibitem[{{Meier}(2001)}]{MEIER2001}
{Meier}, D.~L. 2001, \apjl, 548, L9

\bibitem[{{Miller} \& {Hamilton}(2002)}]{MILLER2002}
{Miller}, M.~C. \& {Hamilton}, D.~P. 2002, \mnras, 330, 232

\bibitem[{{Milosavljevi{\'c}}(2004)}]{MILOSAVLJEVIC2004}
{Milosavljevi{\'c}}, M. 2004, \apjl, 605, L13

\bibitem[{{Mitchell} {et~al.}(2018){Mitchell}, {Lacey}, {Lagos}, {Frenk},
  {Bower}, {Cole}, {Helly}, {Schaller}, {Gonzalez-Perez}, \&
  {Theuns}}]{MITCHELL2018}
{Mitchell}, P.~D., {Lacey}, C.~G., {Lagos}, C. D.~P., {et~al.} 2018, \mnras,
  474, 492

\bibitem[{{Muzzin} {et~al.}(2013){Muzzin}, {Marchesini}, {Stefanon}, {Franx},
  {McCracken}, {Milvang-Jensen}, {Dunlop}, {Fynbo}, {Brammer}, {Labb{\'e}}, \&
  {van Dokkum}}]{MUZZIN2013}
{Muzzin}, A., {Marchesini}, D., {Stefanon}, M., {et~al.} 2013, \apj, 777, 18

\bibitem[{{Nelson} \& {Whittle}(1995)}]{NELSON1995}
{Nelson}, C.~H. \& {Whittle}, M. 1995, \apjs, 99, 67

\bibitem[{{Neumayer} {et~al.}(2020){Neumayer}, {Seth}, \&
  {B{\"o}ker}}]{NEUMAYER2020}
{Neumayer}, N., {Seth}, A., \& {B{\"o}ker}, T. 2020, \aapr, 28, 4

\bibitem[{{Neumayer} \& {Walcher}(2012)}]{NEUMAYER2012}
{Neumayer}, N. \& {Walcher}, C.~J. 2012, Advances in Astronomy, 2012, 709038

\bibitem[{{Neumayer} {et~al.}(2011){Neumayer}, {Walcher}, {Andersen},
  {S{\'a}nchez}, {B{\"o}ker}, \& {Rix}}]{NEUMAYER2011}
{Neumayer}, N., {Walcher}, C.~J., {Andersen}, D., {et~al.} 2011, \mnras, 413,
  1875

\bibitem[{{Nguyen} {et~al.}(2017){Nguyen}, {Seth}, {den Brok}, {Neumayer},
  {Cappellari}, {Barth}, {Caldwell}, {Williams}, \& {Binder}}]{NGUYEN2017}
{Nguyen}, D.~D., {Seth}, A.~C., {den Brok}, M., {et~al.} 2017, \apj, 836, 237

\bibitem[{{Nguyen} {et~al.}(2019){Nguyen}, {Seth}, {Neumayer}, {Iguchi},
  {Cappellari}, {Strader}, {Chomiuk}, {Tremou}, {Pacucci}, {Nakanishi},
  {Bahramian}, {Nguyen}, {den Brok}, {Ahn}, {Voggel}, {Kacharov}, {Tsukui},
  {Ly}, {Dumont}, \& {Pechetti}}]{NGUYEN2019}
{Nguyen}, D.~D., {Seth}, A.~C., {Neumayer}, N., {et~al.} 2019, \apj, 872, 104

\bibitem[{{Nguyen} {et~al.}(2018){Nguyen}, {Seth}, {Neumayer}, {Kamann},
  {Voggel}, {Cappellari}, {Picotti}, {Nguyen}, {B{\"o}ker}, {Debattista},
  {Caldwell}, {McDermid}, {Bastian}, {Ahn}, \& {Pechetti}}]{NGUYEN2018}
{Nguyen}, D.~D., {Seth}, A.~C., {Neumayer}, N., {et~al.} 2018, \apj, 858, 118

\bibitem[{{Parkinson} {et~al.}(2008){Parkinson}, {Cole}, \&
  {Helly}}]{PARKINSON2008}
{Parkinson}, H., {Cole}, S., \& {Helly}, J. 2008, \mnras, 383, 557

\bibitem[{{Pechetti} {et~al.}(2017){Pechetti}, {Seth}, {Cappellari},
  {McDermid}, {den Brok}, {Mieske}, \& {Strader}}]{PECHETTI2017}
{Pechetti}, R., {Seth}, A., {Cappellari}, M., {et~al.} 2017, \apj, 850, 15

\bibitem[{{Pechetti} {et~al.}(2020){Pechetti}, {Seth}, {Neumayer}, {Georgiev},
  {Kacharov}, \& {den Brok}}]{PECHETTI2020}
{Pechetti}, R., {Seth}, A., {Neumayer}, N., {et~al.} 2020, \apj, 900, 32

\bibitem[{{Peterson} \& {Caldwell}(1993)}]{PETERSON1993}
{Peterson}, R.~C. \& {Caldwell}, N. 1993, \aj, 105, 1411

\bibitem[{{Phillipps} \& {Disney}(1986)}]{PHILLIPPS1986}
{Phillipps}, S. \& {Disney}, M. 1986, \mnras, 221, 1039

\bibitem[{{Pinna} {et~al.}(2021){Pinna}, {Neumayer}, {Seth}, {Emsellem},
  {Nguyen}, {B{\"o}ker}, {Cappellari}, {McDermid}, {Voggel}, \&
  {Walcher}}]{PINNA2021}
{Pinna}, F., {Neumayer}, N., {Seth}, A., {et~al.} 2021, \apj, 921, 8

\bibitem[{{Planck Collaboration} {et~al.}(2020){Planck Collaboration},
  {Akrami}, {Arroja}, {Ashdown}, {Aumont}, {Baccigalupi}, {Ballardini},
  {Banday}, {Barreiro}, {Bartolo}, {Basak}, {Benabed}, {Bernard}, {Bersanelli},
  {Bielewicz}, {Bock}, {Bond}, {Borrill}, {Bouchet}, {Boulanger}, {Bucher},
  {Burigana}, {Butler}, {Calabrese}, {Cardoso}, {Carron}, {Challinor},
  {Chiang}, {Colombo}, {Combet}, {Contreras}, {Crill}, {Cuttaia}, {de
  Bernardis}, {de Zotti}, {Delabrouille}, {Delouis}, {Di Valentino}, {Diego},
  {Donzelli}, {Dor{\'e}}, {Douspis}, {Ducout}, {Dupac}, {Dusini}, {Efstathiou},
  {Elsner}, {En{\ss}lin}, {Eriksen}, {Fantaye}, {Fergusson}, {Fernandez-Cobos},
  {Finelli}, {Forastieri}, {Frailis}, {Franceschi}, {Frolov}, {Galeotta},
  {Galli}, {Ganga}, {Gauthier}, {G{\'e}nova-Santos}, {Gerbino}, {Ghosh},
  {Gonz{\'a}lez-Nuevo}, {G{\'o}rski}, {Gratton}, {Gruppuso}, {Gudmundsson},
  {Hamann}, {Handley}, {Hansen}, {Herranz}, {Hivon}, {Hooper}, {Huang},
  {Jaffe}, {Jones}, {Keih{\"a}nen}, {Keskitalo}, {Kiiveri}, {Kim}, {Kisner},
  {Krachmalnicoff}, {Kunz}, {Kurki-Suonio}, {Lagache}, {Lamarre}, {Lasenby},
  {Lattanzi}, {Lawrence}, {Le Jeune}, {Lesgourgues}, {Levrier}, {Lewis},
  {Liguori}, {Lilje}, {Lindholm}, {L{\'o}pez-Caniego}, {Lubin}, {Ma},
  {Mac{\'\i}as-P{\'e}rez}, {Maggio}, {Maino}, {Mandolesi}, {Mangilli},
  {Marcos-Caballero}, {Maris}, {Martin}, {Mart{\'\i}nez-Gonz{\'a}lez},
  {Matarrese}, {Mauri}, {McEwen}, {Meerburg}, {Meinhold}, {Melchiorri},
  {Mennella}, {Migliaccio}, {Mitra}, {Miville-Desch{\^e}nes}, {Molinari},
  {Moneti}, {Montier}, {Morgante}, {Moss}, {M{\"u}nchmeyer}, {Natoli},
  {N{\o}rgaard-Nielsen}, {Pagano}, {Paoletti}, {Partridge}, {Patanchon},
  {Peiris}, {Perrotta}, {Pettorino}, {Piacentini}, {Polastri}, {Polenta},
  {Puget}, {Rachen}, {Reinecke}, {Remazeilles}, {Renzi}, {Rocha}, {Rosset},
  {Roudier}, {Rubi{\~n}o-Mart{\'\i}n}, {Ruiz-Granados}, {Salvati}, {Sandri},
  {Savelainen}, {Scott}, {Shellard}, {Shiraishi}, {Sirignano}, {Sirri},
  {Spencer}, {Sunyaev}, {Suur-Uski}, {Tauber}, {Tavagnacco}, {Tenti},
  {Toffolatti}, {Tomasi}, {Trombetti}, {Valiviita}, {Van Tent}, {Vielva},
  {Villa}, {Vittorio}, {Wandelt}, {Wehus}, {White}, {Zacchei}, {Zibin}, \&
  {Zonca}}]{PLANCK2020}
{Planck Collaboration}, {Akrami}, Y., {Arroja}, F., {et~al.} 2020, \aap, 641,
  A10

\bibitem[{{Portegies Zwart} {et~al.}(2004){Portegies Zwart}, {Baumgardt},
  {Hut}, {Makino}, \& {McMillan}}]{PORTEGIES2004}
{Portegies Zwart}, S.~F., {Baumgardt}, H., {Hut}, P., {Makino}, J., \&
  {McMillan}, S. L.~W. 2004, \nat, 428, 724

\bibitem[{{Portegies Zwart} \& {McMillan}(2002)}]{PORTEGIES2002}
{Portegies Zwart}, S.~F. \& {McMillan}, S. L.~W. 2002, \apj, 576, 899

\bibitem[{{Punturo} {et~al.}(2010){Punturo}, {Abernathy}, {Acernese}, {Allen},
  {Andersson}, {Arun}, {Barone}, {Barr}, {Barsuglia}, {Beker}, {Beveridge},
  {Birindelli}, {Bose}, {Bosi}, {Braccini}, {Bradaschia}, {Bulik}, {Calloni},
  {Cella}, {Chassande Mottin}, {Chelkowski}, {Chincarini}, {Clark}, {Coccia},
  {Colacino}, {Colas}, {Cumming}, {Cunningham}, {Cuoco}, {Danilishin},
  {Danzmann}, {De Luca}, {De Salvo}, {Dent}, {De Rosa}, {Di Fiore}, {Di
  Virgilio}, {Doets}, {Fafone}, {Falferi}, {Flaminio}, {Franc}, {Frasconi},
  {Freise}, {Fulda}, {Gair}, {Gemme}, {Gennai}, {Giazotto}, {Glampedakis},
  {Granata}, {Grote}, {Guidi}, {Hammond}, {Hannam}, {Harms}, {Heinert},
  {Hendry}, {Heng}, {Hennes}, {Hild}, {Hough}, {Husa}, {Huttner}, {Jones},
  {Khalili}, {Kokeyama}, {Kokkotas}, {Krishnan}, {Lorenzini}, {L{\"u}ck},
  {Majorana}, {Mandel}, {Mandic}, {Martin}, {Michel}, {Minenkov}, {Morgado},
  {Mosca}, {Mours}, {M{\"u}ller{\textendash}Ebhardt}, {Murray}, {Nawrodt},
  {Nelson}, {Oshaughnessy}, {Ott}, {Palomba}, {Paoli}, {Parguez},
  {Pasqualetti}, {Passaquieti}, {Passuello}, {Pinard}, {Poggiani}, {Popolizio},
  {Prato}, {Puppo}, {Rabeling}, {Rapagnani}, {Read}, {Regimbau}, {Rehbein},
  {Reid}, {Rezzolla}, {Ricci}, {Richard}, {Rocchi}, {Rowan}, {R{\"u}diger},
  {Sassolas}, {Sathyaprakash}, {Schnabel}, {Schwarz}, {Seidel}, {Sintes},
  {Somiya}, {Speirits}, {Strain}, {Strigin}, {Sutton}, {Tarabrin},
  {Th{\"u}ring}, {van den Brand}, {van Leewen}, {van Veggel}, {van den Broeck},
  {Vecchio}, {Veitch}, {Vetrano}, {Vicere}, {Vyatchanin}, {Willke}, {Woan},
  {Wolfango}, \& {Yamamoto}}]{PUNTURO2010}
{Punturo}, M., {Abernathy}, M., {Acernese}, F., {et~al.} 2010, Classical and
  Quantum Gravity, 27, 194002

\bibitem[{{Querejeta} {et~al.}(2015){Querejeta}, {Meidt}, {Schinnerer},
  {Cisternas}, {Mu{\~n}oz-Mateos}, {Sheth}, {Knapen}, {van de Ven}, {Norris},
  {Peletier}, {Laurikainen}, {Salo}, {Holwerda}, {Athanassoula}, {Bosma},
  {Groves}, {Ho}, {Gadotti}, {Zaritsky}, {Regan}, {Hinz}, {Gil de Paz},
  {Menendez-Delmestre}, {Seibert}, {Mizusawa}, {Kim}, {Erroz-Ferrer}, {Laine},
  \& {Comer{\'o}n}}]{QUEREJETA2015}
{Querejeta}, M., {Meidt}, S.~E., {Schinnerer}, E., {et~al.} 2015, \apjs, 219, 5

\bibitem[{{Quinlan} \& {Shapiro}(1987)}]{QUINLAN1987}
{Quinlan}, G.~D. \& {Shapiro}, S.~L. 1987, \apj, 321, 199

\bibitem[{{Quirk} {et~al.}(2022){Quirk}, {Guhathakurta}, {Gilbert}, {Chemin},
  {Dalcanton}, {Williams}, {Seth}, {Patel}, {Fung}, {Tangirala}, \&
  {Yusufali}}]{QUIRK2022}
{Quirk}, A. C.~N., {Guhathakurta}, P., {Gilbert}, K.~M., {et~al.} 2022, \aj,
  163, 166

\bibitem[{{Rees}(1984)}]{REES1984}
{Rees}, M.~J. 1984, \araa, 22, 471

\bibitem[{{Regan} \& {Downes}(2018)}]{REGAN2018}
{Regan}, J.~A. \& {Downes}, T.~P. 2018, \mnras, 475, 4636

\bibitem[{{Reines} \& {Volonteri}(2015)}]{REINES2015}
{Reines}, A.~E. \& {Volonteri}, M. 2015, \apj, 813, 82

\bibitem[{{Reinoso} {et~al.}(2023){Reinoso}, {Klessen}, {Schleicher}, {Glover},
  \& {Solar}}]{Reinoso2023}
{Reinoso}, B., {Klessen}, R.~S., {Schleicher}, D., {Glover}, S. C.~O., \&
  {Solar}, P. 2023, \mnras, 521, 3553

\bibitem[{{Riaz} {et~al.}(2022){Riaz}, {Hartwig}, \& {Latif}}]{RIAZ2022}
{Riaz}, S., {Hartwig}, T., \& {Latif}, M.~A. 2022, \apjl, 937, L6

\bibitem[{{Rusli} {et~al.}(2011){Rusli}, {Thomas}, {Erwin}, {Saglia}, {Nowak},
  \& {Bender}}]{RUSLI2011}
{Rusli}, S.~P., {Thomas}, J., {Erwin}, P., {et~al.} 2011, \mnras, 410, 1223

\bibitem[{{Saglia} {et~al.}(2016){Saglia}, {Opitsch}, {Erwin}, {Thomas},
  {Beifiori}, {Fabricius}, {Mazzalay}, {Nowak}, {Rusli}, \&
  {Bender}}]{SAGLIA2016}
{Saglia}, R.~P., {Opitsch}, M., {Erwin}, P., {et~al.} 2016, \apj, 818, 47

\bibitem[{{S{\'a}nchez-Janssen} {et~al.}(2019){S{\'a}nchez-Janssen},
  {C{\^o}t{\'e}}, {Ferrarese}, {Peng}, {Roediger}, {Blakeslee}, {Emsellem},
  {Puzia}, {Spengler}, {Taylor}, {{\'A}lamo-Mart{\'\i}nez}, {Boselli},
  {Cantiello}, {Cuillandre}, {Duc}, {Durrell}, {Gwyn}, {MacArthur},
  {Lan{\c{c}}on}, {Lim}, {Liu}, {Mei}, {Miller}, {Mu{\~n}oz}, {Mihos},
  {Paudel}, {Powalka}, \& {Toloba}}]{SANCHEZJANSSEN2019}
{S{\'a}nchez-Janssen}, R., {C{\^o}t{\'e}}, P., {Ferrarese}, L., {et~al.} 2019,
  \apj, 878, 18

\bibitem[{{Sarzi} {et~al.}(2001){Sarzi}, {Rix}, {Shields}, {Rudnick}, {Ho},
  {McIntosh}, {Filippenko}, \& {Sargent}}]{SARZI2001}
{Sarzi}, M., {Rix}, H.-W., {Shields}, J.~C., {et~al.} 2001, \apj, 550, 65

\bibitem[{{Sassano} {et~al.}(2021){Sassano}, {Schneider}, {Valiante},
  {Inayoshi}, {Chon}, {Omukai}, {Mayer}, \& {Capelo}}]{Sassano2021}
{Sassano}, F., {Schneider}, R., {Valiante}, R., {et~al.} 2021, \mnras, 506, 613

\bibitem[{{Satyapal} {et~al.}(2007){Satyapal}, {Vega}, {Heckman}, {O'Halloran},
  \& {Dudik}}]{SATYAPAL2007}
{Satyapal}, S., {Vega}, D., {Heckman}, T., {O'Halloran}, B., \& {Dudik}, R.
  2007, \apjl, 663, L9

\bibitem[{{Schleicher} {et~al.}(2023){Schleicher}, {Reinoso}, \&
  {Klessen}}]{Schleicher2023}
{Schleicher}, D. R.~G., {Reinoso}, B., \& {Klessen}, R.~S. 2023, \mnras, 521,
  3972

\bibitem[{{Schleicher} {et~al.}(2022){Schleicher}, {Reinoso}, {Latif},
  {Klessen}, {Vergara}, {Das}, {Alister}, {D{\'\i}az}, \&
  {Solar}}]{Schleicher2022}
{Schleicher}, D.~R.~G., {Reinoso}, B., {Latif}, M., {et~al.} 2022, \mnras, 512,
  6192

\bibitem[{{Sch{\"o}del} {et~al.}(2014){Sch{\"o}del}, {Feldmeier}, {Kunneriath},
  {Stolovy}, {Neumayer}, {Amaro-Seoane}, \& {Nishiyama}}]{SCHODEL2014}
{Sch{\"o}del}, R., {Feldmeier}, A., {Kunneriath}, D., {et~al.} 2014, \aap, 566,
  A47

\bibitem[{{Schulze} \& {Gebhardt}(2011)}]{SCHULZE2011}
{Schulze}, A. \& {Gebhardt}, K. 2011, \apj, 729, 21

\bibitem[{Scott(2015)}]{SCOTT2015}
Scott, D.~W. 2015, Multivariate density estimation: theory, practice, and
  visualization (John Wiley \& Sons)

\bibitem[{{Scott} \& {Graham}(2013)}]{SCOTT2013}
{Scott}, N. \& {Graham}, A.~W. 2013, \apj, 763, 76

\bibitem[{{S{\'e}rsic}(1963)}]{SERSIC1963}
{S{\'e}rsic}, J.~L. 1963, Boletin de la Asociacion Argentina de Astronomia La
  Plata Argentina, 6, 41

\bibitem[{{Sesana} {et~al.}(2014){Sesana}, {Barausse}, {Dotti}, \&
  {Rossi}}]{SESANA2014}
{Sesana}, A., {Barausse}, E., {Dotti}, M., \& {Rossi}, E.~M. 2014, \apj, 794,
  104

\bibitem[{{Seth} {et~al.}(2008{\natexlab{a}}){Seth}, {Ag{\"u}eros}, {Lee}, \&
  {Basu-Zych}}]{SETH2008}
{Seth}, A., {Ag{\"u}eros}, M., {Lee}, D., \& {Basu-Zych}, A.
  2008{\natexlab{a}}, \apj, 678, 116

\bibitem[{{Seth} {et~al.}(2008{\natexlab{b}}){Seth}, {Blum}, {Bastian},
  {Caldwell}, \& {Debattista}}]{SETH2008B}
{Seth}, A.~C., {Blum}, R.~D., {Bastian}, N., {Caldwell}, N., \& {Debattista},
  V.~P. 2008{\natexlab{b}}, \apj, 687, 997

\bibitem[{{Seth} {et~al.}(2010){Seth}, {Cappellari}, {Neumayer}, {Caldwell},
  {Bastian}, {Olsen}, {Blum}, {Debattista}, {McDermid}, {Puzia}, \&
  {Stephens}}]{SETH2010}
{Seth}, A.~C., {Cappellari}, M., {Neumayer}, N., {et~al.} 2010, \apj, 714, 713

\bibitem[{{Shakura} \& {Sunyaev}(1973)}]{SHAKURA1973}
{Shakura}, N.~I. \& {Sunyaev}, R.~A. 1973, \aap, 24, 337

\bibitem[{{Shapiro} {et~al.}(2006){Shapiro}, {Cappellari}, {de Zeeuw},
  {McDermid}, {Gebhardt}, {van den Bosch}, \& {Statler}}]{SHAPIRO2006}
{Shapiro}, K.~L., {Cappellari}, M., {de Zeeuw}, T., {et~al.} 2006, \mnras, 370,
  559

\bibitem[{{Shen} \& {Gebhardt}(2010)}]{SHEN2010}
{Shen}, J. \& {Gebhardt}, K. 2010, \apj, 711, 484

\bibitem[{{Shlosman} {et~al.}(1990){Shlosman}, {Begelman}, \&
  {Frank}}]{SHLOSMAN1990}
{Shlosman}, I., {Begelman}, M.~C., \& {Frank}, J. 1990, \nat, 345, 679

\bibitem[{{Simien} \& {Prugniel}(2002)}]{SIMIEN2002}
{Simien}, F. \& {Prugniel}, P. 2002, \aap, 384, 371

\bibitem[{{Singh} {et~al.}(2023){Singh}, {Monaco}, \& {Tan}}]{SINGH2023}
{Singh}, J., {Monaco}, P., \& {Tan}, J.~C. 2023, \mnras, 525, 969

\bibitem[{{Spengler} {et~al.}(2017){Spengler}, {C{\^o}t{\'e}}, {Roediger},
  {Ferrarese}, {S{\'a}nchez-Janssen}, {Toloba}, {Liu}, {Guhathakurta},
  {Cuillandre}, {Gwyn}, {Zirm}, {Mu{\~n}oz}, {Puzia}, {Lan{\c{c}}on}, {Peng},
  {Mei}, \& {Powalka}}]{SPENGLER2017}
{Spengler}, C., {C{\^o}t{\'e}}, P., {Roediger}, J., {et~al.} 2017, \apj, 849,
  55

\bibitem[{{Spolaor} {et~al.}(2010){Spolaor}, {Hau}, {Forbes}, \&
  {Couch}}]{SPOLAOR2010}
{Spolaor}, M., {Hau}, G. K.~T., {Forbes}, D.~A., \& {Couch}, W.~J. 2010,
  \mnras, 408, 254

\bibitem[{{Tanaka} {et~al.}(2024){Tanaka}, {Silverman}, {Ding}, {Jahnke},
  {Trakhtenbrot}, {Lambrides}, {Onoue}, {Taufik Andika}, {Bongiorno}, {Faisst},
  {Gillman}, {Hayward}, {Hirschmann}, {Koekemoer}, {Kokorev}, {Liu}, {Magdis},
  {Renzini}, {Casey}, {Drakos}, {Franco}, {Gozaliasl}, {Kartaltepe}, {Liu},
  {McCracken}, {Rhodes}, {Robertson}, \& {Toft}}]{TANAKA2024}
{Tanaka}, T.~S., {Silverman}, J.~D., {Ding}, X., {et~al.} 2024, arXiv e-prints,
  arXiv:2401.13742, submitted to ApJ

\bibitem[{{Tielens}(2010)}]{TIELENS2010}
{Tielens}, A.~G.~G.~M. 2010, {The Physics and Chemistry of the Interstellar
  Medium} (Cambridge University Press)

\bibitem[{{Tremaine} {et~al.}(1975){Tremaine}, {Ostriker}, \&
  {Spitzer}}]{TREMAINE1975}
{Tremaine}, S.~D., {Ostriker}, J.~P., \& {Spitzer}, L., J. 1975, \apj, 196, 407

\bibitem[{{Trinca} {et~al.}(2022){Trinca}, {Schneider}, {Valiante}, {Graziani},
  {Zappacosta}, \& {Shankar}}]{Trinca2022}
{Trinca}, A., {Schneider}, R., {Valiante}, R., {et~al.} 2022, \mnras, 511, 616

\bibitem[{{van den Bosch} \& {de Zeeuw}(2010)}]{VANDENBOSCH2010}
{van den Bosch}, R. C.~E. \& {de Zeeuw}, P.~T. 2010, \mnras, 401, 1770

\bibitem[{{Vergara} {et~al.}(2023){Vergara}, {Escala}, {Schleicher}, \&
  {Reinoso}}]{VERGARA2023}
{Vergara}, M.~C., {Escala}, A., {Schleicher}, D.~R.~G., \& {Reinoso}, B. 2023,
  \mnras, 522, 4224

\bibitem[{{Vergara} {et~al.}(2024){Vergara}, {Schleicher}, {Escala}, {Reinoso},
  {Flammini Dotti}, {Kamlah}, {Liempi}, {Hoyer}, {Neumayer}, \&
  {Spurzem}}]{Vergara2024}
{Vergara}, M.~C., {Schleicher}, D.~R.~G., {Escala}, A., {et~al.} 2024, \aap,
  689, A34

\bibitem[{{Vika} {et~al.}(2009){Vika}, {Driver}, {Graham}, \&
  {Liske}}]{VIKA2009}
{Vika}, M., {Driver}, S.~P., {Graham}, A.~W., \& {Liske}, J. 2009, \mnras, 400,
  1451

\bibitem[{{Volonteri}(2010)}]{VOLONTERI2010}
{Volonteri}, M. 2010, \aapr, 18, 279

\bibitem[{{Volonteri} \& {Rees}(2005)}]{VOLONTERI2005}
{Volonteri}, M. \& {Rees}, M.~J. 2005, \apj, 633, 624

\bibitem[{{Walcher} {et~al.}(2006){Walcher}, {B{\"o}ker}, {Charlot}, {Ho},
  {Rix}, {Rossa}, {Shields}, \& {van der Marel}}]{WALCHER2006}
{Walcher}, C.~J., {B{\"o}ker}, T., {Charlot}, S., {et~al.} 2006, \apj, 649, 692

\bibitem[{{Walcher} {et~al.}(2005){Walcher}, {van der Marel}, {McLaughlin},
  {Rix}, {B{\"o}ker}, {H{\"a}ring}, {Ho}, {Sarzi}, \& {Shields}}]{WALCHER2005}
{Walcher}, C.~J., {van der Marel}, R.~P., {McLaughlin}, D., {et~al.} 2005,
  \apj, 618, 237

\bibitem[{{Wehner} \& {Harris}(2006)}]{WEHNER2006}
{Wehner}, E.~H. \& {Harris}, W.~E. 2006, \apjl, 644, L17

\bibitem[{{Whalen} \& {Fryer}(2012)}]{WHALEN2012}
{Whalen}, D.~J. \& {Fryer}, C.~L. 2012, \apjl, 756, L19

\bibitem[{{Williams} {et~al.}(2017){Williams}, {Dolphin}, {Dalcanton}, {Weisz},
  {Bell}, {Lewis}, {Rosenfield}, {Choi}, {Skillman}, \&
  {Monachesi}}]{WILLIAMS2017}
{Williams}, B.~F., {Dolphin}, A.~E., {Dalcanton}, J.~J., {et~al.} 2017, \apj,
  846, 145

\bibitem[{{Williams}(1977)}]{WILLIAMS1977}
{Williams}, T.~B. 1977, \apj, 214, 685

\bibitem[{{Wise} {et~al.}(2008){Wise}, {Turk}, \& {Abel}}]{Wise2008}
{Wise}, J.~H., {Turk}, M.~J., \& {Abel}, T. 2008, \apj, 682, 745

\bibitem[{{Woods} {et~al.}(2019){Woods}, {Agarwal}, {Bromm}, {Bunker}, {Chen},
  {Chon}, {Ferrara}, {Glover}, {Haemmerl{\'e}}, {Haiman}, {Hartwig}, {Heger},
  {Hirano}, {Hosokawa}, {Inayoshi}, {Klessen}, {Kobayashi}, {Koliopanos},
  {Latif}, {Li}, {Mayer}, {Mezcua}, {Natarajan}, {Pacucci}, {Rees}, {Regan},
  {Sakurai}, {Salvadori}, {Schneider}, {Surace}, {Tanaka}, {Whalen}, \&
  {Yoshida}}]{WOODS2019}
{Woods}, T.~E., {Agarwal}, B., {Bromm}, V., {et~al.} 2019, \pasa, 36, e027

\end{thebibliography}

\begin{appendix}

\section{Mass resolution and convergence tests}\label{MassResolutionTest}

We checked the convergence of the {\sc Galacticus} output running the same models listed in Table~ \ref{InitialParameters} changing the mass resolution.

\begin{figure}[!h]
    \centering
\includegraphics{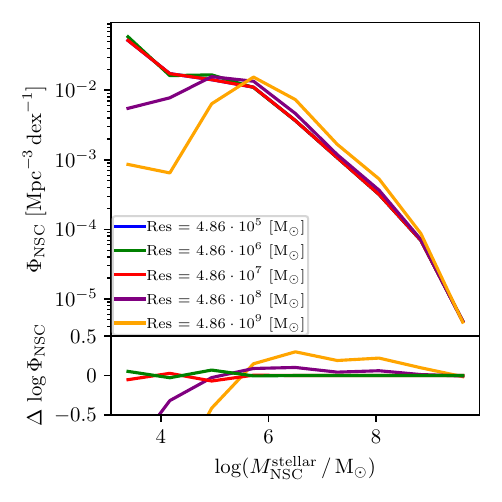}
    \caption{The top panel is  the NSC mass function  for different mass resolutions specified in the legend. The bot panel is the error  relative to the model A4 with the higher mass resolution ($4.86\times 10^{5}$~M$_\odot$ and color blue).   }
    \label{NSCMFResA4}
\end{figure}

In Fig. \ref{NSCMFResA4} top panel shows the NSC mass function for mass resolutions in the order of $10^5$,~$10^6$,~$10^7$,~$10^8$, and  $10^9$~M$_\odot$. We note that mass functions show a different shape in lower resolutions (that means solving DM halos with masses equal or lager than the resolution).

Although the mass function is converged for NSC with stellar masses in the range $10^7-10^{10}$~M$_\odot$, it starts to converge in the low regime in resolutions higher than $\sim10^7$~M$_\odot$. As shown in the bottom panel of the Fig. \ref{NSCMFResA4}, the relative error of $\Phi_\mathrm{NSC}$ (relative to the model A1 with a resolution of $4.86\times 10^{5}$~M$_\odot$) is less than $\Delta \log{\Phi_\mathrm{NSC}} <0.5$, and decreases as the mass resolution increases.

\begin{figure}[!h]
    \centering
\includegraphics{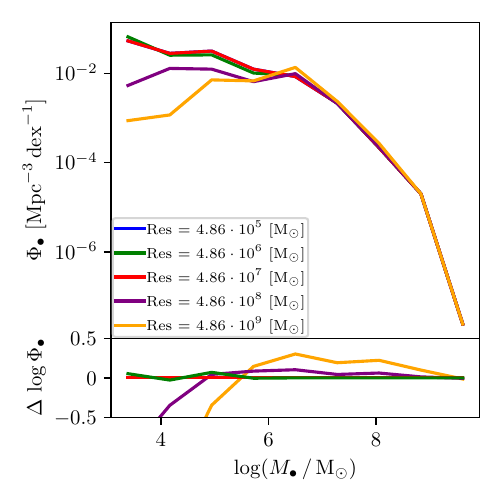}
    \caption{In top panel the BH mass function is shown for different mass resolutions specified in the legend. The error is relative to the model A4 with with the higher mass  resolution ($4.86\times 10^{5}$~M$_\odot$).   }
    \label{BHMFResA4}
\end{figure}

We also check the convergence for the BH mass function. As the model A1 do not form nay seed, we show the results of the convergence test for model A4 in Figs.~\ref{NSCMFResA4},~\ref{BHMFResA4}. We found that the convergence of the BH mass function begins in high mass resolutions, such as $4.86\times 10^7$, $\times 10^6$, and $\times 10^5$~M$_\odot$. Because of the agreement in the convergence of the NSC and BH mass functions we  adopt a resolution equals to $4.86\times10^5~$M$_\odot$.

\FloatBarrier
\section{Galaxy and black hole mass function  deviations}\label{GalaxyMassFunctionComparison}

 We compare the galaxy mass function of \cite{BALDRY2012} with the mass function of galaxies hosting a NSC with stellar mass over $10^{3}$~M$_\odot$ predicted by {\sc Galacticus}.

In order to quantify how deviate from observations are our models  we define $q = \log{\left(\frac{\Phi_{\rm model}}{\Phi_{\rm obs}} \right)}$ at each bin. Afterwards, we compute the average and the maximum (minimum) value of $q$ per model.

\begin{figure}[h!]
    \centering
\includegraphics{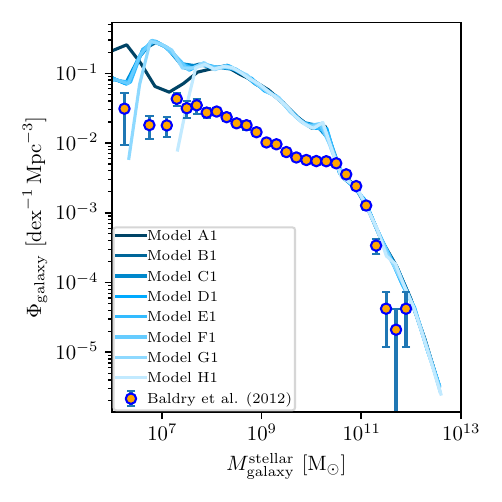}
    \caption{Galaxy stellar mass function at $z=0$ for 8 models run in {\sc Galacticus} (solid lines) and the galaxy stellar mass function of \cite{BALDRY2012} (orange dots). }
    \label{fig:MassfunctionAppendix}
\end{figure}

From Fig. \ref{fig:MassfunctionAppendix} we find (in average) an overestimation of the observed galaxy stellar mass function in the order of $0.5-0.6\pm 0.3$~dex, except for model H1, which  shows the lowest (averaged) overestimation but larger standard deviation ($0.33\pm0.5$~dex).  

It is important to note that the major discrepancy is observed at $M_{\rm galaxy}^{\rm stellar} < 10^{8}$~M$_\odot$. We stress that the observed mass function should be considered as a lower limit due to the {\sc GAMA} survey limitations explained in Sect. \ref{MF}. As a result, our calculation of the deviation in this range represents an upper limit, meaning that any future refinements to the models will only improve the agreement.

For the BHMF we listed the averages and maximum (minimum) values, in Table~\ref{tab:model_deviations}.

\begin{table}[h!]
\centering
\caption{Summary of deviations for various models across different black hole mass ranges. }
\begin{tabular}{cccc}
\hline \hline
 Model &  average$(q)$ &  max($q)$ &  min($q)$   \\
 - & [dex] & [dex] & [dex] \\
 (1) & (2) & (3) & (4) \\ \hline
  A2 & $+0.68 \pm 1.02$ & 2.88  & $-0.64$ \\ 
  A3 & $-0.34\pm 1.32$ & 2.43  & $-1.68$ \\ 
  A4 & $-0.13\pm 1.17$ & 2.45  & $-1.68$ \\ 
  D3 & $+0.05 \pm 1.20$ & 2.67  & $-1.70$ \\ 
  D4 & $-0.19\pm 1.24$ & 2.34  & $-2.08$ \\ 
  D7 & $+0.23 \pm 1.13$ & 2.73  & $-1.34$  \\ 
  D8 & $-0.18\pm 1.24$ & 2.34  & $-2.08$ \\ 
  E2 & $+0.78 \pm 0.99$ & 2.99  & $-0.46$ \\ 
  E3 & $+0.03 \pm 1.19$ & 2.65  & $-1.67$  \\ 
  E4 & $-0.22\pm 1.23$ & 2.35  & $-2.06$ \\ 
  F2 & $+0.55 \pm 1.09$ & 2.99  & $-0.96$ \\ 
  F3 & $+0.30 \pm 1.12$ & 2.83  & $-1.18$ \\ 
  F4 & $+0.30 \pm 1.10$ & 2.23  & $-1.60$ \\
  G2 & $+0.71 \pm 1.00$ & 2.98  & $-0.45$ \\
  G3 & $+0.02 \pm 1.18$ & 2.43  & $-1.68$ \\ 
  G4 & $-0.35\pm 1.04$ & 1.96  & $-1.72$ \\ 
  H2 & $+0.55 \pm 0.93$ & 2.43  & $-0.65$ \\ 
  H3 & $-0.18\pm 1.02$ & 2.00  & $-1.49$ \\ 
  H4 & $-0.30\pm 1.04$ & 2.19  & $-1.55$ \\ \hline\hline
\end{tabular}
\tablefoot{(1) Model name; (2) average value of q of the model; (3) maximum value of q; (4) minimum value of q. All the quantities are in dex units.}
\label{tab:model_deviations}
\end{table}

\section{Probability density function} \label{PDF}

For the different fits made to the  datasets we estimate the Probability Density Function (PDF) using the Kernel Density Estimation (KDE) non-parametric technique using the {\sc SciPy} package available for {\sc Python}.  We encourage readers to review the mathematical formulation in \cite{SCOTT2015}.

\begin{figure}[!h]
    \centering
\includegraphics{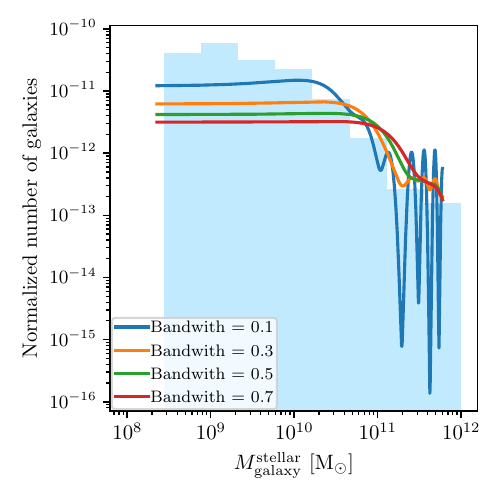}
    \caption{The y-axis shows the normalized number of galaxies as a function of their stellar mass. We plot the different values of the bandwidth parameter which regulates the fit using the KDE technique.}
    \label{KDEGALMASS}
\end{figure}

\begin{figure}
    \centering
    \includegraphics[width=\hsize]{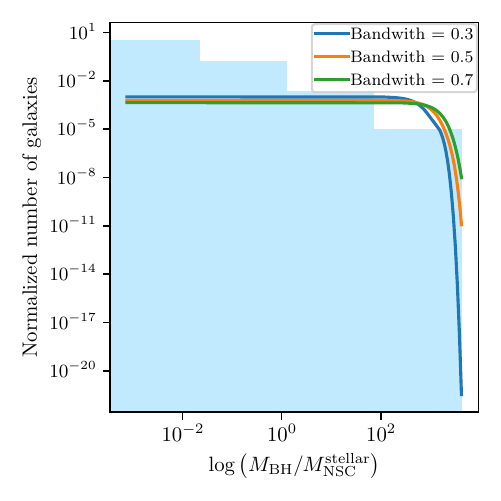}
    \caption{Same as Fig. \ref{KDEGALMASS} but showing the normalized distribution of galaxies as a function of the ratio between the mass of the black hole and the stellar mass of the NSC.}
    \label{KDERATIO}
\end{figure}

We fit the distribution of the stellar mass of the galaxy and the ratio between the $M_\mathrm{BH}$ and the $M_\mathrm{NSC}^\mathrm{stellar}$ assuming a Gaussian kernel. An important parameter to choose when the PDF is estimated is the bandwidth.
In Fig. \ref{KDEGALMASS} and Fig. \ref{KDERATIO}, we show the PDF as a function of the bandwidth for the observed distribution. In both cases we choose the value of $0.5$ as it recovers the shape of the original distribution and provides a smooth fit to the observed data.

\section{Observed properties of galaxies hosting NSCs and SMBHs}\label{appendixD}

In Table~\ref{galaxyprop}, we provide a compilation of the properties of the galaxies hosting NSCs and SMBHs based on the sample of \citet{NEUMAYER2020}, but additionally adding the velocity dispersion from the host galaxy employing literature data.

\begin{table*}[h]
\centering
\small
\sidecaption
 \caption[]{\label{galaxyprop} Properties of galaxies hosting NSCs and SMBHs. }
\begin{tabular}{llllll}
 \hline \hline
\multicolumn{1}{c}{Galaxy} & \multicolumn{1}{c}{$M_\mathrm{BH}$}  & \multicolumn{1}{c}{$M_\mathrm{NSC}^\mathrm{stellar}$}  & \multicolumn{1}{c}{$M_\mathrm{galaxy}^\mathrm{stellar}$} & \multicolumn{1}{c}{$\sigma_\mathrm{e}$} & \multicolumn{1}{c}{References}\\
\multicolumn{1}{c}{-} & \multicolumn{1}{c}{[M$_\odot$]} & \multicolumn{1}{c}{[M$_\odot$]} & \multicolumn{1}{c}{[M$_\odot$]} & \multicolumn{1}{c}{$[\mathrm{km}\,\mathrm{s}^{-1}]$} & \multicolumn{1}{c}{-} \\ 
\multicolumn{1}{c}{(1)} & \multicolumn{1}{c}{(2)} & \multicolumn{1}{c}{(3)} & \multicolumn{1}{c}{(4)} & \multicolumn{1}{c}{(5)} & \multicolumn{1}{c}{(6)} \\ \hline
MW       & $4.30^{+0.36}_{-0.36}\times 10^{6}$ & $2.50^{+0.40}_{-0.40}\times 10^7$  & $6.00^{+1.00}_{-1.00}\times 10^{10}$ & $100^{+20}_{-20}$ & 1;2;3;4 \\
M31      & $1.40^{+0.90}_{-0.30}\times 10^{8}$ & $3.50^{+0.80}_{-0.80}\times 10^{7}$ & $9.00^{+2.00}_{-2.00}\times 10^{10}$ & $169^{+8}_{-8}$     & 5;6;7;5    \\
NGC 205  & $6.80^{+95.6}_{-6.70}\times 10^{3}$ & $1.80^{+0.80}_{-0.80}\times 10^{6}$ & $9.72^{+1.52}_{-1.52}\times 10^{9} $ &  $20^{+1}_{-1}$     & 8;9;9;10   \\
M32      & $2.50^{+0.50}_{-0.50}\times 10^{6}$ & $1.45^{+0.24}_{-0.24}\times 10^{7}$ & $1.00                \times 10^{9} $ &  $77^{+3}_{-3}$     & 9;9;9;11   \\
NGC 404  & $2.00^{+1.25}_{-0.75}\times 10^{5}$ & $1.10^{+0.20}_{-0.20}\times 10^{7}$ & $9.20                \times 10^{8} $ &  $40^{+3}_{-3}$     & 12;13;13;14\\
NGC 5102 & $8.80^{+4.20}_{-6.60}\times 10^{5}$ & $7.30^{+2.34}_{-2.34}\times 10^{7}$ & $5.92^{+0.83}_{-0.83}\times 10^{9} $ &  60                 & 9;9;9;9    \\
NGC 5206 & $6.31_{-2.74}^{+1.06}\times 10^{5}$ & $1.54^{+0.51}_{-0.51}\times 10^{7}$ & $2.41^{+0.47}_{-0.47}\times 10^{9} $ &  $39^{+5}_{-5}$     & 8;9;9;15   \\
NGC 4395 & $4.10^{+8.00}_{-3.00}\times 10^{5}$ & $2.00                \times 10^{6}$ & $7.94^{+7.91}_{-1.09}\times 10^{9} $ &  $30^{+5}_{-5}$     & 16;16;17;18\\
NGC 4578 & $1.90^{+0.60}_{-1.40}\times 10^{7}$ & $5.25                \times 10^{7}$ & $2.82^{+0.18}_{-0.17}\times 10^{10}$ & $116^{+3}_{-3}$     & 19;20;21;10\\
NGC 1023 & $3.90^{+0.40}_{-0.40}\times 10^{7}$ & $4.40                \times 10^{6}$ & $4.27                \times 10^{10}$ & $205^{+0}_{-0}$     & 22;23;17;22\\
NGC 3115 & $0.94^{+1.06}_{-2.90}\times 10^{9}$ & $1.50                \times 10^{7}$ & $7.40                \times 10^{10}$ & $230^{+11}_{-11}$   & 24;23;23;24\\
NGC 3384 & $1.60^{+0.10}_{-0.20}  \times 10^{7}$ & $2.20                \times 10^{7}$ & $1.40                \times 10^{10}$ & $146^{+10}_{-10}$   & 25;23;23;26\\
NGC 4026 & $2.10^{+0.70}_{-0.40}  \times 10^{8}$ & $5.60                \times 10^{6}$ & $2.23^{+0.46}_{-0.38}\times 10^{10}$ & $180^{+9}_{-9}$     & 27;23;28;29\\
NGC 4697 & $1.70^{+0.20}_{-0.10}\times 10^{8}$ & $2.80                \times 10^{7}$ & $1.50                \times 10^{11}$ & $177^{+8}_{-8}$     & 25;23;23;26\\
IC 1459  & $2.48^{+0.49}_{-0.41}\times 10^{9}$ & $4.89                \times 10^{7}$ & $2.52^{+0.48}_{-0.41}\times 10^{11}$ & $331^{+5}_{-5}$     & 28;30;28;31\\
NGC 4552 & $5.00^{+0.62}_{-0.45}\times 10^{8}$ & $9.18                \times 10^{6}$ & $2.63^{+0.74}_{-0.57}\times 10^{11}$ & $275^{+15}_{-15}$   & 28;30;28;32\\
NGC 4558 & $1.03_{-0.83}^{+0.84}\times 10^{8}$ & $1.05                \times 10^{8}$ & $1.07                \times 10^{10}$ & $102.8$             & 33;34;21;35\\
M33      & $1.50                \times 10^{3}$ & $2.00                \times 10^{6}$ & $2.80                \times 10^{9} $ & $16^{+0.3}_{-0.2}$  & 36;23;37;38\\
NGC 4474 & $1.50                \times 10^{6}$ & $1.00                \times 10^{8}$ & $1.55                \times 10^{10}$ &  $87^{+3}_{-3}$     & 19;34;21;10\\
NGC 4551 & $5.00                \times 10^{6}$ & $4.35                \times 10^{7}$ & $1.81                \times 10^{10}$ & $109$               & 19;34;21;35\\
NGC 300  & $1.00                \times 10^{5}$ & $1.05_{-0.45}^{+0.77}\times 10^{6}$ & $2.75                \times 10^{9} $ & $13.3^{+2}_{-2}$    & 39;40;9;40\\
NGC 428  & $7.00                \times 10^{4}$ & $3.24_{-0.90}^{+1.23}\times 10^{6}$ & $8.13                \times 10^{9} $ & $24.4^{+3.7}_{-3.7}$& 39;40;9;40\\
NGC 1042 & $3.00                \times 10^{6}$ & $3.24_{-1.24}^{+2.04}\times 10^{6}$ & $1.51                \times 10^{9} $ & $32^{+4.8}_{-4.8}  $& 39;40;9;40\\
NGC 1493 & $8.00                \times 10^{5}$ & $2.40^{+0.91}_{-0.66}\times 10^{6}$ & $4.27                \times 10^{9} $ & $25^{+3.8}_{-3.8}$  & 39;40;9;39\\
NGC 2139 & $4.00                \times 10^{5}$ & $8.31^{+4.90}_{-3.10}\times 10^{5}$ & $2.00                \times 10^{10}$ & $16.5^{-2.5}_{-2.5}$& 39;40;9;39\\
NGC 3423 & $7.00                \times 10^{5}$ & $3.39^{+1.21}_{-0.93}\times 10^{6}$ & $6.61                \times 10^{9} $ & $30.4^{+4.6}_{-4.6}$& 39;40;9;39\\
NGC 7418 & $9.00                \times 10^{6}$ & $6.03^{+3.30}_{-2.10}\times 10^{7}$ & $1.20                \times 10^{10}$ & $34.1^{+5.1}_{-5.1}$& 39;40;9;39\\
NGC 7424 & $4.00                \times 10^{5}$ & $1.23^{+0.46}_{-3.40}\times 10^{6}$ & $1.41                \times 10^{9} $ & $15.6^{+2.3}_{-2.3}$& 39;40;9;39\\
NGC 7793 & $8.00                \times 10^{5}$ & $7.77^{+2.95}_{-2.19}\times 10^{6}$ & $4.57                \times 10^{9} $ & $24.6^{+3.7}_{-3.7}$& 39;40;9;39\\
VCC 1254 & $9.00                \times 10^{6}$ & $1.10                \times 10^{7}$ & $3.20                \times 10^{9} $ & $46$                & 41;23;23;41\\
NGC 3621 & $3.00                \times 10^{6}$ & $1.00                \times 10^{7}$ & $1.51                \times 10^{10}$ & $102$               & 41;42;43;44\\
IC  342  & $5.00                \times 10^{5}$ & $1.26                \times 10^{7}$ & $1.41                \times 10^{11}$ & $33$                & 45;45;46;45\\
NGC 2778 & $1.77_{-1.77}^{+3.23}\times 10^{8}$ & $2.29                \times 10^{7}$ & $3.16                \times 10^{10}$ & $175$               & 33;30;21;26\\
NGC 4379 & $2.06_{-1.47}^{+1.49}\times 10^{8}$ & $4.79                \times 10^{7}$ & $1.86                \times 10^{10}$ & $110^{+11}_{-11}$    & 33;34;21;47\\
NGC 4387 & $3.89                \times 10^{7}$ & $3.49                \times 10^{7}$ & $1.51                \times 10^{10}$ & $100_{-10}^{+10}$   & 33;34;21;47\\
NGC 4612 & $3.69^{+9.31}_{-0.00}\times 10^{7}$ & $8.91                \times 10^{6}$ & $1.86                \times 10^{10}$ & $86_{-8}^{+8}$      & 33;34;21;47\\
NGC 4623 & $7.91_{-7.41}^{+8.09}\times 10^{7}$ & $1.29                \times 10^{8}$ & $1.48                \times 10^{10}$ &  $77^{+7}_{-7}$     & 33;20;21;47\\
NGC 4486 & $6.60^{+0.40}_{-0.40}\times 10^{9}$ & $2.00                \times 10^{8}$ & $6.00                \times 10^{11}$ & $323^{+32}_{-32}$   & 48;39;39;47\\
NGC 4374 & $1.5_{-0.60}^{+1.10} \times 10^{9}$ & $6.30                \times 10^{7}$ & $3.60                \times 10^{11}$ & $78^{+27}_{-27}$    & 49;39;39;47\\
NGC 1332 & $1.45_{-0.20}^{+0.20}\times 10^{9}$ & $1.40                \times 10^{7}$ & $4.89                \times 10^{11}$ & $328$               & 50;39;9;50\\
NGC 3031 & $7.00^{+2.00}_{-1.00}\times 10^{7}$ & $7.00                \times 10^{6}$ & $6.31                \times 10^{10}$ & $150$               & 51;39;52;53\\
NGC 4261 & $4.90^{+1.00}_{-1.00}\times 10^{8}$ & $1.70                \times 10^{6}$ & $3.60                \times 10^{11}$ & $315_{-15}^{+15}$   & 54;39;39;54\\
NGC 4649 & $2.00_{-0.60}^{+0.40}\times 10^{9}$ & $2.00                \times 10^{6}$ & $4.90                \times 10^{11}$ & $380_{-19}^{+19}$   & 25;39;39;55\\
NGC 3998 & $2.70^{+2.40}_{-2.00}\times 10^{8}$ & $8.50                \times 10^{5}$ & $4.57                \times 10^{10}$ & $333_{-7}^{+7}$     & 56;39;43;57\\
NGC 2787 & $7.00_{-0.90}^{+0.70}\times 10^{7}$ & $1.90                \times 10^{6}$ & $1.38                \times 10^{10}$ & $189_{-10}^{+10}$   & 58;39;43;58\\
NGC 3379 & $1.40^{+2.60}_{-1.00}\times 10^{8}$ & $1.40                \times 10^{4}$ & $6.76                \times 10^{10}$ & $187_{-10}^{+10}$   & 59;39;39;60\\
NGC 4342 & $3.00^{+1.70}_{-1.00}\times 10^{8}$ & $2.50                \times 10^{6}$ & $1.29                \times 10^{11}$ & $225_{-9}^{+9}$     & 61;39;39;61\\
NGC 4291 & $3.10_{-2.30}^{+0.80} \times 10^{8}$ & $5.00                \times 10^{6}$ & $6.39                \times 10^{10}$ & $242_{-12}^{+12}$   & 25;39;39;26\\
 \hline \hline
\end{tabular}
\tablebib{~(1) \citet{GENZEL2010}; (2) \citet{SCHODEL2014}; (3) \citet{BLAND2016}; (4) \citet{FERRARESE2000}; (5) \citet{BENDER2005}; (6) \citet{KORMENDY2013}; (7) \citet{WILLIAMS2017}; (8) \citet{NGUYEN2019}; (9) \citet{NGUYEN2018}; (10) \citet{SIMIEN2002}; (11) \citet{VANDENBOSCH2010}; (12) \citet{NGUYEN2017}; (13) \citet{SETH2010}; (14) \citet{BARTH2002}; (15) \citet{PETERSON1993}; (16) \citet{DENBROK2015}; (17) \citet{REINES2015}; (18) \citet{FILIPPENKO2003}; (19) \citet{KRAJNOVIC2018}; (20) \citet{COTE2006}; (21) \citet{CAPPELLARI2013}; (22) \citet{BOWER2001}; (23) \citet{GRAHAM2009}; (24) \citet{EMSELLEM1999}; (25) \citet{GEBHARDT2003}; (26) \citet{SCHULZE2011}; (27) \citet{GULTEKIN2009}; (28) \citet{SAGLIA2016}; (29) \citet{GULTEKIN2009b}; (30) \citet{LAUER2005}; (31) \citet{CAPPELLARI2002}; (32) \citet{FABER1976}; (33) \citet{PECHETTI2017}; (34) \citet{SPENGLER2017}; (35) \citet{SPOLAOR2010}; (36) \citet{GEBHARDT2001}; (37) \citet{MCCONNACHIE2012}; (38) \citet{QUIRK2022}; (39) \citet{NEUMAYER2012}; (40) \citet{WALCHER2005}; (41) \citet{GEHA2002}; (42) \citet{BARTH2009}; (43) \citet{JARRET2003}; (44) \citet{SATYAPAL2007}; (45) \citet{BOKER1999}; (46) \citet{KARACHENTSEV2013}; (47) \citet{GRAHAM2019}; (48) \citet{GEBHARDT2011}; (49) \citet{BOWER1998}; (50) \citet{RUSLI2011}; (51) \citet{DEVEREUX2003}; (52) \citet{QUEREJETA2015}; (53) \citet{WILLIAMS1977}; (54) \citet{FERRARESE1996}; (55) \citet{SHEN2010}; (56) \citet{DEFRANCESCO2006}; (57) \citet{NELSON1995}; (58) \citet{SARZI2001}; (59) \citet{SHAPIRO2006}; (60) \citet{BURBIDGE1961}; (61) \citet{CRETTON1999}.}
\tablefoot{(1) Galaxy name; (2) black hole mass; (3) stellar mass of the NSC; (4) stellar mass of the host galaxy; (5) velocity dispersion of the bulge/pseudo-bulge/nuclei when corresponding; (6) references. All the columns show the estimated error when available in literature.}
\end{table*}

\end{appendix}
\end{document}